\documentclass[aps,prb,twocolumn,showkeys,longbibliography]{revtex4-2}
\usepackage[plainpages = false, pdfpagelabels, 
                 pdfpagelayout = useoutlines, 
                 bookmarksopen = true,
                 bookmarksnumbered = true,
                 breaklinks = true,
                 linktocpage,
                 colorlinks = true,  
                 linkcolor = blue,
                 urlcolor  = blue,
                 citecolor = blue,
                 anchorcolor = green,
                 hyperindex = true,
                 hyperfigures
                 ]{hyperref} 

\usepackage{enumerate}
\usepackage[caption=false,singlelinecheck=off]{subfig}
\usepackage{slashed}
\usepackage{braket}
\usepackage{subfloat}
\usepackage{bbold}
\usepackage{booktabs}
\usepackage{blindtext}
\usepackage{amsfonts}
\usepackage{bbm}
\usepackage{graphicx}
\usepackage{dcolumn}
\usepackage{bm}
\usepackage{amsmath, amssymb}
\usepackage[T1]{fontenc}
\usepackage[latin1]{inputenc}
\usepackage[normalem]{ulem}

\renewcommand{\vec}[1]{{\boldsymbol #1}}

\newcommand{\diag}{\text{diag}}

\usepackage{amsthm}

\begin{document}

\title{Localization and conductance in fractional quantum Hall edges}
\author{Misha Yutushui}
\affiliation{Department of Condensed Matter Physics, Weizmann Institute of Science, Rehovot 76100, Israel}
\author{Jinhong Park}
\affiliation{Institute for Quantum Materials and Technologies, Karlsruhe Institute of Technology, 76021 Karlsruhe, Germany}
\affiliation{Institut f{\"u}r Theorie der Kondensierten Materie, Karlsruhe Institute of Technology, 76128 Karlsruhe, Germany}
\author{Alexander D. Mirlin}
\affiliation{Institute for Quantum Materials and Technologies, Karlsruhe Institute of Technology, 76021 Karlsruhe, Germany}
\affiliation{Institut f{\"u}r Theorie der Kondensierten Materie, Karlsruhe Institute of Technology, 76128 Karlsruhe, Germany}

\begin{abstract}

The fractional quantum Hall (FQH) effect gives rise to abundant topological phases, presenting an ultimate platform for studying the transport of edge states. Generic FQH edge contains multiple edge modes, commonly including the counter-propagating ones. A question of the influence of Anderson localization on transport through such edges arises. Recent experimental advances in engineering novel devices with interfaces of different FQH states enable transport measurements of FQH edges and edge junctions also featuring counter-propagating modes. These developments provide an additional strong motivation for the theoretical study of the effects of localization on generic edge states. We develop a general framework for analyzing transport in various regimes that also naturally includes localization. Using a reduced field theory of the edge after localization, we derive a general formula for the conductance. We apply this framework to analyze various experimentally relevant geometries of FQH edges and edge junctions.
 
\end{abstract}

\date{\today}
\maketitle

\section{Introduction} 
The fractional quantum Hall (FQH) effect~\cite{Tsui_fqh_1982, Laughlin_fqh_1983, Haldane_fqh_1983, Halperin1984statistics,Halperin_FQHE_2020} 
is a paradigmatic platform for generating a rich variety of topological states of matter. 
Via bulk-boundary correspondence, the topological order of the FQH bulk states is fully encoded in their edges and thus can be measured by edge probes.  Prototypical examples are observation of fractional charge~\cite{Picciotto_fractional_charge_1998, Saminadayar_fractional_charge_1997} and, more recently, of fractional (anyonic) exchange statistics~\cite{Nakamura2020, Bartoleomei2020, Lee2023} of elementary quasiparticles. 
These exotic properties of topological excitations have been measured by exploring FQH edge transport in specially designed geometries; see Ref.~\cite{Heiblum2020edge} for a recent review.
 
Gapless excitations on the edge of an Abelian FQH state are described by an integer-valued matrix $K$ and an integer-valued vector $\vec{t}$
characterizing the bulk topological order~\cite{Wen_Topological_1995}. Most natural transport observables characterizing the edge are the electric and thermal conductances, which have been the subject of intense theoretical studies  \cite{Kane_Randomness_1994,Kane_Impurity_1995,kane_quantized_1997,Rosenow_signatures_2010,Protopopov_transport_2_3_2017,Nosiglia2018,Spanslatt2021}.
It was shown \cite{Protopopov_transport_2_3_2017} that, in the presence of counter-propagating modes, the conductances depend not only on the underlying topology but also on the transport regime. In particular, for the $\nu=2/3$ state, the electric conductance (in units of $e^2/h$) was predicted to be $4/3$ for a clean (ballistic) edge, and $2/3$ in the regime of inelastic equilibration (also studied in Ref.~\cite{Nosiglia2018}), with both limiting values reflecting the topology of the state. 

On the experimental side, recent years have witnessed major advances in investigations of electric and thermal transport in various FQH edges with counter-propagating modes in GaAs and graphene samples~\cite{Banerjee_Observed_2017, Banerjee_observation_2018, Melcer_Absent_2022,Dutta_Isolated_2022,Srivastav2021May,
Srivastav2022Sep, Breton2022}. 
Furthermore, over the last few years, there has been significant progress in fabricating novel structures involving FQH edge states and measuring their transport properties. For instance, in GaAs samples, interfaces of different FQH states have been engineered~\cite{Grivnin2014, Wang2021, Hashisaka2021, Dutta_novel_2022, Hashisaka2023}, as well as for the case when they originate from different FQH layers~\cite{Ronen2018, Cohen2019}. This experimental approach gives rise to a new versatile platform for engineering ``artificial'' FQH edges with multiple edge modes, including counter-propagating modes. 
Controlling experimental tuning knobs, such as the magnetic field and the gate voltage, allows one to create a variety of FQH edges at the interface. Furthermore, the quantum-point-contact geometry provides another route to FQH state engineering; see, e.g., Ref.~\cite{Spanslatt_condplateau} for more details. Recently, such a novel structure has also been achieved in graphene \cite{Cohen2023}, triggered by the development of quantum point contacts in this material~\cite{Ronen2021,Deprez2021}. 

FQH edges are multi-mode one-dimensional (1D) structures. On the other hand, it is well known that Anderson localization has dramatic implications for low-temperature transport in conventional 1D wires. This leads to a question: does Anderson localization affect low-temperature (coherent) transport properties of FQH edges in the presence of disorder? Clearly, localization can only be operative for edges that allow for a back-scattering, i.e., involve counter-propagating modes. Indeed, it has been known since the work \cite{Kane_Randomness_1994} on $\nu=2/3$ edge that disorder may play an important role on edges with counter-propagating modes. At the same time, the coherent transport of a $\nu=2/3$ edge is not characterized by localization; rather, the electric conductance exhibits mesoscopic fluctuations \cite{Protopopov_transport_2_3_2017}. Does this apply to other FQH states as well? The answer has been known since the seminal paper by Haldane \cite{Haldane_Stability_1995}, who introduced the notion of topological stability of FQH states. The $\nu=2/3$ state is stable, as well as any other FQH state with only two edge modes. At the same time, FQH states with three or more counter-propagating edge modes are frequently topologically unstable. This implies that the edges of these FQH phases are prone to localization. 

Implications of topological instability for the edge physics were addressed by using an example of the $\nu=9/5$ state in Ref.~\cite{Kao_Binding_1999}, where the notion of binding transition was introduced. Since we consider a disordered edge, with random tunneling between the modes, the physics of this transition on the edge will be that of localization. Specifically, counter-propagating edge modes may undergo a localization phase transition driven by tunneling processes between the modes. As a result, two counter-propagating modes are effectively eliminated by this localization process and, hence, do not contribute to the low-energy dynamics. Upon localization, topological low-energy excitations are governed by a certain reduced theory $(K_{\text{red}}, \vec{t}_{\text{red}})$ involving only the remaining modes (those that do not participate in the localization process).

While Ref.~\cite{Kao_Binding_1999} focused on tunneling to the edge, manifestations of localization transition in transport properties of the $\nu=9/5$ edge were studied in Ref.~\cite{Spanslatt_binding_2023}. The localization transition in topologically unstable edges takes place in the parameter space spanned by tunneling strengths and interaction couplings. This transition is expected to be of Berezinskii-Kosterlitz-Thouless type, in analogy with the case of conventional interacting disordered wires \cite{Giamarchi1988Anderson}. In this paper, we will not consider the physics near the transition but rather will focus on the properties of the two systems before and after localization. 

Recent experimental progress reviewed above paves the way for the engineering of a wealth of multi-mode edges and edge junctions, thus providing a strong motivation for the investigation of the effect of localization on FQH transport. This makes it highly desirable to develop a general theoretical framework for understanding and evaluating transport properties in various setups with multiple FQH edge modes, which is the central goal of this work.


In this paper, we develop a general framework for the evaluation of conductance of a multi-mode FQH edge in various transport regimes, including (i) ballistic transport, (ii) incoherent equilibration, (iii) localization, and (iv) localization in a junction in combination with incoherent equilibration in edge segments connecting the junction to the leads. 
 In particular, in the presence of localization, we determine the reduced theory $(K_{\text{red}}, \vec{t}_{\text{red}})$ and derive a general result for the conductance within this theory. Furthermore, we show how this general framework actually works in a number of relatively simple examples: (a) $\nu = 9/5$ edge, (b) $\nu = 8/3$ edge, (c) spin-polarized and spin-unpolarized $\nu = 2/3$ edges, (d) interface between spin-polarized and spin-unpolarized $\nu = 2/3$ states. 

The paper is organized as follows. Section~\ref{sec:edgetheory} presents the basics of the edge theory. In Sec.~\ref{sec:ballistic}, we describe contacts by employing a line-junction model and find the conductance in the clean limit (ballistic transport). Section~\ref{sec:incoherentequilibration} deals with the fully equilibrated case where the conductance always acquires a universal value. In Sec.~\ref{sec:localization}, we finally include the localization in the formalism and derive a formula for the conductance through a localized region. This section further considers an interplay of localization and inelastic equilibration. Specifically, we explore the geometry of a junction that undergoes localization transition and is connected by equilibrated segments to the leads.  Section~\ref{sec:summary} contains a summary of the results of this work and a discussion of future research directions.

Throughout the paper, we use units of $e = \hbar = k_B = 1$. The dimensional conductance $G$ is measured in units of $e^2/h$, according to the common convention. Technical details of our analysis are presented in several appendices.

\section{Edge theory} 
\label{sec:edgetheory}

\subsection{Effective action and excitations}
\label{sec:action-and-excitations}

We begin by briefly summarizing relevant aspects of Wen's $K$-matrix formalism \cite{Wen_edge_1991,Wen_Topological_1995}.  The effective theory of an edge with $d$ modes  is described by the action 
\begin{align}
\label{eq.L0}
    S_0 = \frac{1}{4\pi}\int dt dx\; \sum_{a,b=1}^d\partial_x\phi_a( K_{ab}\partial_t\phi_b-V_{ab}\partial_x\phi_b),
\end{align}
where $K_{ab}$ is an integer-valued symmetric matrix, with $\text{dim}(K)=d$, which is determined by the topological order of the FQH state. The signature of $K$, i.e., the number of positive ($n_R$) and negative ($n_L$) eigenvalues of $K$, determine the number of right and left-moving modes, respectively. 
Further, $V_{ab}$ is a positive-definite velocity matrix, which is non-universal (depending on the strengths of interactions between the modes). 

The particle density and current of the $a$'th edge mode are $\rho_a= \partial_x\phi_a/2\pi$ and $I_a=\partial_t\phi_a/2\pi$, respectively. The charges of the modes are specified by an integer-valued charged vector $t_a$; thus, the charge current carried by $a$'th mode is given by $J_a = t_a I_a$, where we set the electron charge to be $e=1$. The bulk filling factor $\nu$ is related to $K$-matrix and $\vec{t}$ vector via
\begin{equation}
\nu=t_aK^{-1}_{ab}t_b \,,
\end{equation}
where $K^{-1}_{ab}$ is the $ab$ matrix element of the inverse matrix $K^{-1}$.
(The sum over repeated indices is assumed unless stated otherwise).


Upon canonical quantization, $\rho_a$ and $\phi_b$  obey commutation relations 
\begin{align}
\label{eq.commutation}
    [\rho_a(x),\phi_{b}(x')] = i K^{-1}_{ab}\delta(x-x').
\end{align}
The operators that annihilate quasiparticles, 
\begin{equation}
\chi_{\vec m} (x)\propto e^{ i \sum_{a}m_a\phi_a (x)}\,,
\label{eq:quasiparticle-operator}
\end{equation}
are parameterized by integer-valued vectors $m_a$. From the commutation relations Eq.~\eqref{eq.commutation}, one sees that $\chi^\dag_m(x)$ creates a charge 
\begin{equation}
Q_{\vec m} = t_aK^{-1}_{ab}m_b
\label{eq:Qm}
\end{equation}
at the position $x$. The exchange statistics of quasiparticles $\chi_{\vec m^{(1)}}$ and $\chi_{\vec m^{(2)}}$ reads $\theta_{12} = \pi m^{(1)}_aK^{-1}_{ab}m^{(2)}_b$. 

\subsection{Basis change}
\label{sec:basis_change}

The action Eq.~\eqref{eq.L0} has the same form under field redefinition $\tilde{\phi} = O^{-1}\phi$ where $O\in \text{GL}(d,\mathbb{R})$, with the transformed matrices $\tilde{K}=O^{T}KO$ and $\tilde{V}=O^{T}VO$. This transformation also implies the transformation of the charge vector, $\tilde{\vec{t}} = O^{T}\vec{t}$, and of quasiparticle excitation vectors,
$\tilde{\vec{m}} = O^{T}\vec{m}$. However, the integer-valued characters of $K$, $\vec t$, and $\vec m$ are, in general, not preserved by such a transformation. 

The theory fully retains its form 
 if we restrict transformations $\tilde{\phi} = W^{-1}\phi$ to $W\in \text{SL}(d,\mathbb{Z})$ (the group of matrices with integer entries and unit determinant). Then, quasiparticle vectors transform as $\tilde{m} = W^{T}m$ and remain integer-valued. Furthermore, the topological order is given, as before, by an integer-valued $\tilde{K}$-matrix and an integer-valued charge vector $\tilde{\vec{t}}$,
\begin{align} \label{eq:sldztransformation}
    \tilde{K} = W^{T}KW,\qquad \tilde{\vec{t}} = W^{T}\vec{t}\,.
\end{align} 
The currents and the chemical potentials transform under such a transformation as
$\tilde{\vec I}=W^{-1} \vec I$ and $\tilde{\vec \mu}=W^T\vec \mu$.

\subsection{Interfaces}
\label{sec:interfaces}

When two FQH states described by $(K_\text{A},\vec{t}_\text{A})$ and $(K_\text{B},\vec{t}_\text{B})$ are interfaced, the resulting edge theory is described by a block-diagonal $K$-matrix and by a combined $\vec t$-vector, 
\begin{align}
    K_\text{AB} = \begin{pmatrix}
        K_\text{A}&0\\0&-K_\text{B}
    \end{pmatrix},\qquad
    \vec{t}_\text{AB} = (\vec{t}_\text{A},\vec{t}_\text{B}).
\end{align}
The velocity matrix, however, is not block-diagonal, reflecting the density-density interactions between the edges.

\subsection{Disorder}

In realistic systems, the disorder breaks translation symmetry, facilitating momentum non-conserving tunneling processes between different modes comprising the edge \cite{Moore_Classification_1997}. Random tunneling is described by an action 
\begin{align}\label{eq.tunneling}
    S_\text{tun} =- \int dt dx\;  \xi(x)\exp\left[i\sum_{a}M_a\phi_a\right] +\text{H.c.},
\end{align}
where $\vec M$ is an excitation with zero charge, $Q_{\vec M} = 0$. 
In the interface geometry, Sec.~\ref{sec:interfaces}, we have $\vec M =(\vec m_\text{A}, - \vec m_\text{B})$, and the tunneling process is allowed if the quasiparticle charges are equal, $Q_{\vec m_\text{A}}=Q_{\vec m_\text{B}}$.
Furthermore, if the two edges, A and B, are separated by vacuum, only electrons are allowed to tunnel between them, leading to an extra constraint:  $\vec m_i = K_i \vec L_i$ with $\vec L_i$ being an integer vector and $i = \text{A}, \text{B}$. 

The back-scattering induced by random tunneling Eq.~\eqref{eq.tunneling} may lead to the localization of counter-propagating modes. Such edges are referred to as topological unstable~\cite{Haldane_Stability_1995}. 

\subsection{Topological stability}
\label{sec:topo-stability}

As was discovered by Haldane in Ref.~\cite{Haldane_Stability_1995}, if for a given $(K,\vec{t})$, there exists an integer-valued null-vector $M_a$ such that the corresponding excitation is (a) charge neutral,
\begin{equation}
t_aK^{-1}_{ab}M_b=0 \,,
\end{equation}
and (b) non-chiral (characterized by zero topological spin),
\begin{equation}
M_aK^{-1}_{ab}M_b=0 \,,
\end{equation}
the theory is then topologically unstable. Depending on the velocity matrix, the term \eqref{eq.tunneling} in the action of a topologically unstable edge may be relevant or irrelevant in the renormalization group sense. In the former case, the tunneling term flows to strong coupling and becomes dominant at a certain scale, determining the localization length $\xi$. In such a case, two counter-propagating modes get localized and do not contribute to transport at distances longer than $\xi$. In this regime, the edge theory is effectively described by a reduced matrix $K_\text{red}$ with $n_R-1$ right-moving and $n_L-1$ left-moving modes, respectively. 

After this brief review of key aspects of the edge theory (serving as a starting point for this work), we are ready to turn to the analysis of charge transport in the edge.

\section{Ballistic transport} 
\label{sec:ballistic}
In this section, we derive a set of kinetic equations describing transport in the edge coupled to the contacts and solve these equations to obtain the two-terminal conductance in the ballistic regime. 

\subsection{Contacts}
\label{sec:contacts}

The equation of motion of Eq.~\eqref{eq.L0},
\begin{align} 
    \partial_x(K_{ab} I_b - V_{ab} \rho_b)=0,   
\end{align}
is interpreted as a continuity equation for the current of $a$'th mode $I_a = (1/2\pi)\partial_t\phi_a$. The solution imposes the relation between the current and the density: $I_a = K^{-1}_{ab}V_{bc}\rho_c$. By definition, the chemical potential of $a$'th mode $\mu_a$ reflects the energy change due to a change in the density $\rho_a$
\begin{align}\label{eq.mua}
    \mu_a = \frac{\delta H}{\delta \rho_a} = 2\pi V_{ab}\rho_b,
\end{align}
where $H = \pi \int dx \rho_a V_{ab}\rho_b$ is the Hamiltonian corresponding to Eq.~\eqref{eq.L0}. Thus, currents and chemical potentials are related by
\begin{align} \label{eq:relationcurrentchemicalpot}
    I_a = \frac{1}{2\pi} K^{-1}_{ab}\mu_b.
\end{align}
Importantly, the $V$-matrix does not enter Eq.~\eqref{eq:relationcurrentchemicalpot}. This indicates that the interaction between the modes and the mode velocities do not affect the transport properties. 

\begin{figure}[t!]
\includegraphics[width =0.7\columnwidth]{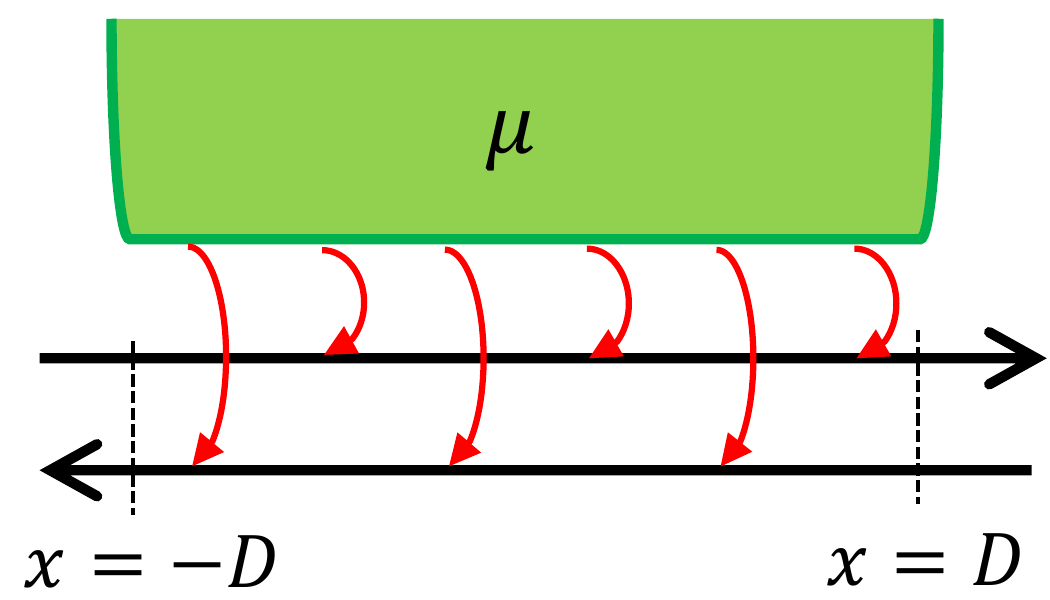}
\caption{ {\bf Line-junction model of a contact}. A contact of length $2D$ (depicted as a green region) at a chemical potential $\mu$ is coupled via electron tunneling (red arrows) to $n_R$ right-moving and $n_L$ left-moving edge modes $\phi_a$ with $a=1,2,\ldots, d$ (black arrows). The contact is sufficiently long to ensure the equilibration of respective modes to a chemical potential $\mu$; see text for detail.
} \label{Fig:contact}
\end{figure}

We adopt the line junction model for a contact employed in Refs.~\cite{Kane1995, Spanslatt2021}, which involves tunneling processes between electrons in the reservoir (a contact) at a chemical potential $\mu$ and electron-like excitations on the edge underneath the contact $x\in[-D,D]$, see Fig.~\ref{Fig:contact}. It is crucially important that the tunneling Hamiltonian involves the tunneling of  {\it electrons} (rather than some fractionally charged excitations) and that $\mu$ is the chemical potential of {\it electrons} in the contact. There are (infinitely) many excitations that are ``made of electrons'' in the edge; they are created by operators $\chi_{\vec l}^\dagger$ (see 
Eq.~\eqref{eq:quasiparticle-operator}) with  $\vec l=K\vec{m}$, where $\vec{m}$ is an integer-valued vector whose components have no common divisor. The corresponding charge is $Q_{K\vec m} = \vec t^T \vec m \in \mathbb{Z}$,  according to Eq.~\eqref{eq:Qm}. 
%
%
We assume that the most dominant tunneling processes involve linearly independent vectors $\vec m^{(a)}$ with non-zero integer charge, with $a=1, \ldots, d$. (The subleading tunneling terms result in the back-scattering between the modes and are considered in Appendix.~\ref{app.gen_sol}.)
We keep only these dominant processes and write the equations of motion in the basis determined by them. In this basis, the modes coupled to the contact are $m^{(a)}_b = \delta_{ab}$, and the charges are $Q_{\vec{l}^{(a)}} = t_a$. If originally a different basis was chosen, this can be achieved by performing a $W\in \text{SL}(d,\mathbb{Z})$ transformation, see Sec.~\ref{sec:basis_change}. The tunneling current to mode $a$ is proportional to the difference between chemical potentials
$\mu_a-t_a\mu$, which is the energy difference of removing $t_a$ electrons from the contact and adding one particle of charge $t_a$ to the edge $a$ so that the current in mode $a$ obeys
\begin{align}
\label{eq.equilb1a}
     \partial_x  I_a (x) = - \frac{\Gamma_a}{2\pi t_a} (\mu_a (x) - t_a\mu),
\end{align}
where $\Gamma_a >0$ are (non-universal) charge tunneling strengths per unit length. Using Eq.~\eqref{eq:relationcurrentchemicalpot}, we obtain the current-balance equation of the form 
\begin{align}
\label{eq.equilb1}
     \partial_x  I_a (x) =  -  \frac{\Gamma_a}{t_a} \sum_b K_{ab} (I_b (x) - I^{(0)}_b(\mu)),
\end{align}
where $I^{(0)}_b(\mu) = (\mu/2\pi)\sum_c K^{-1}_{bc}t_c$ is the equilibrated particle current.  
We absorb this constant term by shifting the current, $\delta I_a (x)\equiv I_a(x)-I^{(0)}_a(\mu)$,  and define a diagonal matrix of tunneling strengths, $\Upsilon_{ab} = \delta_{ab} \Gamma_a/ t_a$. Equation \eqref{eq.equilb1} takes then the following matrix form
\begin{align} 
\label{eq:differentialeqcontacts}
    \partial_{x}\delta\vec{I} = -\Upsilon K \delta \vec{I}.
\end{align}
Since $\Gamma_a>0$, the matrix $\Upsilon K$ has the same signature as $K$, i.e.,
$n_R$ positive and $n_L$ negative eigenvalues. 
(This follows from the Sylvester's law of inertia.)
The positive (negative) eigenvalues correspond to the solutions that decay in the right (respectively, left) direction. 
We assume that the contact length $2D$ is sufficiently long so that all these decay rates (governed essentially by $\Gamma_a$) are much larger than $1/2D$. 
Then, up to exponentially small corrections, the contact can be viewed as imposing boundary conditions at $x= D$ ($x=-D$) on $n_R$ (respectively, $n_L$) combinations of currents corresponding to the positive (respectively, negative) eigenvectors of $\Upsilon K$.

Equation ~\eqref{eq:differentialeqcontacts} can be solved by performing a basis change that diagonalizes $\Upsilon K$. Similarly to Ref.~\cite{Moore_Classification_1997}, we first define the basis change $\vec{I} \rightarrow \vec{I}'=M_K\vec{I}$ that diagonalizes $K$, such that $K = M_{K}^{T} \Lambda M_{K}$, where $\Lambda$ is the $(n_R,n_L)$ pseudo-identity matrix with $\Lambda_{aa}\equiv\lambda_a=+1$ for $a\leq n_R$ and $\Lambda_{aa}=-1$ for $a> n_R$.  Upon this basis change, the tunneling matrix $\Upsilon$ transforms to a positive definite, symmetric matrix $\Upsilon'=M_K \Upsilon M_K^T$. In a second step, we diagonalize $\Upsilon'$ by a pseudo-orthogonal matrix $O_\Upsilon \in\text{SO}(n_R,n_L)$ so that $O_\Upsilon \Lambda O_\Upsilon^T = \Lambda$ and  $[O_\Upsilon \Upsilon' O_{\Upsilon}^T]_{ab}=\delta_{ab}\tau_a$, with $\tau_a>0$. Thus, we have 
\begin{equation}
K=U^T \Lambda U \qquad \text{and} \qquad
U \Upsilon K U^{-1} = \hat{\tau} \Lambda
\label{eq:U-diag-Upsilon-K}
\end{equation}
with $U= O_\Upsilon M_K$, where $\hat{\tau}$ is a diagonal matrix with eigenvalues $\tau_a$, i.e., the transformation $U$ diagonalizes $\Upsilon K$. 
To avoid confusion, we note that, in a general case, $U$ does not belong to $\text{SL}(d,\mathbb{Z})$. 

We now define $\vec{f} =U \delta \vec{I}$ so that currents and chemical potentials can be expressed in terms of $\vec f$ as
\begin{align}\label{eq.I_in_f}
    \vec{I}(x) &=\frac{\mu}{2\pi}K^{-1}\vec{t} + U^{-1} \vec{f}(x),\\
    \vec{\mu}(x) &=\mu \vec{t} + U^T \Lambda \vec{f}(x).
\end{align} 
The matrix equation \eqref{eq:differentialeqcontacts} decouples when written in terms of the components $f_a$ of $\vec f$: 
\begin{align}
   \partial_x f_a = - \tau_{a} \lambda_a f_a.
\end{align}
We immediately see that the solutions $f_a(x)=C_a e^{-\tau_a(D+\lambda_a x)}$ exponentially decay in positive-$x$ direction for right-moving modes, $\lambda_a>0$, and in negative-$x$ direction for left-moving modes, $\lambda_a<0$. We define the projection matrices $\chi_R = \text{diag}(1,\ldots1,0,\ldots,0)$ and $\chi_L = \text{diag}(0,\ldots0,1,\ldots,1)$, such that $\Lambda=\chi_R-\chi_L$. Discarding exponentially small corrections, we thus obtain
\begin{align}
\label{eq:bc_for_f}
    \vec{f}(D) = \chi_L \vec{C},\qquad
    \vec{f}(-D) = \chi_R \vec{C}
\end{align}
at the boundaries of the contact. In other words, the first $n_R$ components of $\vec f$ (those corresponding to right-moving modes) vanish at the right boundary, $x=D$, and the remaining $n_L$ components vanish at the left boundary, $x=-D$.

The following comment is in order here. We considered here an idealized contact: the only tunneling processes that we retained in the contact region are the electron tunneling between the contact and $d$ modes in the edge. What we discarded are tunneling processes between these modes. As a result, the tunneling matrix $\Upsilon$ in the above analysis was diagonal in a certain basis that can be obtained by a $\text{SL}(d,\mathbb{Z})$ transformation. This is fully justified in many cases, in particular, when the modes forming the edge are spatially separated (like 1 and 1/3 modes for the spin-polarized $\nu=2/3$ state; see also below) so that tunneling between them is much weaker than the tunneling between any of them and the contact. At the same time, in some situations, there might also be a sizeable tunneling between the modes in the contact. The above analysis can be straightforwardly adjusted to this case; see Appendix~\ref{app.gen_sol}.

\begin{figure}[t!]
\includegraphics[width =.9\columnwidth]{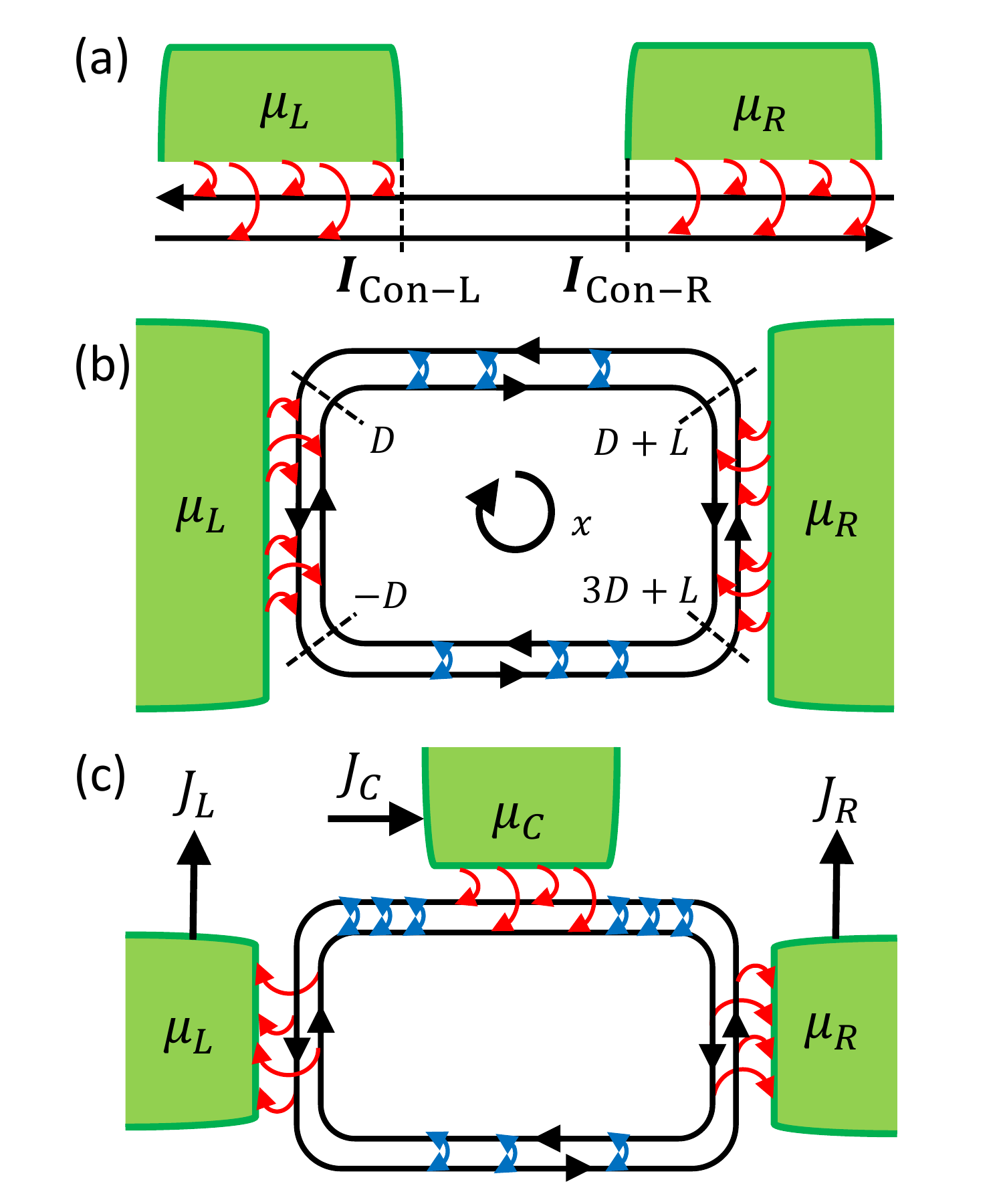}
\caption{
{\bf (a) Schematic setup} for the transport within an edge segment between the left and right contacts (green regions) at chemical potentials $\mu_L$ and $\mu_R$. Here, 
$\vec{I}_{\text{Con-L}}$ and $\vec{I}_{\text{Con-R}}$ represent the currents on the edge segment boundaries with the left and right contacts, respectively. 
{\bf (b) Two-terminal conductance setup} with the edge coordinate $x \in [-D, 3D + 2L]$, where the points $x=-D$ and $x=3D+2L$ are identified. In the 
contact regions, $x \in [-D, D] \cup  [D+L, 3D +L]$, charge tunnels between the contacts and the edge modes (red arrows). In the edge segments between the contacts, $x \in [D, D+L] \cup [3D + L, 3D+2L]$, the charge in the edge is conserved. 
If the charge tunneling between the modes in these regions (blue arrows) is allowed, inelastic equilibration or (for topologically unstable edges) localization may take place. In the absence of such tunneling, the transport is ballistic. 
{\bf (c) Three-terminal setup} to measure the conductance. The potential $\mu_C$ is applied to the central contact, while the left and right contacts are usually grounded, i.e., $\mu_R = \mu_L = \mu_0$. Then, charge current $J_C$ is inserted to the central contact, and it is split to $J_L$ and $J_R$, which are collected at the left and right contacts. The two-terminal conductance $G$ is obtained as $G = J_C/ (\mu_C - \mu_0) = (J_R + J_L) / (\mu_C- \mu_0)$; see text for more detail. 
}
\label{Fig:setup1}
\end{figure}

\subsection{Clean limit: Ballistic conductance}
\label{sec:ballistic-cond}

After having analyzed the equilibration between the contact and the edge modes, we turn to the 
calculation of the conductance of an edge coupled to two contacts at chemical potentials $\mu_L$ and $\mu_R$, see Fig.~\ref{Fig:setup1}(a) for the schematic setup. We consider first the clean limit, i.e., the absence of inter-mode scattering in the edge, which implies the ballistic transport. For simplicity, we assume that the contacts are equivalent so that the modes coupled to the contact and the tunneling strengths $\Gamma_a$ (and, consequently, also the matrices $U$) are the same for both contacts. In the clean limit, there is no back-scattering, so the current emanating from the left contact $\vec{I}_\text{Con-L} \equiv\vec{I} (D)$ propagates unimpeded to the right contact $\vec{I}_\text{Con-R} \equiv\vec{I}(D + L)$, see Fig.~\ref{Fig:setup1}(a). According to Eqs.~\eqref{eq.I_in_f} and \eqref{eq:bc_for_f}, the continuity equation $\vec{I}_\text{Con-L} =\vec{I}_\text{Con-R}$ takes the form
\begin{align} \label{eq.continutity}
& \vec{I}_\text{Con-L} =   \frac{\mu_L}{2\pi} K^{-1}\vec{t} + U^{-1}\chi_L \vec{C}_L
 \nonumber    \\   
  = \ & \vec{I}_\text{Con-R} =
    \frac{\mu_R}{2\pi} K^{-1}\vec{t} + U^{-1}\chi_R \vec{C}_R\,.
\end{align}
When written in components, this equation represents a set of $n_R+n_L$ equations for $n_R$ variables $[C_R]_{a\leq n_R}$ and $n_L$ variables  $[C_L]_{a> n_R}$. Multiplying Eq.~\eqref{eq.continutity} by $\chi_p U$ 
with $p =R,L$
and using that $\chi_R\chi_L=0$ and $\chi_p^2=\chi_p$, we get the solution:
\begin{align}
    \chi_R\vec{C}_R &= \frac{\mu_L-\mu_R}{2\pi} \chi_R U K^{-1} \vec{t},\\
    \chi_L\vec{C}_L &= - \frac{\mu_L-\mu_R}{2\pi} \chi_L U K^{-1} \vec{t}.
\end{align}
Using these solutions, we obtain the total charge current along the edge, $J_\text{tot} = \vec{t}^T\vec{I}_\text{Con-L} = \vec{t}^T\vec{I}_\text{Con-R}$, as 
\begin{align}
    J_\text{tot} &= \frac{\mu_L}{2\pi}\nu -
    \frac{\mu_L-\mu_R}{2\pi}
    \vec{t}^TU^{-1} \chi_L U K^{-1}\vec{t}
    \nonumber \\ &= 
    \frac{\mu_R}{2\pi}\nu +
    \frac{\mu_L-\mu_R}{2\pi}
    \vec{t}^TU^{-1} \chi_R U K^{-1}\vec{t},
\end{align}
where we used that $\nu=\vec{t}K^{-1}\vec{t}$.
Averaging over these equations, we get
\begin{align}
\label{eq.Jtot}
    J_\text{tot} = (\mu_L+\mu_R) \frac{\nu}{4\pi}
    + (\mu_L-\mu_R) \frac{1}{4\pi} G,
\end{align}
with the two-terminal conductance $G$
\begin{align}
\label{eq:condclean-a}
    G= \vec{t}^TU^{-1} \Lambda U K^{-1}\vec{t},
\end{align}
where we used $\chi_R-\chi_L=\Lambda$.

Using $KU^{-1}=U^T\Lambda $, we can equivalently write  this equation as
\begin{align} \label{eq:condclean}
    G= \vec{t}^T(U^{T} U)^{-1} \vec{t}.
\end{align}
It should be emphasized that the total current Eq.~\eqref{eq.Jtot} is anomalous in the sense that it is not a function of $\mu_L - \mu_R$.
This is a manifestation of the fact there is a current on the boundary of a FQH sample at equilibrium and that we have included, up to now, only one edge segment between the two contacts. Consider now a two-terminal setup shown in Fig.~\ref{Fig:setup1}(b). The total current between the contacts now includes contributions of the upper and lower edges, yielding
\begin{eqnarray}
J_\text{two-term} &=& J_\text{tot}(\mu_L,\mu_R) -J_\text{tot}(\mu_R,\mu_L) \nonumber \\
&=& (\mu_L-\mu_R) \frac{1}{2\pi}G.
\label{eq:J-two-term}
\end{eqnarray}
We see that the anomaly cancels here as expected, and $G$ is indeed the two-terminal conductance. 

According to common convention, we define the dimensionless conductance $G$ as measured in units of $e^2/h$. This explains the factor $1/2\pi$ in front of $G$ in Eq.~\eqref{eq:J-two-term}. (We recall that we set $e=\hbar = 1$ in all formulas.)

When writing Eq.~\eqref{eq:J-two-term}, we assumed that the upper and lower edges of the two-terminal device shown in Fig.~\ref{Fig:setup1}(b) are identical (and thus characterized by the same $G$). In practice, one frequently uses a three-terminal scheme to measure the conductance shown in Fig.~\ref{Fig:setup1}(c). In a typical experiment, the left and right contacts are grounded, $\mu_L = \mu_R = \mu_0$, while the central contact is biased and has a potential $\mu_C$. The current $J_C$ enters the system through the central contact and is split into $J_L$ and $J_R$ measured at left and right contacts, with $J_L + J_R = J_C$, see Fig.~\ref{Fig:setup1}(c). Let us denote by $G_L$ the two-terminal conductance corresponding to the segment between the left and central contacts (i.e., the conductance of the device shown in Fig.~\ref{Fig:setup1}(b) with both edges identical to this segment). Similarly, let $G_R$ be the two-terminal conductance corresponding to the segment between the central and right contacts. Then, applying Eq.~\eqref{eq.Jtot} to all three segments of the edge in Fig.~\ref{Fig:setup1}(c) and using the current conservation at contacts, we get 
\begin{eqnarray}
J_L &=& \frac{1}{4\pi} (G_L - \nu) (\mu_C - \mu_0) \,; \\
J_R &=& \frac{1}{4\pi} (G_R + \nu) (\mu_C - \mu_0) \, ; \\
J_C &=& \frac{1}{4\pi} (G_R + G_L) (\mu_C - \mu_0) \,. 
\label{eq:three-terminal-JC}
\end{eqnarray}
For a symmetric device, $G_L = G_R = G$, Eq.~\eqref{eq:three-terminal-JC} becomes equivalent to Eq.~\eqref{eq:J-two-term} (with $\mu_L \to \mu_C$ and $\mu_R \to \mu_0$). The three-terminal scheme allows one also to measure $G_L$ and $G_R$ separately in non-symmetric devices where $G_L$ and $G_R$ may be different (e.g., due to different lengths of segments LC and CR leading to different transport regimes, including localization and inelastic equilibration that are studied below in this paper). This is in particular relevant to measurements of the conductance in interface structures, see Sec.~\ref{sec:localizationandequilibration}.

Equation \eqref{eq:condclean} is the sought general result for the two-terminal conductance $G$ in the ballistic regime. It expresses $G$ in terms of the charge vector $\vec{t}$ and the matrix $U$ defined by \eqref{eq:U-diag-Upsilon-K}. The conductance is determined by the topology of the edge ($K$ and $\vec t$) and by the way the modes are fed from the contact (matrix $\Upsilon$). Importantly, the conductance \eqref{eq:condclean} does not depend on the interactions in the edge: the matrix $V_{ab}$ of mode velocities and inter-mode interactions have dropped completely from the result. In order to better understand the result \eqref{eq:condclean}, we now discuss some special cases and examples.

As the simplest case, consider first a fully chiral edge, with all modes propagating in the same direction: $n_L = d$  and $n_R=0$. In this case $\Lambda=\mathbb{1}$, Eq.~\eqref{eq:condclean-a} reduces to $G= \vec{t}^TK^{-1}\vec{t} = \nu$. The conductance for such edges is thus fully universal and is equal to the filling factor $\nu$. We turn now to edges with counter-propagating modes.

\subsection{Commuting $K$ and $\Upsilon$ matrices. Example: Spin-polarized $\nu=\frac{2}{3}$ edge}

\label{sec:commuting-K-Upsilon}

For an edge with counter-propagating modes, the conductance \eqref{eq:condclean}  depends on how contacts feed electrons to edge modes. This includes two aspects: (i) the basis of edge modes that are fed (which is the basis in which the matrix $\Upsilon$ is diagonal and which we are using here) and (ii) values of the corresponding tunneling strengths $\Gamma_a$. In general, the conductance will continuously change as a function of $\Gamma_a$ (more precisely, of their ratios) and will be thus non-universal (see Sec.~\ref{sec:non-commuting-K-Upsilon}). There is, however, an important class of FQH edges for which a high degree of universality is restored. These are systems for which the matrix $K$ is diagonal in the basis determined by the coupling to contacts, i.e., the matrices $K$ and $\Upsilon$ commute and are thus diagonal in the same basis. In this case, the matrix $U$ defined above is diagonal so that $U^{-1} \Lambda U =  \Lambda$ and Eq.~\eqref{eq:condclean-a} for the conductance reduces to
\begin{align}
\label{G-commuting-K-Upsilon}
    G =  \vec{t}^T \Lambda K^{-1} \vec{t} \,.
\end{align}
We see that the conductance does not depend on relations between the strengths $\Gamma_a$, implying its universality announced above. To avoid confusion, it is worth recalling that the form of Eq.~\eqref{G-commuting-K-Upsilon} is not preserved under $\text{SL}(d,\mathbb{Z})$ transformations; it is written in the preferential basis in which the matrices $K$ and $\Upsilon$ are diagonal. Nevertheless, the results are invariant under the basis change, as can be seen from the transformation properties of  $\Upsilon$, see Appendix~\ref{app.gen_sol}.

What is the physical situation in which $K$ and $\Upsilon$ matrices commute so that Eq.~\eqref{G-commuting-K-Upsilon} applies?
This is, in particular, the case when there is a basis of modes $a=1,\ldots, d$ that satisfy two conditions. First, these modes should be spatially separated so that one can argue that these are the modes that are (predominantly) fed from the contact. Second, the matrix $K$ should be diagonal in this basis.

A paradigmatic example of such a situation is the edge of a spin-polarized $\nu=2/3$ state. This edge consists of counter-propagating 1 and 1/3 modes, which are spatially separated: the 1 mode is closer to the physical boundary of the FQH system, and the 1/3 mode is closer to its interior~\cite{Johnson1991composite}. 
In the basis of 1 and 1/3 modes, the $K$-matrix and the $\vec{t}$-vector are
\begin{align} \label{eq:Kmat23-polar}
    K=\begin{pmatrix}
1 & 0 \\ 0& -3
\end{pmatrix},\qquad 
\vec{t}=\begin{pmatrix}
1  \\ 1
\end{pmatrix},
\end{align} 
and Eq.~\eqref{G-commuting-K-Upsilon} yields the conductance
\begin{align}
    G = \frac{4}{3} \,.  
    \label{G-23-polarized}
\end{align}
The universal value \eqref{G-23-polarized} of the ballistic conductance of the 2/3 state was derived in Ref.~\cite{Protopopov_transport_2_3_2017}. 
It is worth contrasting this result to an interaction-dependent formula proposed in Ref.~\cite{Kane_Randomness_1994}, which was based on a naive application of an infinite-system Kubo formula without a proper treatment of contacts. 

On the experimental side, observation of the ballistic charge transport in the 2/3 edge is a challenging task in view of a rather short inelastic equilibration length in typical structures at experimentally relevant temperatures. However, recent fabrication progress has allowed achieving the ballistic transport regime (and exploring the full ballistic-to-equilibrated crossover) in an engineered structure in 
Ref.~\cite{Cohen2019}, which experimentally confirmed the value $G=4/3$ of the ballistic conductance of the 2/3 edge.

The above spatial structure of edge modes of the $\nu=2/3$ state is related to the fact that this state is a daughter state of the $\nu=1$ state in the hierarchy picture. Thus, the 2/3 state can be viewed as resulting from a $\delta \nu = -1/3$ condensate on top of a $\nu=1$ condensate, implying the corresponding two edges. This argument can be extended to other spin-polarized single-layer hole-conjugated FQH states. For example, the $\nu = 3/5$ state is obtained on the next step of the hierarchy from the $2/3$ by the formation of an additional $\delta \nu = - 1/15$ condensate. Its edge thus consists of three modes (counting from the boundary of the sample towards its interior): the downstream 1 mode, the upstream 1/3 mode, and the upstream 1/15 mode. Arguing again that they are sufficiently well separated spatially, we obtain the ballistic conductance $G = 1 + 1/3 + 1/15 = 7/5$.

\subsection{Non-commuting $K$ and $\Upsilon$ matrices. Example: Spin-unpolarized $\nu=\frac{2}{3}$ edge}

\label{sec:non-commuting-K-Upsilon}

We now consider the situation of the $K$-matrix being non-diagonal in the basis of modes coupled to the contact. As discussed above, the conductance \eqref{eq:condclean} depends, in this case, on ratios between tunneling strengths $\Gamma$. To illustrate this, it is instructive to consider the model of a spin-unpolarized $\nu=\frac{2}{3}$ state. The corresponding $K$-matrix and $\vec{t}$-vector, when written in the basis of two opposite spin projections (parallel and anti-parallel to magnetic field), read
\begin{align} 
\label{eq:Kmat23}
    K=\begin{pmatrix}
1 & 2 \\ 2& 1
\end{pmatrix},\qquad 
\vec{t}=\begin{pmatrix}
1  \\ 1
\end{pmatrix}.
\end{align} 
Assuming that spin conservation holds, we can expect that the spin-up and spin-down modes are those that are fed by the contact, i.e., that the matrix $\Upsilon$ is diagonal in this basis. 
We parametrize the corresponding two tunneling strengths as $\Gamma_1 = (1-\alpha)\Gamma$ and $\Gamma_2 = (1+\alpha)\Gamma$ with $0 \le |\alpha| \le 1$. 

The $M_K$ and $O_\Upsilon$ that diagonalize 
Eq.~\eqref{eq:differentialeqcontacts} (equilibration with contacts) are now found to be
\begin{align}
    M_K= \begin{pmatrix}
\sqrt{\frac{3}{2}} & \sqrt{\frac{3}{2}} \\ -\sqrt{\frac{1}{2}} & \sqrt{\frac{1}{2}} 
\end{pmatrix},\quad
O_\Upsilon
=  \begin{pmatrix}
\cosh \theta & \sinh \theta\\ \sinh \theta & \cosh \theta
\end{pmatrix}, 
\end{align}
where 
\begin{equation}
\theta = \frac{1}{4}\log \frac{2-\alpha\sqrt{3}}{2+\alpha\sqrt{3}} \,.
\end{equation}
Substituting $U= O_\Upsilon M_K$ into Eq.~\eqref{eq:condclean}, we obtain
the ballistic two-terminal conductance 
\begin{align}
   G = \frac{4}{3\sqrt{4-3\alpha^2}}.
   \label{G-23-unpolarized}
\end{align}
We see that the conductance is non-universal and bounded between $G= 2/3$ for a symmetric ($\alpha=0$) coupling to the contact and $G= 4/3$ for the maximally asymmetric ($\alpha \to \pm 1$) case. 

Let us emphasize that the edge theory \eqref{eq:Kmat23} is topologically equivalent to \eqref{eq:Kmat23-polar}, i.e., they are related by a $\text{SL}(2,\mathbb{Z})$ transformation.  If we perform such a transformation in the spin-unpolarized case, the $K$-matrix will acquire the form \eqref{eq:Kmat23-polar}, but the matrix $\Upsilon$ of coupling to the contact will become non-diagonal. This leads to a remarkable 
difference between the results \eqref{G-23-polarized} and  \eqref{G-23-unpolarized} for the ballistic conductance despite their topological equivalence.

\section{Incoherent equilibration} \label{sec:incoherentequilibration}

In this section, we consider the effect of
equilibration between the edge modes in the region between two contacts with potentials $\mu_L$ and $\mu_R$; see Fig.~\ref{Fig:setup1}(b). Such equilibration results when tunneling processes between edge modes are added incoherently. The corresponding inelastic equilibration length (above which the equilibration within the edge becomes operative) diverges in the zero-temperature limit but is rather short at typical experimental temperatures in most structures. We first consider the upper edge of Fig.~\ref{Fig:setup1}(b), assuming for definiteness that $\nu = \vec{t}^T K^{-1} \vec{t} > 0$ so that the dominant (``downstream'') direction of charge flow is the positive $x$ direction.

The tunneling {\it charge} current $\delta J^{T}_{ab}$ between the $a$'th and $b$'th channels is proportional to the
energy difference of taking the charge $\delta Q_T = t_a\delta \rho_a$ from $a$'th channel and adding $\delta Q_T=t_b\delta \rho_b$ to $b$'th channel. This energy difference reads, in terms of the chemical potentials \eqref{eq.mua} of edge modes,  $\mu_a\delta \rho_a - \mu_b\delta\rho_b=(t^{-1}_a\mu_a - t^{-1}_b\mu_b)\delta Q_T$. Thus, the current balance equation in the edge subjected to incoherent equilibration has the form
\begin{align}\label{eq:incoherent_eq}
    \partial_xI_a =  -\sum_b  \frac{\gamma_{ab}}{2\pi} (\mu_a t_a^{-1}t_b - \mu_b),
\end{align} 
where we introduced a charge tunneling strength per unit length $\gamma_{ab} >0$ between modes $a$ and $b$. Equation \eqref{eq:incoherent_eq} can be rewritten as
\begin{align}\label{eq:incoherent_eq1}
    \partial_x \vec{I} = - \mathcal{T} K \vec{I},
\end{align}
with a symmetric matrix 
\begin{equation}
\label{eq:inelastic-equil-T-matrix}
\mathcal{T}_{ab} =  \delta_{ab}\sum_{c}t_a^{-1}\gamma_{ac}t_c - \gamma_{ab} \,.
\end{equation}
Clearly, the two terms in Eq.~\eqref{eq:inelastic-equil-T-matrix} are the scattering-out and scattering-in terms, respectively. Due to the charge conservation, the matrix $\mathcal{T}$ has a zero eigenvalue corresponding to eigenvector $\vec t$. All other eigenvalues of $\mathcal{T}$ are positive. It follows (under the above assumption $\nu = \vec{t}^T K^{-1} \vec{t} > 0$) that the matrix $\mathcal{T}K$ has one zero, $n_R-1$ positive, and $n_L$ negative eigenvalues. We will provide proof of this important statement at the end of this section.

 We denote the eigenvalues of $\mathcal{T}K$ by $\tilde{\tau}_j$ and order them according to $\tilde{\tau_1} > \tilde{\tau}_2 > \ldots$ so that the first $n_R -1$ eigenvalues are positive, $\tilde{\tau}_{n_R}=0$, and the last $n_L$ eigenvalues are negative. We further denote by  $\vec{v}_j$ the corresponding eigenvectors of $\mathcal{T}K$. The eigenvector corresponding to the zero eigenvalue is
$\vec{v}_{n_R} = K^{-1}\vec{t}$ (``charge mode''); the remaining eigenvectors ($j \ne n_R$) are neutral
since $\vec t^T \vec{v}_j =
\tilde{\tau}_j^{-1} \vec{t}^T \mathcal{T} K \vec{v}_j = 0$ in view of $ \vec t^T {\cal T}= 0$.  A general solution of Eq.~\eqref{eq:incoherent_eq1} reads
\begin{align} \label{eq:solutionincoherent2}
   \vec{I} = & \sum_{j<n_R} A_j  \vec{v}_j  e^{- \tilde{\tau}_j  (x - D)} + A_{n_R} \vec{v}_{n_R} 
 \nonumber \\   
   &+  \sum_{j>n_R} A_j  \vec{v}_j  e^{-(D+ L-x) |\tilde{\tau}_j|}\,,
\end{align}
with $n_R + n_L = d$ coefficients $A_j$. These coefficients should be determined from the boundary conditions at the contacts; see Sec.~\ref{sec:contacts}. Specifically, Eq.~\eqref{eq:bc_for_f} provides exactly $d$ conditions needed to determine the $d$ coefficients $A_j$.

Let us consider the limit of strong equilibration, $L \gg |\tilde{\tau}_j|^{-1}$ for all non-zero eigenvalues ($j \ne n_R$).  This means that the edge length $L$ is sufficiently large, such that the edge modes are fully equilibrated with each other, and the edge reaches a steady state in the region far from the contacts. Then all contributions in Eq.~\eqref{eq:solutionincoherent2} with $\tau_j\neq0$ are exponentially suppressed away from the contacts, and the only contribution comes from the zero mode $\vec{v}_{n_R}$, yielding
\begin{align}
    \vec{I} = A_{n_R} \vec{v}_{n_R} = \left(\frac{\mu_\text{eq}}{2\pi} \right) K^{-1} \vec{t}.
\end{align}
Here, $A_{n_R} = \mu_{\text{eq}} / 2 \pi$ is defined in terms of the chemical potential in the equilibrated region $\mu_{\text{eq}}$. Consequently, the total charge current reads 
\begin{align} \label{eq:totalcurrenteq}
    J_\text{tot} = \vec{t}^T \vec{I}  = \frac{\mu_\text{eq}\nu}{2\pi}\,,
\end{align} 
which solely depends on the value of $\mu_{\text{eq}}$. 
To find $\mu_{\text{eq}}$, we use Eqs.~\eqref{eq.I_in_f} and \eqref{eq:bc_for_f} for the current
on the left (upstream) contact $\vec{I}(x = D) = \vec{I}_{\text{Con-L}}$. Equating it to
Eq.~\eqref{eq:solutionincoherent2} and discarding exponentially small terms [the second sum in Eq.~\eqref{eq:solutionincoherent2}], we obtain
\begin{align} \label{eq:currentconserincoh}
     \frac{\mu_L-\mu_{\text{eq}}}{2\pi} K^{-1} \vec{t} - \sum_{j < n_R} A_j \vec{v}_j + U^{-1} \chi_L \vec{C}_L = 0\,.
\end{align}
The left-hand side of Eq.~\eqref{eq:currentconserincoh}
represents a linear combination of $d$ vectors in the $d$-dimensional space. (Or, equivalently, Eq.~\eqref{eq:currentconserincoh} is a system of $d$ homogeneous linear equations for $d$ variables.)
These vectors are linearly independent so that the only solution is trivial: $A_{j<n_R}=0$, $ \chi_L \vec{C}_L=0$, and $\mu_{\text{eq}}=\mu_L$. 
(An explicit proof of this statement is provided in Appendix \ref{app.relat}.)
Thus, the total charge current is 
\begin{align} \label{eq:solutioncontact2}
   J_{\text{tot}} = \frac{\nu \mu_L}{2 \pi}\,. 
\end{align}
This result reflects the fully chiral nature of the edge that emerges in the equilibrated limit. Note also that the value of the total current, Eq.~\eqref{eq:solutioncontact2}, does not depend on the microscopic details of contacts. The universality of transport properties is fully restored in this incoherent regime.

We turn now to the full two-terminal geometry of Fig.~\ref{Fig:setup1}(b) involving the upper and the lower edge. The total current from the left to the right terminal is then
\begin{equation}
J_{\rm two-term} = \frac{\nu (\mu_L - \mu_R)}{2 \pi} \,,
\end{equation}
resulting in the fully universal value of the two-terminal charge conductance in the equilibrated regime,
\begin{align} \label{eq:twoterminalcond}
    G = 2\pi \, \frac{J_{\text{two-term}}}{\mu_L - \mu_R} = \nu \,.
\end{align}
It follows from the above analysis that the conductance $G$ approaches exponentially to the value $G=\nu$ in the incoherent (equilibrated) regime as  $L$ grows. 

The fact that the inelastic relaxation between the edge modes tends to establish universal quantized values of conductances in quantum Hall devices was pointed out in early works \cite{Buettiker1988absence,
Kane1995}.  More recently, the incoherent transport regime was studied in detail in the context of the $\nu=2/3$ edge \cite{Sen2008line,Rosenow_signatures_2010,Protopopov_transport_2_3_2017,Nosiglia2018,Spanslatt2021}.

We return now to the proof of the signature of the matrix $\mathcal{T}K$. If the matrix $\mathcal{T}$ would be positive definite, the signature of $\mathcal{T}K$ would be the same as that of $K$, i.e., $n_R$ positive and $n_L$ negative eigenvalues, by virtue of Sylvester's inertia law. The tricky point is that while all but one eigenvalue of $\mathcal{T}$ are positive, there is one zero eigenvalue with eigenvector $\vec t$.  This obviously means that $\mathcal{T}K$ also has a zero eigenvalue corresponding to the eigenvector $\vec v_0 = K^{-1}\vec t$. The question is, what are the signs of remaining $n_R+n_L-1$ eigenvalues? We will now prove that for $\nu \equiv \vec t^T K^{-1} \vec t >0$, there are $n_R-1$ positive and $n_L$ negative eigenvalues, while for $\vec t^T K^{-1} \vec t <0$, there are $n_R$ positive and $n_L-1$ negative eigenvalues. An alternative proof of this statement is presented in Appendix \ref{app.signature}. The case $\vec t^T K^{-1} \vec t = 0$ is thus special, and we will also comment on it.

We consider first the special case when the matrices $K $ and $\mathcal{T}$ commute and thus can be simultaneously diagonalized. The statement of the signature of $\mathcal{T}K$ that we have just made then immediately follows for both cases $\vec t^T K^{-1} \vec t > 0$ and $\vec t^T K^{-1} \vec t <0$. We now continuously vary the matrix $\mathcal{T}$  (preserving its signature), so that its eigenvector $\vec t$ corresponding to zero eigenvalue changes, and thus $\vec t^T K^{-1} \vec t$ changes as well, to interpolate between the cases $\vec t^T K^{-1} \vec t > 0$ and $\vec t^T K^{-1} \vec t <0$. The signature of $\mathcal{T}K$ should change at some point on such an interpolating trajectory from $(n_R-1, 1, n_L)$ to $(n_R, 1, n_L-1)$, where $(m,p,q)$ means $m$ positive, $n$ zero, and $q$ negative eigenvalues. Obviously, in the transition point we should have two zero eigenvalues. On the other hand, we know that $\mathcal{T}K$  has only one eigenvector corresponding to a zero eigenvalue. (If it would have a second one, then $\mathcal{T}$ would also have a second one, which contradicts our assumptions.)  It follows that, at the transition point, the matrix $\mathcal{T}K$ has a $2 \times 2$ Jordan block with zero eigenvalue. There is thus a vector $\vec u_0$ such that  $\mathcal{T}K \vec u_0 = \vec v_0$. Multiplying this equation on the left by $\vec t^T$ and using $\vec v_0 = K^{-1}\vec t$ and $\vec t^T\mathcal{T} = 0$, we obtain $\vec t^TK^{-1}\vec t = 0$.

We have thus proven that the signature of $\mathcal{T}K$ is $(n_R-1, 1, n_L)$  for $\nu > 0$, 
$(n_R, 1, n_L-1)$ for $\nu < 0$, and $(n_R - 1, 2, n_L-1)$ for $\nu =0$, where $\nu = \vec t^TK^{-1}\vec t$. 
Before closing the section, we briefly comment on the special case $\nu=0$. While this situation is obviously not relevant to FQH edges, it can be realized in edge junctions. The second zero eigenvalue leads to the appearance of a linear-in-$x$ term in Eq.~\eqref{eq:solutionincoherent2} and consequently to a linear dependence of the equilibrated potential $\mu_{\rm eq}$ on $x$. This corresponds to a diffusive transport, with the conductance scaling as $G \propto 1/L$. The simplest realization of this incoherent transport regime at $\nu=0$ is the conventional disordered Luttinger liquid [corresponding to $K=\text{diag}(1,-1)$ and $\vec t^T = (1,1)$] at elevated temperatures \cite{Gornyi2007electron}.

\section{Localization} 
\label{sec:localization}

\subsection{Basics} \label{subsec:localzationbasictheory}

The localization can be viewed as a coherent counterpart of equilibration. It may happen on an edge with counter-propagating modes if the edge is not topologically stable; see Sec.~\ref{sec:topo-stability}. 

It is instructive to begin with a simple example of
 localization between counter-propagating integer electron modes. Consider, for instance, a coherent transport via an edge with $K=\diag(1,1,-1)$ and $\vec{t}^T = (1,1,1)$
 (which can be obtained as a junction between $\nu=2$ and $\nu=1$ edges). In the absence of inter-mode tunneling, we have a ballistic conductance $G=3$. Now, we ``switch on'' random inter-mode tunneling. For a repulsive interaction (or a not-too-strong attraction), the tunneling term between the second and third modes, $e^{i (\phi_2+\phi_3)} \equiv e^{i \vec M_1 \vec \phi}$ with $\vec M_1^T = (0,1,1)$, is relevant and thus flows to strong coupling. This results in the localization of two counter-propagating modes, $\phi_2$ and $\phi_3$. In such a case, they do not contribute to the transport on the scale beyond the localization length $\xi_\text{loc}$, where only the mode $\phi_1$ remains. As a result, the conductance is reduced down to $G=1$. There is an alternative tunneling process, $e^{i (\phi_1+\phi_3)} \equiv e^{i \vec M_2 \vec \phi}$ with $\vec M_2^T = (1,0,1)$, which may lead to the localization of counter-propagating modes $\phi_1$ and $\phi_3$.  As a result of this localization, the modes 
$\phi_1$ and $\phi_3$ would not propagate at distances larger than the localization length, and we would stay only with the mode $\phi_2$, again yielding the conductance $G=1$.  Obviously, the localization channels corresponding to $\vec M_1$ and $\vec M_2$ are competing: the localization may take place only in one of these channels but not in both. After having inspected this simple example, we turn to a general situation of localization in FQH edges.

Localization of counter-propagating edge modes is driven by tunneling processes between the edge modes, given by 
\begin{align} \label{eq:tunnelingprocessloc}
      S_\text{tun} = - \int dx dt\; g(x) \cos \Big (\vec M^T \vec \phi  + \zeta(x) \Big )\,,
\end{align}
where the tunneling strength $g(x)=|\xi(x)|$ and the phase $\zeta(x)=\text{arg}[\xi(x)]$. This tunneling process permits localization, provided that the tunneling vector $\vec{M}$ satisfies Haldane topological-instability conditions: the charge neutrality condition 
\begin{equation}
 \vec{t}^T K^{-1} \vec{M}= 0
\label{eq:localization-neutrality}
\end{equation}
and the null-vector condition 
\begin{equation}
\vec{M}^T K^{-1} \vec{M} = 0 \,. 
\label{eq:localization-null}
\end{equation}
When such a tunneling process is relevant in the renormalization-group sense, the tunneling term flows to a strong-coupling regime, and the phase of the cosine in Eq.~\eqref{eq:tunnelingprocessloc} is locked to a minimum $ \vec M^T \vec \phi(x,t)+\zeta(x)=0$ and is thus time-independent. This implies that 
\begin{equation}
\label{eq:loc-current-condition}
\partial_t \vec M^T \vec \phi/2\pi= \vec M^T \vec I(x) = 0 \,,
\end{equation}
i.e., a certain combination of current components $I_a$ is zero. 

If localization is possible, then one frequently finds that it is possible in more than one channel. In other words, if there exists a neutral null-vector, then one frequently finds multiple neutral null-vectors. In most of the examples that we will consider below, there will be multiple competing null-vectors $\vec M_j$ such that the localization cannot happen in two channels simultaneously; see the beginning of this section for a simple example of such a situation with two competing null-vectors $\vec M_1$ and $\vec M_2$. On the other hand, in the most general case, an edge may also permit multiple {\it independent} neutral null-vectors $\vec{M}_{j}$, with $j=1, \ldots, m$, that fulfill the condition 
\begin{equation}
\label{eq:many-null-vectors-condition}
\vec{M}_{j}^T K^{-1} \vec{M}_{k} = 0 \,,
\end{equation}
which generalizes Eq.~\eqref{eq:localization-null}.
In this case, the localization may take place simultaneously in all $m$ independent channels governed by these null-vectors $\vec M_j$. 

 To analyze a topologically unstable edge, we perform a basis transformation by a matrix $W\in {\rm SL}(d, \mathbb{Z})$ such that
\begin{align} \label{eq:wtransformation}
&  W^T K W  =  K_{\text{red}} \oplus \underbrace{q_1 \sigma_z \oplus \cdots \oplus q_m \sigma_z}_{m\,\, \text{copies}} \,, \nonumber \\
&  W^T \vec{t}  =  \vec{t}_{\text{red}} \oplus [ (t_{\text{loc}}^1,t_{\text{loc}}^1)^T]\oplus \cdots  [ (t_{\text{loc}}^m,t_{\text{loc}}^m)^T]  \,,
\end{align}
where $q_1, \ldots, q_m$ are odd integers.
This implies that, upon transformation to a new basis,
the reduced, topologically stable sector governed by $K_\text{red}$ and $\vec t_\text{red}$  is decoupled from the modes undergoing localization. 
The procedure for deriving a suitable $W$ transformation is detailed in Appendix.~\ref{appen:bindingtransition}. The matrix $W^{-1}$, which transforms currents  according to $\tilde{\vec{I}} = W^{-1}\vec{I}$, can be written as 
\begin{align} \label{eq:inversew}
  W^{-1}=\begin{pmatrix}\vec{e}_1^{\text{red}}, & \cdots, & \vec{e}_{d-2m}^{\text{red}},
  & \vec{w}_1^{\text{loc}},  & \cdots, & \vec{w}_m^{\text{loc}}
    \end{pmatrix}^{T}\,,
\end{align}
where $\vec{e}_i^{\text{red}}$ are basis vectors of the reduced theory, and $\vec{w}_j^{\text{loc}}= (\vec{e}_j^{\text{loc}}, \vec{M}_{j} - \vec{e}_j^{\text{loc}})$ represents a pair of counter-propagating modes. 

In the simple example at the beginning of this section, with $K=\diag(1,1,-1)$ and $\vec{t}^T = (1,1,1)$, only one pair of modes can get localized. For the localization governed by the null-vector $\vec M^T = (0,1,1)$, the theory is in the form \eqref{eq:wtransformation} with $m=1$ and $q_1=1$ without any basis change needed (i.e., $W=\mathbb{1}$). The pair of modes $\vec{w}^{\rm loc}$ consists of vectors $\vec e^{\rm loc} = (0,1,0)^T$ and 
$\vec M - \vec e^{\rm loc} = (0,0,1)^T$. The reduced theory has a single mode, with 
$\vec e^{\rm red} = (1,0,0)^T$, $K=1$, and $t=1$.

It is instructive to compare  Eq.~\eqref{eq:wtransformation} with the form of the $K$-matrix of a topologically unstable theory as derived in the pioneering paper \cite{Haldane_Stability_1995} by Haldane. 
In Eq. (4) of Ref.~\cite{Haldane_Stability_1995}, the block that represents the topologically unstable sector (and thus can undergo localization) has the form
\begin{align} 
\label{eq:Haldane-K}
    K_0=\begin{pmatrix}
0 & 1 \\ 1 & k
\end{pmatrix},\qquad 
\vec{t}_0=\begin{pmatrix}
0  \\ b
\end{pmatrix},
\end{align} 
with either (i) $k = 1$ and odd $b$ or (ii) $k=0$ and even $b$. In the first case, $k=1$, it is easy to see that there is a $\text{SL}(2,\mathbb{Z})$ rotation that transforms $K_0 \to \sigma_z$ and $\vec{t}_0 \to (b,b)^T$ in correspondence with our Eq.~\eqref{eq:wtransformation} with $q_j=1$. The $q_j$ factor is absent in Eq. (4) of Ref.~\cite{Haldane_Stability_1995} because it is written for a ``primitive-form'' theory (as defined in Ref.~\cite{Haldane_Stability_1995}), while our Eq.~\eqref{eq:wtransformation} does not involve this assumption. In the second case, $k=0$, we have $K_0 = \sigma_x$ in Eq.~\eqref{eq:Haldane-K}. While $\sigma_x$ cannot be $\text{SL}(2,\mathbb{Z})$-transformed to $\sigma_z$, we should take into account that there is also the remaining sector of the theory $K_{\rm red}$. As was shown in \cite{Cano_Bulk_edge_2014}, for any odd matrix $K_{\rm red}$, there exists $W \in \text{SL}(d,\mathbb{Z})$  such that 
\begin{equation}
W^T ( K_{\rm red} \oplus \sigma_x) W = 
K_{\rm red} \oplus \sigma_z \,.
\label{eq:loc-sigmax-to-sigmaz}
\end{equation}
Here, the matrix $K$ is called odd if at least one of its diagonal elements is odd; otherwise (all diagonal elements are even), it is called even. The matrix $K$ representing a fermionic system must be odd (since otherwise, it would not have fermionic excitations). Correspondingly, if in some basis it has a form  $K_{\rm red} \oplus \sigma_x$, the matrix $K_{\rm red}$ must be odd. Thus, we can use Eq.~\eqref{eq:loc-sigmax-to-sigmaz} to trade $\sigma_x$ to $\sigma_z$. For several independent localization channels, as in Eq.~\eqref{eq:wtransformation}, this can be done consecutively for all $\sigma_x$ involved. Therefore, Eq.~\eqref{eq:wtransformation} is fully generic for the case of a fermionic (odd) $K$-matrix that we assume.

\subsection{Conductance of an edge with localization}

\begin{figure}[t!]
\includegraphics[width =\columnwidth]{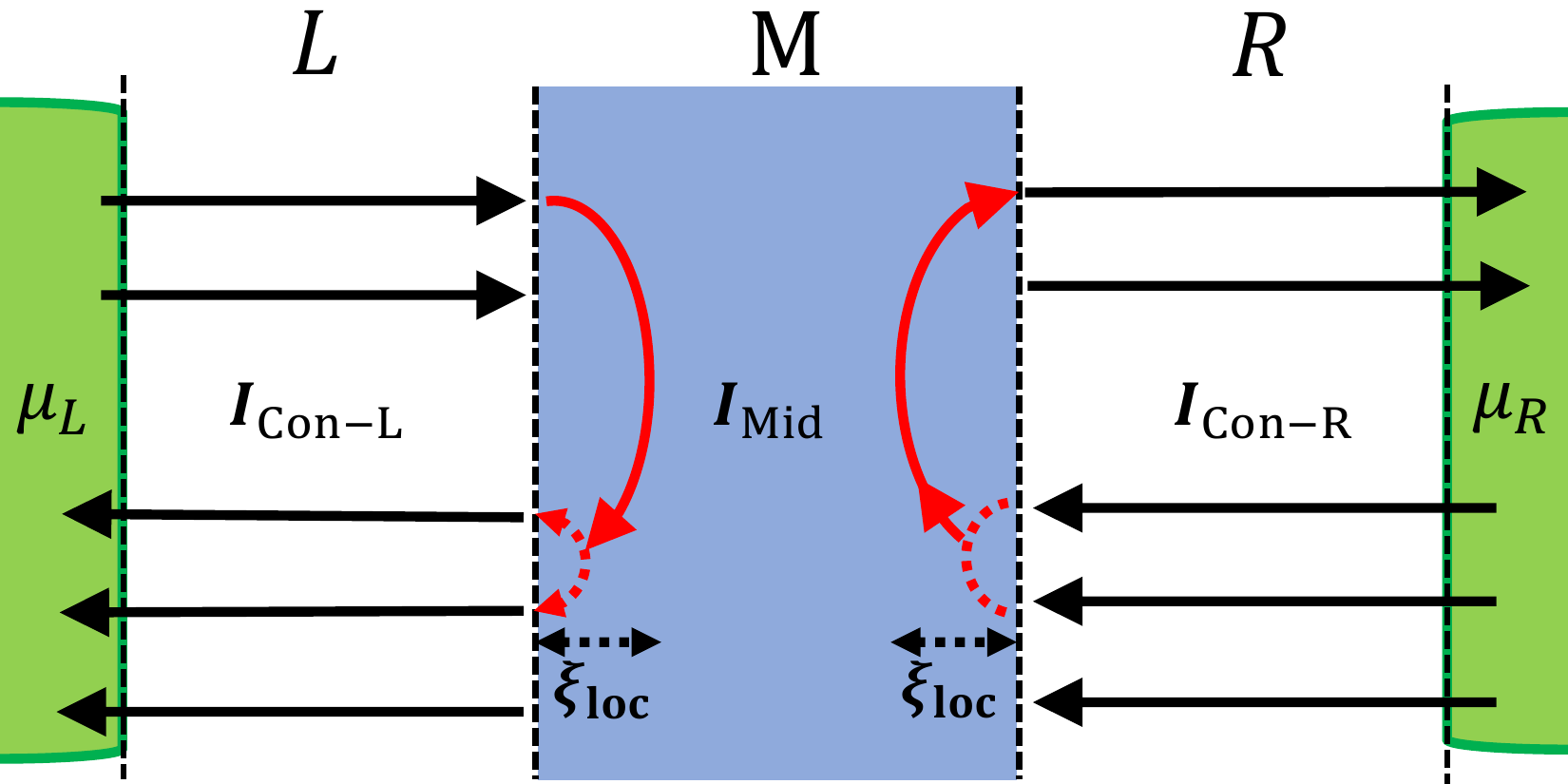}
\caption{
{\bf Schematic setup for transport through a localization region}, depicted as blue-shaded. 
Vertical dashed lines on the left and right depict interfaces with contacts (green regions). 
Edge segments L and R are clean so that each individual component of the current vector $\vec{I}_{\text{Con-L}}$ ($\vec{I}_{\text{Con-R}}$) flowing through the cross-section with the left (right) contact is conserved along the segment L (respectively, R). The disordered region M has a length much larger than the localization length $\xi_{\rm loc}$. The current vector in the central part of this region is denoted $I_{\rm Mid}$.
} 
\label{Fig:localization}
\end{figure}

We now consider the conductance of a disordered edge that undergoes localization, Fig.~\ref{Fig:localization}. Specifically, we consider 
an edge with $n_R$ right-moving and $n_L$ left-moving modes coupled to two contacts with potentials $\mu_L$ and $\mu_R$. In the central part M (blue, shaded region) of the edge, the random inter-mode scattering is operative and leads to localization, as discussed in Sec.~\ref{subsec:localzationbasictheory}. 
The length of this segment is much larger than the localization length $\xi_{\rm loc}$. For convenience, we assumed in Fig.~\ref{Fig:localization} that there are clean regions L and R that connect the disordered region M to the contact regions (green). The presence of these ballistic regions is immaterial; the conductance does not depend on their lengths. 


 The currents emanating from the left and right contacts $\vec I_\text{Con-L}$ and $\vec I_\text{Con-R}$ satisfy the corresponding boundary conditions 
given by Eqs.~\eqref{eq.I_in_f}, \eqref{eq:bc_for_f}. 
The boundary conditions yield $n_R$ equations for $\vec I_\text{Con-L}$ and $n_L$ equations for $\vec I_\text{Con-R}$, i.e., in total $n_R+n_L=d$ equations for $2d$ current components $\vec I_\text{Con-L}$ and $\vec I_\text{Con-R}$. We will now determine the remaining $d$ equations that are needed to find the currents.

In the middle region, the localization governed by $m$ independent null-vectors $\vec{M}_j$ with $1\leq j \leq m$ takes place. According to Eq.~\eqref{eq:loc-current-condition}, this implies that
\begin{equation}
\label{eq:localization-condition-current-1}
\vec{M}_j^T \vec {I_{\rm Mid}}=0 \,, \qquad j =1, \ldots, m \,,
\end{equation}
where $\vec I_{\rm Mid}$ is the current in the localized region. Let us recall that tunneling operators that we consider are $e^{i \vec M_j^T \vec \phi}$. This tunneling conserves the density combinations $\vec M_i^T \vec \rho$ in view of
\begin{equation}
[\vec M_i^T \vec \rho(x), \vec M_j^T \vec \phi(x')] = i \vec M_i^T K^{-1} \vec M_j \delta(x-x') = 0 \,,
\end{equation}
where we used Eqs.~\eqref{eq.commutation}
and \eqref{eq:many-null-vectors-condition}. Thus, $\partial_t \vec M^T_i \vec \rho = 0$ and, consequently,
\begin{equation}
\label{eq:localization-condition-current-2}
\partial_x \vec M^T_i \vec I (x) = 0 \,, 
\qquad i =1, \ldots, m\,.
\end{equation}
It follows from Eqs.~\eqref{eq:localization-condition-current-1} and \eqref{eq:localization-condition-current-2} that $\vec{M}_j^T\vec I=0$ holds everywhere in the edge, including the ballistic regions L and R adjacent to the contacts:
\begin{align} \label{eq:nullvectorconditions}
\vec{M}_j^T \vec{I}_\text{Con-L}= \vec{M}_j^T \vec{I}_\text{Con-R}= 0
\,,
\end{align}
which yields $2m$ equations for the current components.
Furthermore, the currents of the reduced theory $\vec{I}_\text{red} = [W^{-1} \vec{I}]_{a\leq d-2m}$ are not affected by tunneling in the localized sector and, therefore, are continuous. This imposes $d-2m$ conditions
\begin{align} \label{eq:reducedcurrent}
(\vec{e}^\text{red}_a)^T \vec{I}_\text{Con-L} = (\vec{e}^\text{red}_a)^T \vec{I}_\text{Con-R} \,,
\end{align}
where we used the form  Eq.~\eqref{eq:inversew} of the matrix $W$. In total, the boundary conditions Eqs.~\eqref{eq.I_in_f}, \eqref{eq:bc_for_f} in combination with Eqs.~\eqref{eq:nullvectorconditions} and \eqref{eq:reducedcurrent} provide the required
$2d$ equations for $2d$ current components $\vec{I}_\text{Con-L}$ and $\vec{I}_\text{Con-L}$. 
By solving them, we can find the current components and thus the total charge current and the conductance.

To derive a general formula for the conductance, it is convenient to parameterize the currents $\vec{I}_{\text{Con-s}}$, with $s= L, R$, in terms of the chemical potentials $\vec{\mu}_{\text{Con}}^s$  
of modes in the ballistic regions L and R
via the relation~\eqref{eq:relationcurrentchemicalpot}:
\begin{align} \label{eq:Iconleftright}
 \vec{I}_{\text{Con-L}} = \frac{1}{2\pi} K^{-1} \vec{\mu}^L_{\text{Con}}\,, \quad 
 \vec{I}_{\text{Con-R}} = \frac{1}{2\pi} K^{-1} \vec{\mu}^R_{\text{Con}}\,. 
\end{align}
The boundary conditions on interfaces with contacts, Eqs.~\eqref{eq.I_in_f} and \eqref{eq:bc_for_f}, imply
\begin{align}
\label{eq.continutity:localization}
    \frac{\mu_L}{2\pi} K^{-1}\vec{t} + U^{-1}\chi_L \vec{C}_L &=  \frac{1}{2\pi} K^{-1} \vec{\mu}^L_{\text{Con}}\,, \nonumber \\ 
    \frac{\mu_R}{2\pi} K^{-1}\vec{t} + U^{-1}\chi_R \vec{C}_R &=  \frac{1}{2\pi} K^{-1} \vec{\mu}^R_{\text{Con}}\,.
\end{align}
Acting with the matrix $\chi_R U$ on the first of these equations and with the matrix $\chi_L U$ on the second one, we eliminate the unknown coefficients $\vec{C}_L$ and $\vec{C}_R$ and obtain
\begin{align}
\label{eq.continutity:localization2}
    \frac{\mu_L}{2\pi} \chi_R U K^{-1}\vec{t}&= \frac{1}{2\pi} \chi_R U K^{-1} \vec{\mu}^L_{\text{Con}}\,, \nonumber \\ 
    \frac{\mu_R}{2\pi} \chi_L U K^{-1}\vec{t} &= \frac{1}{2\pi} \chi_L U K^{-1} \vec{\mu}^R_{\text{Con}}\,.
\end{align}
Next, according to the conditions \eqref{eq:nullvectorconditions}, we can expand the chemical potentials $\vec{\mu}^{s}_{\text{Con}}$ as
\begin{align} 
\label{eq:localchemicalpotential}
    \vec{\mu}^s_{\text{Con}} = \sum_{a =1}^{d - 2m}\mu^{s, a}_{\text{red}} \vec{e}_{a}^{\text{red}} + \sum_{j =1}^{m}\mu^{s}_{j} \vec{M}_j\,, \qquad s=L,R \,.
\end{align}
The first term on the right-hand side of Eq.~\eqref{eq:localchemicalpotential} corresponds to the reduced-theory sector, spanned by 
the modes $\vec{e}_a^{\text{red}}$ that do not participate in the localization processes and remain conducting through the localization region. The second term involves modes that get localized.
 The current conservation of the reduced part, Eq.~\eqref{eq:reducedcurrent}, results in 
\begin{align} 
\label{eq:localcontinuity}
    \mu_{\text{red}}^{L,a} =  \mu_{\text{red}}^{R,a} \equiv \mu_{\text{red}}^a\,. 
\end{align}
We have thus in total $d$ unknown chemical potentials: $d-2m$  potentials  $\mu_{\text{red}}^a$ of the reduced modes and $2m$ potentials $\mu_j^L$, $\mu_j^R$ in the localized sector. They are determined from the system of $d$ equations, Eq.~\eqref{eq.continutity:localization2}. 
We present only the result here; details of the derivation can be found in Appendix~\ref{app:condutanceformula}. The total current reads
\begin{align} \label{eq:totalcurrentloc}
     J_{\text{tot}} = \frac{\nu}{4 \pi } (\mu_L + \mu_R) +   \frac{G}{4\pi} (\mu_L - \mu_R)\,,
\end{align}
where we have defined the two-terminal conductance as before [see Eqs.~\eqref{eq.Jtot} and \eqref{eq:J-two-term}].  The final result for the conductance reads
\begin{align} 
\label{eq:twoterminalcondlocalization}
    G =  \sum_{a, b =1}^{d -2m} t_\text{red}^a (W^T U^T U W)^{-1}_{ab} t_{\text{red}}^b - \sum_{j, j' = 1}^{m } B_{j} C^{-1}_{jj'} B_{j'}\,, 
\end{align}
with
\begin{align}
 B_j &= \sum_{a = 1}^{d -2m} t_{\text{red}}^a ((U^T U W)^{-1} \vec{M}_j)_a \,, \\ 
  C_{jj'} &= \vec{M}_j^T (U^T U)^{-1}\vec{M}_{j'}\,. 
\end{align}

As a sanity check, in the absence of localization ($m=0$), we have $ \vec{t}_{\text{red}} = W^T \vec{t} $ according to Eq.~\eqref{eq:wtransformation} so that Eq.~\eqref{eq:twoterminalcondlocalization} reproduces the ballistic conductance Eq.~\eqref{eq:condclean}.

If the theory becomes fully chiral (i.e., all modes propagate in the same direction) after localization, the resulting conductance should be $G = \nu$. We show in  Appendix~\ref{app:condutanceformula} that Eq.~\eqref{eq:twoterminalcondlocalization} indeed reduces to $G = \nu$ for this special class of edges.  

\subsection{Conductance of edge junctions with localization. Interplay of localization and equilibration.} 
\label{sec:localizationandequilibration}

\begin{figure}[t!]
\includegraphics[width =\columnwidth]{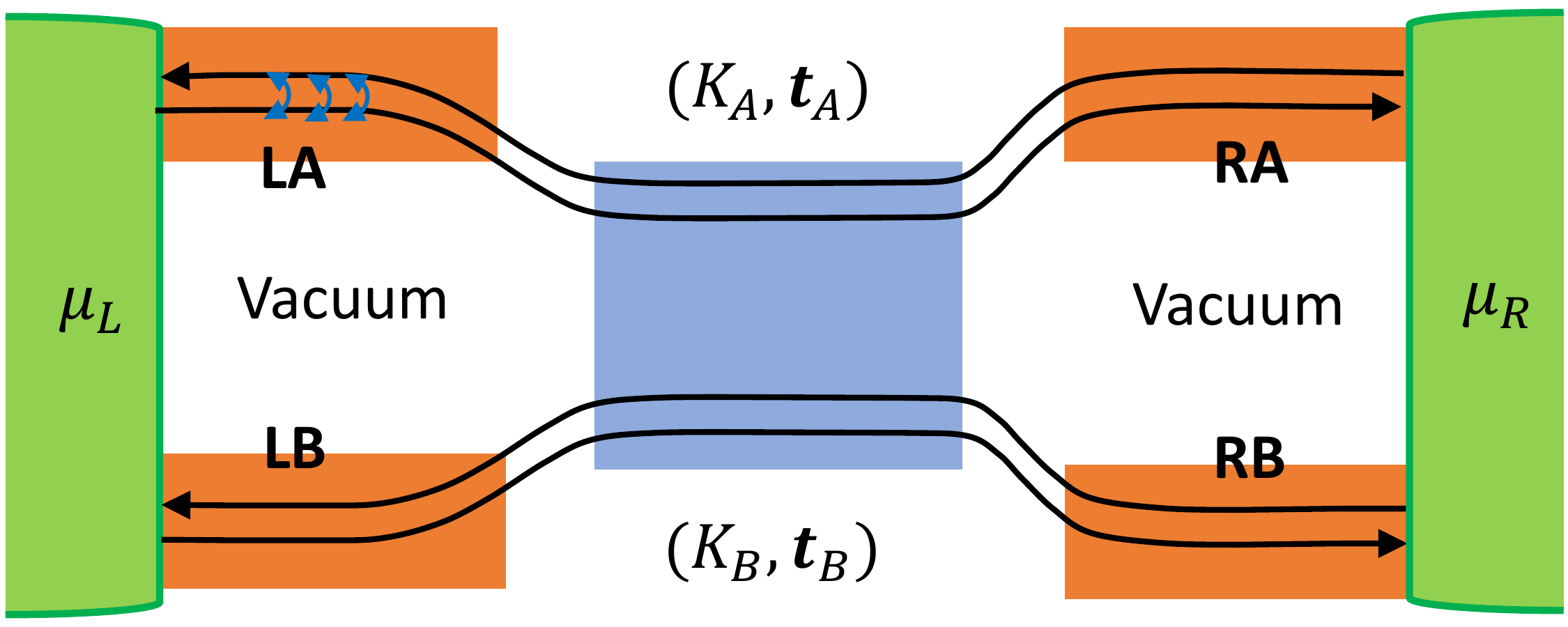}
\caption{ {\bf The junction at the interface between two different FQH bulk states} characterized by the respective sets of $K$-matrix and $\vec{t}$-vector, $(K_A, \vec{t}_A)$ and  $(K_B, \vec{t}_B)$. The edges A and B are in proximity to each other in the middle region (shaded blue), where the localization may occur. The edge segments LA, LB, RA, and RB bridge the localization region to the contacts that undergo the incoherent equilibration. The transport in these segments may be either ballistic (coherent) or equilibrated (incoherent). 
} \label{Fig:interface}
\end{figure}

The question of localization is particularly relevant to interfaces of two FQH phases, naturally giving rise to multi-mode edge junctions with counter-propagating modes. We now consider the conductance of a junction of two FQH edges as shown schematically in Fig.~\ref{Fig:interface}.  Specifically, we consider an interface between two different FQH states, A and B. We assume that the edges A and B are in close proximity of each other in the central part of the device (blue shaded region in Fig.~\ref{Fig:interface}) so that 
the localization processes in the combined edge become operative. This combined edge is described by the $K$-matrix and $\vec t$-vector given by
\begin{align}
     K_\text{AB} = K_\text{A} \oplus (-K_\text{B})\,, \quad 
    \vec{t}_\text{AB} = \vec{t}_\text{A} \oplus \vec{t}_\text{B}\,.
    \label{eq:KAB-tAB}
\end{align}
The transport in this AB junction is assumed to be coherent, i.e., the length of the junction is much shorter than the inelastic equilibration length. 

In the remaining part of the device, the edges A and B are assumed to be far apart so that they do not affect each other. The transport in the corresponding segments LA, LB, RA, and RB may be either coherent (ballistic) or incoherent (equilibrated) depending on whether the length of these segments is smaller or larger than the inelastic equilibration length. 
To study the case with some degree of incoherent equilibration present in these segments, we can equivalently shift this equilibration in the contact region by modifying the corresponding boundary conditions, see Appendix \ref{app.gen_sol}.

Only an integer charge is allowed to tunnel between the edges A and B. Nevertheless, it turns out that localization is not only possible in such junctions but is, in fact, ubiquitous when the total number of counter-propagating modes in the combined edge is four or larger. As a result of localization, a reduced theory $(K_{\rm red}, \vec t_{\rm red})$ will emerge in the junction, thus making the two edges A and B ``entangled''. In general, this localization will also modify the two-terminal conductance $G$ of the device. 

We recall that the two-terminal conductance $G$ is defined for a setup including two identical edge segments, or two identical edge interfaces of the type shown in Fig.~\ref{Fig:interface}. Experimentally, it can be also measured in a three-terminal setup,  
see a discussion in Sec.~\ref{sec:ballistic-cond}, which allows one to determine $G$ in a structure with a single junction.

\section{Examples of edges with localization and their conductances}

We now proceed by presenting several characteristic examples of FQH edges and edge junctions that are topologically unstable and thus prone to localization. For each of them, we will use our general formalism to calculate the form of the reduced theory as well as the resulting conductance. 

\subsection{$\nu = 9/5$ state} \label{sec:localizationexample-95}

The $\nu = 9/5$ state is arguably the simplest example of a three-mode edge that exhibits binding transition, as was shown in Ref.~\cite{Kao_Binding_1999}. Manifestations of localization in transport properties of the 9/5 edge were studied in Ref.~\cite{Spanslatt_binding_2023}. We use the $\nu = 9/5$ state as a first test for our general theory. The $\nu = 9/5$ edge consists of two right-moving integer modes and one left-moving 1/5 mode and is characterized by 
\begin{align}
    K = \begin{pmatrix} 1 & 0 & 0 \\ 0&1&0 \\ 0&0&-5 \end{pmatrix}\,, \quad \vec{t} =  \begin{pmatrix} 1  \\ 1 \\1 \end{pmatrix}\,.
\end{align}
The above three modes are spatially well-separated so that the matrix $\Upsilon$ of coupling to contacts is diagonal in the same basis. The ballistic conductance of the edge in the absence of localization is thus 
\begin{equation}
G = \frac{11}{5}
\label{G-95-ball-no-loc}
\end{equation}
according to Eq.~\eqref{G-commuting-K-Upsilon}.

The theory possesses two competing null-vectors $\vec{M}_1 = (-2, 1, -5)^T$ and $\vec{M}_2 = (1, -2, -5)^T$ (they are related by an exchange of the two integer modes and, in this sense, equivalent). After localization in any of these channels, the reduced theory possesses a single conducting mode and is characterized by 
\begin{equation}
K_{\text{red}} = 3 \,, \qquad t_{\text{red}}= 5 \,.
\end{equation}
Assuming for definiteness that the localization is driven by $\vec{M}_1$, the $W^{-1}$ matrix brings the theory to the form
\eqref{eq:wtransformation} (with $m=1$ and $q_1=1$) is
\begin{align} \label{eq:winv95}
 W^{-1} = \begin{pmatrix}
 0 & 1 & 2 \\ 1 & 0 & 0 \\ 0 & -2 & -5 
\end{pmatrix},
\end{align}
see Appendix~\ref{appen:bindingtransition} for more details on the scheme to find $K_{\text{red}}$, $\vec{t}_{\text{red}}$, and $W$.

Since the reduced theory is fully chiral, the conductance after localization is
\begin{align}
    G = t_{\text{red}} K_{\text{red}}^{-1} t_{\text{red}} = \nu = \frac95 \,.
\label{G-loc-95}
\end{align} 
We have verified that our formula 
\eqref{eq:twoterminalcondlocalization}
for $G$ indeed yields this result. For the generality of this test, we have also allowed for an arbitrary degree of equilibration between the modes in the contact region
(which would modify the value of the conductance \eqref{G-95-ball-no-loc} in the absence of localization; see Appendix \ref{app.gen_sol}).  We recall that the matrix $U$ entering \eqref{eq:twoterminalcondlocalization} is  $U = O_{\Upsilon} M_K$, where $M_K = \text{diag} (1, 1, \sqrt{5})$ and $O_{\Upsilon}$ in general depends on the coupling strengths and inter-mode equilibration at contacts. The matrix $O_{\Upsilon} \in \text{SO}(2, 1)$ can be written as $O_{\Upsilon} = B R$ with a boost matrix $B$ and a rotation matrix $R$. While $R$ is parameterized by a single angle $\theta$, the boost matrix $B$ is parameterized by a two-component vector $\vec{p} = (p_1, p_2)$ as~\cite{Moore_Classification_1997}
\begin{align}
    B = \begin{pmatrix} \frac{p_2^2+ \gamma p_1^2}{p^2} & \frac{p_1 p_2 (\gamma-1)}{p^2} & p_1\\ \frac{p_1 p_2 (\gamma-1)}{p^2}  &\frac{p_1^2+ \gamma p_2^2}{p^2}&p_2 \\ p_1&p_2&\gamma \end{pmatrix} \!,\,\,R= \begin{pmatrix} \cos \theta &  -\sin \theta & 0\\ \sin \theta&\cos \theta &0 \\ 0&0&1 \end{pmatrix}\!,
\end{align}
with $\gamma = \sqrt{1 + p^2}$.  
We have checked that, with this most general form of coupling to the contacts, the two-terminal conductance formula~\eqref{eq:twoterminalcondlocalization} indeed yields the universal conductance \eqref{G-loc-95} of the 9/5 edge upon localization.

\subsection{$\nu = 8/3$ state} \label{sec:localizationexample-83}

While localization is a rare occurrence for three-mode edges, it becomes more of a rule than an exception for edges with $d \ge 4$ modes. As an example, we now consider the  $\nu = 8/3$ edge, which consists of three integer right-moving modes and one left-moving 1/3 mode. The corresponding $K$-matrix and $\vec t$-vector read
\begin{align}
    K = \text{diag} (1, 1, 1, -3)\,, \qquad 
    \vec t^T = (1,1,1,1)\,. 
\end{align}
These four modes are spatially well-separated so that the matrix $\Upsilon$ of coupling to contacts is diagonal in the same basis. Therefore, the ballistic conductance of the edge in the absence of localization is  
\begin{equation}
G = \frac{10}{3}
\label{G-83-ball-no-loc}
\end{equation}
according to Eq.~\eqref{G-commuting-K-Upsilon}.

This edge is topologically unstable and, moreover, there are multiple competing null-vectors. To see this, consider first the general form of $\vec{M}$ satisfying the neutrality condition $\vec{t}^T K^{-1} \vec{M}= 0$:
 \begin{equation}
 \vec{M}_{n_c, n_1, n_2} = (n_c + n_1 + n_2, -n_1, -n_2, 3 n_c)^T \,,
 \end{equation}
 with three integers $(n_c, n_1, n_2)$.
Here, $n_c$ denotes the number of electrons tunneling to the $1/3$ mode, whereas $n_1$ and $n_2$ describe the reshuffling of electrons within the integer modes. By imposing the null-vector condition $\vec{M}_{n_c, n_1, n_2}^T K^{-1} \vec{M}_{n_c, n_1, n_2} = 0$, we arrive at the equation 
\begin{align} \label{eq:nullvectorcondition83}
    (n_c+n_1+n_2)^2 + n_1^2 + n_2^2= 3 n_c^2. 
\end{align}
This equation has many (presumably infinitely many) integer-valued solutions. Note that 
we focus on {\it primitive} null-vectors, i.e., such that the components $\vec M_a$ do not have a common divisor. Obviously, a multiplication of a null-vector by any integer yields again a null-vector. We are not interested in such non-primitive null-vectors since they are always less relevant in the renormalization-group sense than the corresponding primitive null-vectors and, also, the corresponding tunneling processes have smaller amplitudes.

It is easy to see from Eq.~\eqref{eq:nullvectorcondition83} that $n_c$ should be an odd number.
Among the multiple competing null-vectors, we consider those with the smallest $n_c$: single-electron tunneling ($n_c = 1$) to the $1/3$ mode and three-electron tunneling ($n_c = 3$).  For the electron tunneling, we find three null-vectors: $\vec{M}_{1, -1, -1} = (-1, 1, 1, 3)^T$, $\vec{M}_{1, -1, 1} = (1, 1, -1, 3)^T$, and $\vec{M}_{1, 1, -1}  = (1, -1, 1, 3)^T$. Clearly, they are related by permutation of integer modes, and thus, the localization in any of these channels will result in the same reduced theory. Using the formalism presented in Appendix \ref{appen:bindingtransition}, we find:
\begin{align} \label{eq:reducedtheoryelectron}
    K_{\text{red}} = \begin{pmatrix} 2 & 1  \\ 1&2 \end{pmatrix}\,, \quad \vec{t}_{\text{red}} =   \begin{pmatrix} 2   \\ 2 \end{pmatrix}\,,
\end{align}
which can be interpreted as a bosonic Jain state~\cite{Wu2013} of charge-two particles. This state can be constructed by starting from the fermionic integer quantum Hall state with $K=\text{diag}(1,1)$ and attaching a flux quantum to each particle.

Note that the reduced theory that resulted from localization has a bosonic character: the diagonal elements of the $K$-matrix and the components of the $\vec{t}$-vector are even. This theory does not contain fermionic (electron) excitations: all integer-charged excitations are bosons with an even charge.

Similarly, we find null-vectors involving the tunneling of three electrons to the 1/3 mode:  $\vec{M}_{3, 1, -5} = (-1,-1,5,9)^T$,
$\vec{M}_{3, -5, 1} = (-1,5,-1,9)^T$, and $\vec{M}_{3, 1, 1}= (5,-1,-1,9)^T$. The localization in any of these channels results in a reduced theory that turns out to be characterized (in an appropriately chosen basis) by the same $(K_{\text{red}}, \vec{t}_{\text{red}})$ as given by Eq.~\eqref{eq:reducedtheoryelectron}. 


While the reduced theory contains two modes, they are co-propagating. Thus, the theory becomes fully chiral after localization so that the conductance of the edge is
\begin{equation}
G = \nu = \frac83 \,,
\label{G-83-after-loc}
\end{equation}
independent of any details of the coupling to contacts. Indeed, we have verified that for any matrix $O_{\Upsilon}$ describing coupling to contacts (which in this case can be parameterized by a boost matrix with three parameters and a rotation matrix with three parameters), application of Eq.~\eqref{eq:twoterminalcondlocalization} gives $G = 8/3$ for the reduced theory, Eq.~\eqref{eq:reducedtheoryelectron}. 

In the following subsections, we will consider examples of four-mode edge junctions for which the two modes remaining after localization propagate in opposite directions, i.e., $K_{\text{red}}$ has the signature  $(1,-1)$. We will see that this leads to a still richer physics. In particular, the value of the conductance $G$ after localization will depend on which null-vector wins the competition for governing the localization.

\subsection{Interface between $\nu = 2/5$ and $\nu = 2/9$ states} \label{sec:localizationexamples2529}

We now consider an edge junction that is formed at the interface between the $\nu = 2/5$ and $\nu = 2/9$ FQH states; see the setup in Fig.~\ref{Fig:interface}. The 2/5 and 2/9 edges (that we label by A and B, respectively) are characterized by $K$-matrices
   \begin{align}
     K_\text{A} = \begin{pmatrix} 3 & 2  \\ 2& 3 \end{pmatrix}\,, \quad  K_\text{B} = \begin{pmatrix} 5 & 4  \\ 4& 5 \end{pmatrix}\,,
\end{align}
both with $\vec t = (1,1)^T$. Taken separately, each of these edges consists of two co-propagating modes. 
In view of this, and since it is assumed that the 2/5 and 2/9 edges are far apart in the contact regions
(see Fig.~\ref{Fig:interface}), the conductance is independent of the form of coupling of these edges to the contacts. When the 2/5 and 2/9 edges are fully decoupled (i.e., there is no junction in the middle region), the two-terminal conductance is 
\begin{equation}
G = \frac25 + \frac29 = \frac{28}{45}.
\end{equation}

When a junction is formed in the central region, the composite four-mode edge is described by $K_{\text{AB}} = K_{\text{A}} \oplus (- K_{\text{B}})$ and $\vec t = (1,1,1,1)^T$, with two right-moving and two left-moving modes. Only electrons, or integer multiple of electrons, are allowed to tunnel across the vacuum region between A and B so that allowed vectors $\vec{M}$ satisfying the charge neutrality condition have the form 
\begin{align} \label{eq:nullvector2529}
    \vec{M} _{n_c, n_A, n_B} = K_{\text{AB}} (n_A, n_c-n_A, -n_B, -n_c + n_B)^T
\end{align}
with three integers $(n_c, n_A, n_B)$. Here, $n_c$ denotes the charge that tunnels between A and B, while $n_A$ and $n_B$ represent the reshuffling of charge within the edges of A and B, respectively. The null-vector condition $\vec{M}_{n_c, n_A, n_B}^T K_{\text{AB}}^{-1} \vec{M}_{n_c, n_A, n_B} =0$ yields the equation
\begin{align} 
\label{eq:nullvectorsol}
   (n_A - n_B) (n_A + n_B - n_c) = n_c^2\,.
\end{align}

It is easy to see that Eq.~\eqref{eq:nullvectorsol} possesses infinitely many primitive solutions and, moreover, that there is at least one such solution for any even $n_c$.  (It is also easy to see that there are no solutions for odd $n_c$.) Among these multiple competing null-vectors, we consider those with the two smallest numbers of electrons tunneling between A and B: $n_c=0$ (neutral tunneling) and $n_c=2$ (two-electron tunneling). For the neutral tunneling, Eq.~\eqref{eq:nullvectorsol} has two solutions $n_A = n_B = 1$ and $n_A = -n_B = 1$. The corresponding two null-vectors $\vec{M}_{0, 1, 1} = (1,-1,1,-1)^T$ and $\vec{M}_{0, 1, -1} = (1,-1,-1,1)^T$ are related by a simple permutation of modes within the edge B. Localization driven by any of them yields the reduced theory
\begin{align} \label{eq:interface2529}
    K_{\text{red}} = \begin{pmatrix} 22 & 23  \\ 23& 22 \end{pmatrix}\,, \quad  \vec{t}_{\text{red}} = \begin{pmatrix} 2  \\ 2 \end{pmatrix}\,, 
\end{align}
which can be viewed as a hole-like bosonic Jain state of charge-two particles. This state can be constructed by starting with a fermionic theory with $K= \text{diag}(-1,-1)$ and adding 23 flux quanta to each particle.

Similarly to the above case of the 8/3 edge, the localization thus converts a four-mode fermionic theory into a two-mode bosonic theory.   
Evaluating the conductance $G$ according to Eq.~\eqref{eq:twoterminalcondlocalization}, we find
\begin{align}
    G = \frac{28}{45}\,,
\end{align}
so that the localization in this neutral channel does 
not affect the conductance. 

We turn now to the case of two-electron tunneling, $n_c = 2$.  Equation \eqref{eq:nullvectorsol} yields two primitive solutions $(n_A = -1, n_B = 1)$ and $(n_A = 3, n_B = 1)$, which yield the null-vectors $\vec{M}_{2,-1,1}= (3,7,9,9)^T$ and $\vec{M}_{2,3,1} = (7,3,9,9)^T$ (related by a permutation of modes in the 2/5 edge). The localization due to any of those null-vectors results in the same reduced theory as Eq.~\eqref{eq:interface2529}.
At the same time,  the conductance takes upon localization a very different value 
\begin{align} \label{eq:interface2529cond}
    G = \frac{8}{45}\,. 
\end{align}

This analysis of the (2/5, 2/9) junction explicitly shows that when the reduced theory involves counter-propagating modes, different localization processes may give rise to different values of $G$. Remarkably, this happens in the above example of two localization channels (with $n_c=0$ and $n_c=2$) despite the fact that the resulting reduced theories $(K_{\text{red}}, \vec t_{\text{red}})$ are equivalent.

\subsection{Interface between spin-polarized and spin-unpolarized $\nu = 2/3$ states}  \label{sec:interfaces:spinpolspinunpol}

As another example of localization in a four-mode junction,  we consider an interface between spin-polarized and spin-unpolarized $\nu = 2/3$ states. 
This example is inspired by Refs.~\cite{ponomarenko2023unusual, Wang2021},
which considered tunneling contact between two such states. Our setup, shown in Fig.~\ref{Fig:interface}, is, however, in general different, although a direct relation emerges in the limit of strong equilibration in the segments LA, LB, RA, and RB. We will return to this relation at the end of this subsection. 

As before, the composite edge in the junction is described by $K_{\text{AB}} = K_\text{A} \oplus (- K_\text{B})$,
with $K_\text{A}$ and $K_\text{B}$ representing now edge theories of $\nu =2/3$ spin-unpolarized and spin-polarized states, respectively,
\begin{align}
     K_\text{A} = \begin{pmatrix} 1 & 2  \\ 2&1 \end{pmatrix}\,, \quad  K_\text{B} = \begin{pmatrix} 1 & 0  \\ 0&-3 \end{pmatrix}\,. 
\end{align}
The corresponding $\vec{t}$-vector is $\vec{t}_{\text{AB}} = (1, 1, 1, 1)^T$. 
The transport in spin-polarized and spin-unpolarized 2/3 edges separately is discussed above in Sec.~\ref{sec:commuting-K-Upsilon} and \ref{sec:non-commuting-K-Upsilon}. 

Similarly to the example of the interface of $(2/5,2/9)$ states in the previous subsection, only electrons, or integer multiple of electrons, are allowed to tunnel across the vacuum region between A and B. Thus, allowed vectors $\vec{M}$ satisfying the charge-neutrality condition have the same form as \eqref{eq:nullvector2529}. At variance with the previous example, the theories of A and B edges in the present case are, however, topologically equivalent. Thus, if we allow for localization without any further restrictions, all four modes can get localized, leaving an empty theory with no propagating modes (and, obviously, zero conductance).  We impose such a restriction by assuming spin conservation, which implies the following condition on $\vec{M}$: 
\begin{align} \label{eq:spinconservation}
    \vec{s}^T K_{\text{AB}}^{-1} \vec{M}_{n_c, n_A, n_B} = 0\,,
\end{align}
where $\vec{s} = (1, -1, 1, 1)^T$ is the vector that encodes the spin of individual edge components similarly to $\vec{t}$ that encodes their charge. This form of the spin vector reflects the fact that $K_A$ is written in the basis of spin-up and spin-down modes, while the edge B is fully spin-polarized. 
This condition leads to $n_A = n_c$  for the neutral vectors \eqref{eq:nullvector2529}. The null-vector condition $\vec{M}_{n_c, n_A, n_B} ^T K_{\text{AB}}^{-1} \vec{M}_{n_c, n_A, n_B} = 0$ further requires 
\begin{align}
    (n_B - n_c) (n_B - 2 n_c) = 0\,,
\end{align}
resulting in two competing primitive solutions $(n_A, n_B, n_c) = (2,1,1)$ and $(2,2,1)$, with the null-vectors $\vec{M}_{2,1,1} = (1,2,1,0)^T$ and $\vec{M}_{2,2,1} = (1,2,2,3)^T$. 

Localization driven by any of these two competing null-vectors yields the reduced theory 
\begin{align}
    K_{\text{red}} = \begin{pmatrix} - 3 & 0  \\ 0& 3 \end{pmatrix}\,, \quad \vec{t}_{\text{red}} =  \begin{pmatrix} 1 \\ 1  \end{pmatrix}\,. 
\end{align}
Clearly, this theory is, in general, not topologically stable. However, in the present case, further localization is forbidden by the spin conservation constraint.

We now calculate the conductance $G$.  For generality, we assume a certain degree of incoherent equilibration in the segments LA, LB, RA, and RB. As discussed in Sec.~\ref{sec:localizationandequilibration}, we incorporate it by using the general contact boundary conditions; see Appendix \ref{app.gen_sol}.  The matrix $U = O_{\Upsilon} M_K$ that enters the boundary conditions and the final result for the conductance then has the form 
\begin{align}
   O_{\Upsilon} &= \begin{pmatrix} \cosh \theta_{\text{u}} & \sinh \theta_{\text{u}}  \\ \sinh \theta_{\text{u}} & \cosh \theta_{\text{u}} \end{pmatrix} \oplus
    \begin{pmatrix} \cosh \theta_{\text{p}} & \sinh \theta_{\text{p}}  \\ \sinh \theta_{\text{p}} & \cosh \theta_{\text{p}} \end{pmatrix} \,, \nonumber \\ 
   M_K &= \begin{pmatrix}
\sqrt{\frac{3}{2}} & \sqrt{\frac{3}{2}} \\ -\sqrt{\frac{1}{2}} & \sqrt{\frac{1}{2}} 
\end{pmatrix} \oplus
    \begin{pmatrix}
1& 0 \\ 0 & \sqrt{3} 
\end{pmatrix}\,, 
\end{align}
parameterized by two real numbers, $\theta_{\text{u}}$ and $\theta_{\text{p}}$. Evaluation of the conductance $G$ by using Eq.~\eqref{eq:twoterminalcondlocalization} yields 
\begin{align} \label{eq:condinterfacespinpolspinunpol}
    G =  \frac{2}{3}\, \frac{1 + 2 \cosh [2 (\theta_{\text{u}} +\theta_{\text{p}})] - \sqrt{3} \sinh [2 (\theta_{\text{u}} +\theta_{\text{p}})]}{2 \cosh (2\theta_{\text{u}}) + \cosh(2\theta_{\text{p}}) - \sqrt{3} \sinh(2\theta_{\text{u}})}\,.
\end{align}
The parameter $\theta_{\text{u}}$, which characterizes the spin-unpolarized edge, depends on the asymmetry of mode coupling and on the degree of incoherent equilibration, see Appendix \ref{app.gen_sol}. The parameter $\theta_{\text{p}}$ that refers to the spin-polarized edge is zero in the absence of inelastic equilibration (since the matrix $\Upsilon$, in this case, commutes with $K_{\text{B}}$, see  Sec.~\ref{sec:commuting-K-Upsilon}) but becomes non-zero if equilibration is included. We now analyze various limiting cases of the result for $G$.

 First, we consider the case of no incoherent equilibration in the spin-polarized edge modes
 (i.e., in the segments LB and RB). Then  $\theta_{\text{p}} = 0$, as we have just pointed out. This yields the conductance 
  \begin{equation}
  G = \frac 23 \,,
  \end{equation}
 regardless of the value of $\theta_{\text{u}}$, i.e., independently of the asymmetry of contact couplings of the modes of the spin-unpolarized edge A and of the degree of equilibration in the segments LA and LB.
Interestingly, localization restores universality in this regime: in the absence of localization, the conductance of the spin-unpolarized edge B (and thus of the whole system) was dependent on $\theta_{\text{u}}$, see Sec.~\ref{sec:non-commuting-K-Upsilon}
and Appendix \ref{app.gen_sol}.
 
Second, we consider the case when both spin-polarized and spin-unpolarized edge modes 
(segments LA, LB, RA, and RB) are fully equilibrated. As discussed in Sec.~\ref{sec:localizationandequilibration}, the effect of equilibration can be mimicked by including the inter-mode equilibration $\gamma_{ab}$ into the $\Upsilon$ matrix; see Appendix \ref{app.gen_sol} for details. To describe the fully equilibrated case, we take the limit $\gamma_{ab} \gg \Gamma_{a}$ in the ${\Upsilon}$ matrix~\eqref{Appeq.Ups}, which leads to $\theta_{\text{u}} = 0$ and $\theta_{\text{p}} = \frac{1}{2} \log (2 + \sqrt{3})$. Inserting those values in Eq.~\eqref{eq:condinterfacespinpolspinunpol}, we obtain 
 \begin{equation}
  G = \frac 13 \,.
  \label{eq:G-23sp-23su-equil}
  \end{equation}
We remind that, in the absence of coupling between the edges A and B, the conductance of the combined system in this equilibrated limit was $G = 2/3 + 2/3 = 4/3$. Interestingly, the spin-conserving localization reduces it down to 1/3. 

We comment now on a relation between the result  \eqref{eq:G-23sp-23su-equil} and that of Ref.~\cite{ponomarenko2023unusual}. In the paper, the tunneling transport between the spin-polarized and spin-unpolarized 2/3 systems was considered, which corresponds to applying a bias between systems A and B, i.e., between top and bottom in our Fig.~\eqref{Fig:interface}. In general, this is essentially different from our setup, where the bias is applied between left and right since this corresponds to different conductances of a device with four terminals. However, the situation simplifies in the limit of strong equilibration in the segments LA, LB, RA, and RB, which implies only ``downstream'' transport in these segments. It is not difficult to see that the top-to-bottom conductance $G_{AB}$ is then given by $G_{AB} = (G_0 - G)/2$, where $G_0 = 4/3$ is the value of conductance in the absence of coupling between the edges A and B. Substituting the result  \eqref{eq:G-23sp-23su-equil}, we obtain $G_{AB} = 1/2$, which is the value that was found in Ref.~\cite{ponomarenko2023unusual}. It is worth reiterating that we obtained this result under the conditions that (i) the transport in the junction is coherent, with the junction length being larger than the localization length, and (ii) the transport in the remaining segments of the edges (bridging the junction with the contacts) is incoherent.



An alternative scenario of localization in a junction of two 2/3 edges may be realized at an interface of two $\nu = 2/3$ FQH states with opposite spin polarizations, where neutral tunneling across the interface is allowed and may lead to superconductivity of neutral modes~\cite{Jukka2022}. Results for the reduced theories and conductances that emerge out of this type of localization will be published elsewhere~\cite{Tobepublished}.

\section{Summary and outlook} \label{sec:summary}

To summarize, we have investigated the impact of localization on the electrical conductance through FQH edges. We have derived a reduced edge theory involving only the remaining modes (i.e., those that do not participate in the localization process) and have determined a general formula for the conductance. We have further shown how this general analysis and the resulting formula for the conductance can be applied to setups involving junctions of FQH edges, with the localization taking place in the junction. Such setups may be fully coherent or else, with incoherent equilibration occurring in edge segments bridging the junction to contacts. We have demonstrated the application of our formalism to a number of examples of FQH edges and edge junctions that undergo localization. 

The general framework developed in this paper is directly applicable to a variety of novel experimentally relevant structures involving engineered complex FQH edges. These include, in particular, structures with multiple subbands as well as interfaces or different FQH states. The localization in complex FQH edges and edge junctions is one of the most remarkable hallmarks of the coherent transport in these devices. Experimental studies of coherent transport in the FQH regime are a subject of major interest currently. We hope that the theoretical framework developed in our work will provide a further boost to experimental activities in the field, including the fabrication of novel FQH devices and investigations of their transport properties. 

While the present paper focuses on Abelian edges, our analysis can also be extended to non-Abelian edges with Majorana modes. We briefly outline in Appendix~\ref{appen:Majorana} how the presence of Majorana modes may make a FQH edge topologically unstable, thus making it prone to localization. We leave a detailed analysis of transport in non-Abelian edges with localization to future work.

\begin{acknowledgments}
MY regards to David F. Mross for the insightful discussions. J.P. and A.D.M. acknowledge support by the DFG Grant MI 658/10-2 and the German-
Israeli Foundation Grant I-1505-303.10/2019. 
\end{acknowledgments}

\appendix

\section{Contacts with inter-mode equilibration}
\label{app.gen_sol}

In Sec.~\ref{sec:ballistic}, we considered an idealized contact by keeping only processes of tunneling between the contact and $d$ edge modes. This yielded a tunneling matrix $\Upsilon$ that is diagonal in a certain basis than can be obtained by an $\text{SL}(d,\mathbb{Z})$ transformation. The corresponding assumption was that the contact is strongly coupled to the edge so that the tunneling rates between the modes and the contact are much larger than the tunneling rates between the modes so that the inter-mode equilibration can be discarded. Following Refs.~\cite{Kane1995, Spanslatt2021}, we consider a more general situation by retaining also the inter-mode equilibration in the contact region. 

As in Sec.~\ref{sec:contacts},  we assume that $d$ modes of the edge equilibrate with the contact (that has a chemical potential $\mu$) with charge tunneling rates $\Gamma_a> 0$, see Eq.~\eqref{eq.equilb1a}.
In addition, we assume that the edge modes equilibrate among themselves with charge tunneling rates $\gamma_{ab}=\gamma_{ba}>0$.
Equation \eqref{eq.equilb1a} is then modified to 
\begin{align} \label{app:currentcontinuity}
    \partial_x I_a =-\frac{1}{2\pi} \frac{\Gamma_a}{t_a} (\mu_a-t_a\mu) -\frac{1}{2\pi} \sum_{b}\gamma_{ab}( t_a^{-1} \mu_a t_b- \mu_b).
\end{align}
We offset chemical potentials and currents to absorb the constant term, $\Delta\mu_a=\mu_a-t_a\mu$ and $\Delta I_a = I_a - \frac{\mu}{2\pi}\sum_b K^{-1}_{ab}t_b$, which yields
\begin{align} 
    \partial_x \Delta I_a =- \frac{1}{2\pi} \frac{\Gamma_a}{t_a} \Delta\mu_a - \frac{1}{2\pi}\sum_{b}\gamma_{ab}(t_a^{-1}\Delta\mu_a t_b  - \Delta\mu_b).
    \label{eq:generalized-contacts-1}
\end{align}
Further, we define a symmetric matrix 
\begin{align}\label{Appeq.Ups}
    \Upsilon_{ab} = \delta_{ab} \left(\frac{\Gamma_{a}}{t_a}+ \sum_{c} t_a^{-1} \gamma_{ac} t_c \right)  - \gamma_{ab}\,,
\end{align}
so that Eq.~ \eqref{eq:generalized-contacts-1} becomes
\begin{align}
    \partial_x \Delta I_a = -\frac{1}{2\pi}\sum_{b}\Upsilon_{ab} \Delta \mu_{b} = -\sum_{b,c}\Upsilon_{ab}K_{bc} \Delta I_{c} \,,
     \label{eq:generalized-contacts-2}
\end{align}
where we used the relation \eqref{eq:bc_for_f} between the currents and the chemical potentials. 

Equation  \eqref{eq:generalized-contacts-2}  has the same form as Eq.~\eqref{eq:differentialeqcontacts} with the only difference in the form of matrix $\Upsilon$. The subsequent derivation of the boundary condition proceeds as in Sec.~\ref{sec:contacts}, and the calculation of the conductance as in Sec.~\ref{sec:ballistic-cond}, with the result Eq.~\eqref{eq:condclean}. The details of the contact are incorporated in the matrix $U$, more specifically in $O_\Upsilon$. 

As an example, we consider the edge of a spin-unpolarized $\nu=\frac{2}{3}$ state, described by $(K, \vec{t})$ given by Eq.~\eqref{eq:Kmat23}. 
The conductance of such edge in the absence of equilibration between the spin-up and spin-down modes in the contact was calculated in Sec.~\ref{sec:non-commuting-K-Upsilon}; now we include the inter-mode equilibration.

The equilibration matrix \eqref{Appeq.Ups} reads
\begin{align}
    \Upsilon = \begin{pmatrix}
\Gamma_1+\gamma_{12} & -\gamma_{12} \\ -\gamma_{12} & \Gamma_2+\gamma_{12} 
\end{pmatrix}.
\end{align}
The $M_K$ and $O_\Upsilon$ matrices that diagonalize the equilibration equations are now given by 
\begin{align}
    M_K= \begin{pmatrix}
\sqrt{\frac{3}{2}} & \sqrt{\frac{3}{2}} \\ -\sqrt{\frac{1}{2}} & \sqrt{\frac{1}{2}} 
\end{pmatrix},\,\,\,\,
O_\Upsilon
=  \begin{pmatrix}
\cosh \theta & \sinh \theta\\ \sinh \theta & \cosh \theta
\end{pmatrix}, 
\end{align}
where 
\begin{equation}
\theta = \frac{1}{4}\log \frac{2-\alpha\sqrt{3} + \beta}{2+\alpha\sqrt{3}+\beta}
\end{equation}
and we parametrized the matrix $\Upsilon$ as follows: $\Gamma_1 = (1-\alpha)\Gamma$, $\Gamma_2 = (1+\alpha)\Gamma$ and $\gamma_{12}=\beta \Gamma$. The two-terminal conductance evaluates to
\begin{align}
\label{eq:G-23-su-with-eq}
    G = \frac{2(2+\beta)}{3\sqrt{(2+\beta)^2-3\alpha^2}}.
\end{align}
In the absence of inter-mode equilibration, $\beta=0$, this reproduces  Eq.~\eqref{G-23-unpolarized}.
For a very strong inter-mode equilibration $\beta\to\infty$, Eq.~\eqref{eq:G-23-su-with-eq} yields $G \to 2/3$, in correspondence with the general result $G \to \nu$ in the case of strong incoherent equilibration in the edge, see Sec.~\ref{sec:incoherentequilibration}. 

Returning to contacts without inter-mode equilibration, it is worth recalling that
Eq.~\eqref{eq:differentialeqcontacts} was written in a preferential basis of modes coupled to the contact so that the matrix $\Upsilon$ is diagonal. One can, of course, perform a transformation to another basis by a matrix $W\in \text{SL}(d,\mathbb{Z})$. 
Under this transformation, the matrix $ \Upsilon$ will be transformed as 
$ \Upsilon \to \tilde{\Upsilon} = W^{-1}K(W^T)^{-1}$ and will in general become off-diagonal. Equation \eqref{eq:differentialeqcontacts} will retain its form in the new basis, with the transformed $\tilde{\Upsilon}$, $\tilde{K}$ and $\tilde{I}$. Clearly, the result for the conductance $G$ does not depend on the basis in which it is evaluated.

\section{Signature of $\mathcal{T}K$}
\label{app.signature}
In this appendix, we prove the statement about the spectrum of the matrix ${\cal T} K$ for different filling factors $\nu=\vec{t}K^{-1}\vec{t}$.  The signature of ${\cal T} K$ is
\begin{align}
\text{sgn}({\cal T} K) = 
\begin{cases}
     (n_R-1,   \;\; 1,\; n_L),   &\text{for } \nu > 0 \\
    (n_R-1,   \; \; 2,\; n_L-1), &\text{for } \nu = 0 \\
    (n_R\qquad,\;1,\;n_L-1), &\text{for } \nu < 0 \,,
\end{cases},
\label{eq.app_sgn}
\end{align}
where $(p,m,q)$ are numbers of positive, zero, and negative eigenvalues, respectively.  This statement played an important role in the analysis of transport in the presence of incoherent equilibration in Sec.~\ref{sec:incoherentequilibration}. We have proven this statement at the end of Sec.~\ref{sec:incoherentequilibration}; here, we provide an alternative proof based on Sylvester's inertia law. We also perform a related analysis of properties of eigenvectors of  $\mathcal T K$, which will be used in Appendix \ref{app.relat}.

Throughout, we assume that: (i) $K$ is not degenerate (in the sense that $\text{det} \, K \neq0$) since it describes a topological order; (ii) ${\cal T}$ is symmetric, positive-semi-definite with a single zero eigenvalue. A zero eigenvalue of ${\cal T}$ is a direct consequence of charge conservation since ${\cal T}\vec{t}=0$. We assume that no other conserved quantities exist, implying that there is a single zero eigenvalue. 

We begin by considering the case $\nu \neq 0$ when $\mathcal T K$ is diagonalizable. (The special case $\nu = 0$ will be analyzed in the end.)
First, we prove Eq.~\eqref{eq.app_sgn} for $\nu>0$, the $\nu<0$ case is then immediately obtained by a substitution $K \to -K$, which implies $n_R \leftrightarrow n_L$.

From the charge conservation condition, $\vec{t}^{T} \mathcal{T} = 0$, we identify the right eigenvector $\vec{v}_{0} = K^{-1} \vec{t}$ and the left eigenvector $\vec{u}^T_{0} = \vec{t}^T / \nu $ of  $\mathcal T K$ corresponding to the zero eigenvalue. Here, the $1/\nu$ factor in $\vec{u}_{0}$ is introduced to ensure the normalization condition $\vec{u}_0^{T} \vec{v}_0 = 1$. 
Then, the diagonalization equation is written as  
\begin{align} \label{appeq:diagonalequation}
    P^{-1} (\mathcal{T} K) P = \begin{pmatrix} 0 & \vec{0} \\ \vec{0} & D_n \end{pmatrix}\,, 
\end{align}
where $P$ is a matrix with columns given by the right eigenvectors $\vec{v}_j$ of $\mathcal T K$ and $P^{-1}$ is a matrix with the rows given by the left eigenvectors $\vec{u}^T_j$ of $\mathcal T K$, with 
$\vec u_j ^T \propto \text{sgn}(\tilde{\tau}_j)\vec v_j^T K$
and the normalization is chosen such that $\vec u^T_i \vec v_j = \delta_{ij}$.  Further, $D_n$ is a diagonal matrix of dimension $d-1$. Physically, $\vec{v}_0$ and $\vec{u}_0$ represent the charge mode that is uniquely determined by the charge conservation and satisfies
\begin{align} \label{appeq:chargecon}
\vec{v}_0^T K \vec{v}_0 = \nu \,, \quad \vec{u}_j^T \mathcal{T} \vec{u}_0 = 0\,.
\end{align}
The remaining eigenvectors, $\vec{v}_j$ and $\vec{u}_j$ with $j \neq 0$, represent neutral modes that satisfy the charge neutrality conditions 
\begin{align}\label{appeq:neutrality}
    \vec{t}^T \vec{v}_j = 0 \quad \rightarrow \quad  \vec{v}_j^T K \vec{v}_0 = 0\,.
\end{align}

The left hand side of Eq.~\eqref{appeq:diagonalequation} can be written as a product of two matrices  $\tilde{\mathcal{T}}$ and $\tilde{K}$ that are related to $\mathcal{T}$ and $K$ by transformations characteristic for quadratic forms:
\begin{align}
    P^{-1} (\mathcal{T} K) P =   [ P^{-1} \mathcal{T} (P^{-1})^T ] 
    [P^T K P] \equiv \tilde{\mathcal{T}} \tilde{K}\,, 
\end{align}
From the properties \eqref{appeq:chargecon} and \eqref{appeq:neutrality}, we find that  $\tilde{\mathcal{T}}$ and $\tilde{K}$ have a block diagonal form, with the zero-mode decoupling from the rest:
\begin{align}
    \tilde{\mathcal{T}} = \begin{pmatrix} 0 & \vec{0} \\ \vec{0} & \mathcal{T}_n\end{pmatrix}\,, \quad \tilde{K} = \begin{pmatrix} \nu & \vec{0} \\ \vec{0} & K_n\end{pmatrix}\,.
\end{align}
According to Sylvester's inertia law, the signature of $\tilde{\mathcal{T}}$ and $\tilde{K}$ is fully inherited from $\mathcal{T}$ and $K$. Thus, $\mathcal{T}_n$ has to be positive-definite, and  $K_n$ has $n_R -1 $ positive eigenvalues and $n_L$ negative eigenvalues (under our assumption that $\nu > 0$). Since $\mathcal{T}_n$ is positive definite, this holds also for $D_n = \mathcal{T}_n K_n$. 
 Therefore, $D$ and hence $\mathcal{T} K$ has one zero, $n_R - 1$ positive, and $n_L$ negative eigenvalues, which completes the proof. 


We now show that the eigenvectors $\vec v_j$ have properties reminiscent of special relativity
if one views $K$ as a metric defining the scalar product by $\vec{v}^TK\vec{v}$. In this language, the 
signature of $K$ determines the number of space-like (right-moving) and time-like (left-moving) directions. 
We now prove that the eigenmodes $\vec{v}_j$ of ${\cal T}K$  can be normalized to obey
\begin{align}
\label{appeq:right_left}
\vec{v}^T_i K \vec{v}_j    &=0, &\text{for }& i \neq j \,, \nonumber \\
   \vec{v}^T_i K \vec{v}_j &= \delta_{ij}, &\text{for }& i,j\leq n_R~,  \nonumber \\
   \vec{v}^T_i K \vec{v}_j &= -\delta_{ij}, &\text{for }& i,j>n_R~,
\end{align}
where the sign is dictated by the sign of the eigenvalues $\tilde{\tau}_j$ of $\mathcal T K$. Specifically, we now order the eigenvectors $\vec v_j$
as in Sec.~\ref{sec:incoherentequilibration},
such that $\tilde{\tau}_j > 0$ for $j=1, \ldots n_R -1$, the zero mode corresponds to $j=n_R$, and 
$\tilde{\tau}_j < 0$ for $j = n_R -1, \ldots, d$. 
For any pair of indices $i$ and $j$, we have
\begin{align}
    \vec{v}^T_i K {\cal T}  K\vec{v}_j=\tilde{\tau}_i \vec{v}^T_i K\vec{v}_j =\tilde{\tau}_j\vec{v}^T_i K \vec{v}_j,
    \label{eq:modes-orthogonality}
\end{align}
For $i \ne j$, and thus $\tilde{\tau}_i \ne \tilde{\tau}_j$, this implies orthogonality of modes in the sense that $\vec{v}^T_i K\vec{v}_j=0$. Further, using Eq.~\eqref{eq:modes-orthogonality} for $i=j \ne n_R$, we get
\begin{align}
    0 < \vec{v}^T_j K {\cal T}  K \vec{v}_j = \tilde{\tau}_j \left[\vec{v}^T_j K \vec{v}_j\right],
\end{align}
where we used the positive-definiteness $\vec{u}^T {\cal T} \vec{u}\geq0$ for $\vec{u}=K\vec{v}_j$. 
For the not normalized zero mode $\tilde{\vec v}_{n_R} \equiv K^{-1}\vec{t}$, we have $\tilde{\vec v}_{n_R}^T K \tilde{\vec v}_{n_R} = \nu > 0$.
This shows that the eigenvectors $\vec{v}_j$ can indeed be normalized to obey Eq.~\eqref{appeq:right_left}.

We now consider a special case $\nu=\vec{t}^TK^{-1}\vec{t}=0$ and prove that ${\cal T}K$ has a two-dimensional Jordan block corresponding to a zero eigenvalue. The eigenvector $\vec{v}_{n_R}=K^{-1}t$ corresponds to zero eigenvalue of ${\cal T}K$. We now find $\vec{u}$ such that ${\cal T}K \vec{u} =\vec{v}_{n_R}$. (Notice, that if $\nu\neq0$, multiplying by $\vec{t}^T$ we would arrive at contraction $0=\vec{t}^T{\cal T}K \vec{u} =\vec{t}\vec{v}_{n_R} =\nu$.) From (ii), it follows that eigenvectors $\vec{r}_j$ of ${\cal T}$ form a complete basis. We choose $\vec{r}_0=\vec{t}$ corresponding to a single zero eigenvalue $\alpha_0=0$. We can expand 
\begin{align}
     \vec{u}  = a_0K^{-1}\vec{t} + \sum_{j=1}^{d-1} a_j K^{-1}\vec{r}_j
\end{align}
since $K$ is not degenerate under assumption (i). We now act with ${\cal T}K$ on the expansion and find
\begin{align}\label{eq.appJordan}
     {\cal T}K\vec{u}  = \sum_{j=1}^{d-1} a_j \alpha_j \vec{r}_j =\vec{v}_{n_R}= K^{-1}\vec{t}.
\end{align}
The vector $K^{-1}\vec{t}$ can always be expanded in the complete basis $\vec{r}_j$. However, only for $\nu=0$, the coefficient of $\vec{r}_0\equiv\vec{t}$ is zero and Eq.~\eqref{eq.appJordan} have the solution $a_{j>0}=~\alpha_j^{-1}\vec{r}_j^TK^{-1}\vec{t}$. 
Finally, we verify that the Jordan block is two-dimensional. Let us assume that there is a solution $\vec{w}$ such that  ${\cal T} K \vec{w}=\vec{u}$. Multiplying both sides by $\vec{t}^T$
\begin{align}
    0=\vec{t}^T{\cal T} K \vec{w}=\vec{t}^T\vec{u}=\sum\limits_{j=1}a_j^{2}\alpha_j>0
\end{align}
we arrive at a contradiction, which proves that $\{\vec{u},\vec{v}_{n_R}\}$ form a Jordan chain of length two. This concludes our proof that the signature of ${\cal T} K$ is given by Eq.\eqref{eq.app_sgn}.

\section{``Relativistic'' proof for the equilibrated value of incoherent conductance}
\label{app.relat}

In Sec.~\ref{sec:incoherentequilibration}, we showed that, in the equilibrated regime, the current is given by Eq.~\eqref{eq:solutioncontact2} so that the two-terminal conductance is given by Eq.~\eqref{eq:twoterminalcond}. This was based on the fact that the equation \eqref{eq:currentconserincoh} has only the trivial solution $A_{j\leq n_R}=[\chi_L\vec{C}_L]_j=0$.
 (Here, we use the definition $A_{n_R}\equiv (\mu_\text{eq}-\mu_L) / 2\pi$, which is shifted by a constant from that in the main text.) 
  In this appendix, we provide a rigorous proof of this result. 

The proof uses the idea reminiscent of the special-relativity theory. As noted in Appendix~\ref{app.signature}, the $K$-matrix can be thought of as a metric of the space with $n_R$ space-like and $n_L$ time-like directions, respectively. We recall that $K$-matrix can be diagonalized as $K=M_K^T\Lambda M_K$, see Sec.~\ref{sec:contacts}. Multiplying Eq.~\eqref{eq:currentconserincoh} with $U=O_\Upsilon M_K$, we obtain
\begin{align}
\label{appeq:rel_1}
    O_\Upsilon \vec{V}_R \equiv O_\Upsilon\left[\sum_{j\leq n_R} A_j  \vec{w}_j\right] = \chi_L\vec{C}_L \equiv \vec{V}_L \,,
\end{align}
where $\vec{w}_j= M_K\vec{v}_j$, and we introduced the notation $\vec{V}_L$ for the right-hand side and $\vec{V}_R$ for the sum in square brackets on the left-hand side.  It is easy to see that $\vec{w}_j$ are the eigenvectors of the matrix ${\cal T}'\Lambda$ with eigenvalues $\tilde{\tau}_j$, where ${\cal T}' = M_K{\cal T}M_K^T$ is a symmetric matrix. 

When written in terms of $\vec{w}_j$, equations \eqref{appeq:right_left} read
\begin{align}\label{appeq:right_left-1}
\vec{w}^T_i\Lambda\vec{w}_j &= 0,&\text{for }& i \neq j\,, \nonumber \\
   \vec{w}^T_i\Lambda\vec{w}_j &= \delta_{ij},&\text{for }& i,j\leq n_R~, \nonumber \\
   \vec{w}^T_i\Lambda\vec{w}_j &= -\delta_{ij},&\text{for }& i,j>n_R~.
\end{align}
Therefore, $V_R$ in Eq.~\eqref{appeq:rel_1} 
satisfies $\vec{V}_R^T\Lambda\vec{V}_R\geq 0$, i.e., 
it is a ``space-like'' vector. On the other hand, by construction, $\vec{V}_L$ is a ``time-like'' vector, i.e.,  $\vec{V}^T_L\Lambda\vec{V}_L\leq0$.

The boost $O_\Upsilon\in \text{SO}(n_R,n_L)$ cannot transform the space-like vector $\vec{V}_R$ into a time-like $\vec{V}_L$ since
\begin{align}
\vec{V}^T_R O_\Upsilon^{T}\Lambda O_\Upsilon \vec{V}_R=\vec{V}^T_R\Lambda\vec{V}_R\geq 0.
\end{align}
Thus, the only solution 
of Eq.~\eqref{appeq:rel_1}
is trivial, $\vec{V}_L=0$ and $\vec{V}_R = 0$. Finally, since the eigenvectors $\vec{w}_j$ are linearly independent, the only solution to $\vec{V}_R=0$  is $A_{j\leq n_R}=0$, which completes the proof.  In particular, $A_{n_R}=0$ implies that the chemical potential of the middle region $\mu_\text{eq}$ is given by the chemical potential of the upstream-side contact, $\mu_\text{eq}=\mu_L$.

\section{Reduced theory after localization}
\label{appen:bindingtransition}

In this appendix, we provide a general framework for determining the reduced theory $(K_{\text{red}}, \vec{t}_{\text{red}})$ of an edge that undergoes  localization transition,
Sec.~\ref{subsec:localzationbasictheory}. 

\begin{figure}[t!]
\includegraphics[width =\columnwidth]{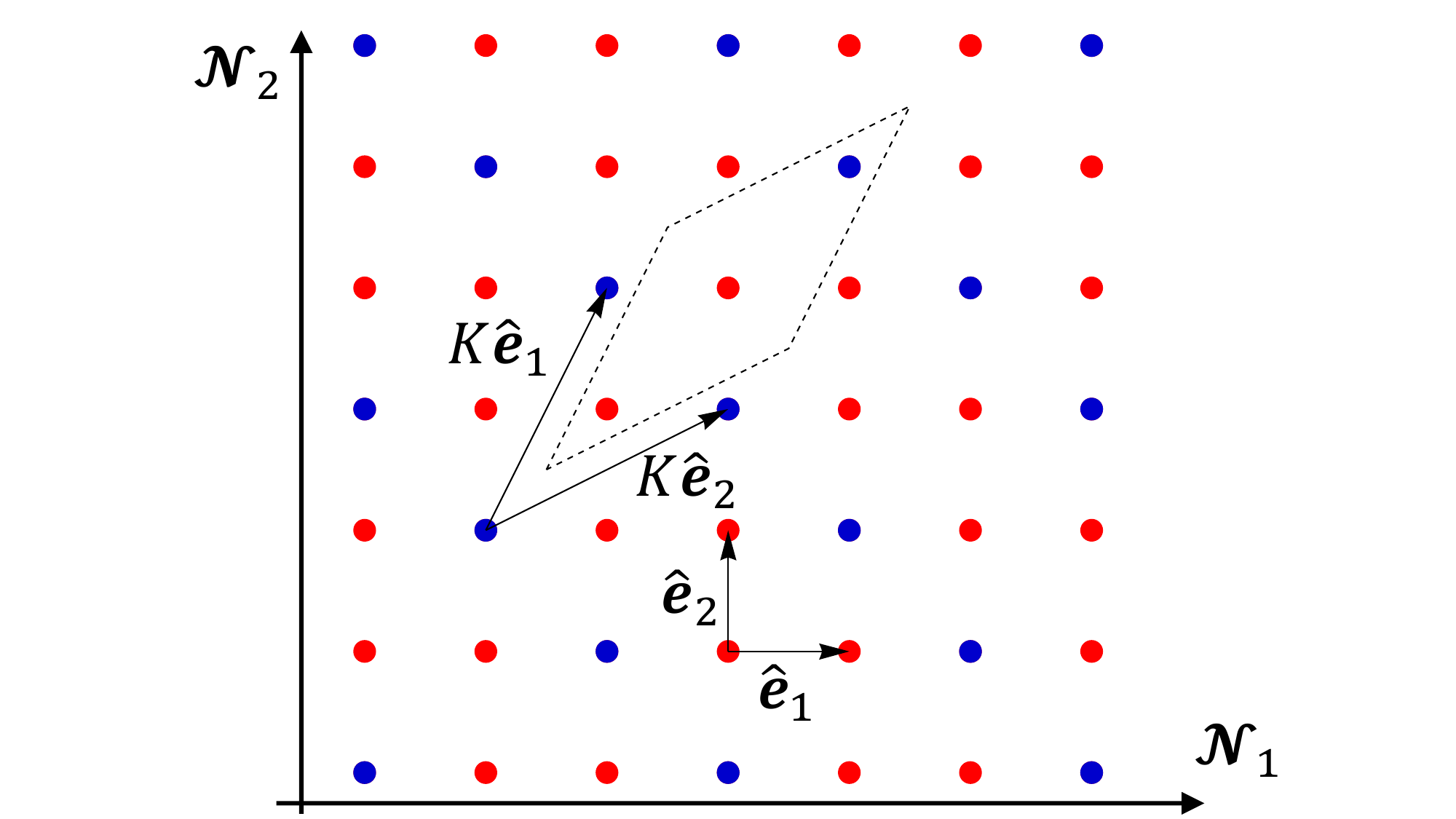}
\caption{{\bf Lattice} ($\mathcal{N}_1, \mathcal{N}_2$) for the $\nu = 2/3$ spin-unpolarized state given by $K 
=\begin{pmatrix}
1 & 2 \\ 2& 1
\end{pmatrix} $ and 
$\vec{t}=\begin{pmatrix}
1  \\ 1
\end{pmatrix}$. The quasiparticle spectrum is fully characterized by a lattice consisting of blue and red dots; $\hat{\vec{e}}_1$ and $\hat{\vec{e}}_2$ are two basis vectors for this lattice. On the other hand, a sublattice (consisting of only blue dots) represents electron excitations spanned by $K \hat{\vec{e}}_1$ and $K \hat{\vec{e}}_2$. The unit cell of the electron sublattice depicted as the black dashed parallelogram includes three lattice points, reflecting the fact that an elementary electron excitation is made out of three quasiparticles. } \label{Fig:lattice}
\end{figure}

We begin with the effective edge theory of $d = \text{dim}(K)$ bosonic modes, described by the action 
\begin{align}\label{appeq:L0}
    S_0 = \frac{1}{4\pi}\int dt dx\; \sum_{a,b=1}^d\partial_x\phi_a( K_{ab}\partial_t\phi_b-V_{ab}\partial_x\phi_b)\,,
\end{align}
with an integer-valued symmetric matrix $K$ and a positive-definite matrix $V$. We first introduce the particle density of the modes, given by $\rho_a = \partial_x \phi_a / (2 \pi)$. Upon the canonical quantization, $\phi_a$ and $\rho_a$ obey the commutation relations  
\begin{align}
    [\rho_a (x), \phi_b (x')] = i (K^{-1})_{ab} \delta (x - x')\,. 
\end{align}
From those commutation relations, one identifies $d$ electron operators $\psi_a^{\text{el}} (x) \sim e^{i K_{ab} \phi_b (x)}$ and $d$ elementary quasiparticle operators with charge $Q_a = \sum_{b} (K^{-1})_{ab}$, $\psi_a^{\text{qp}} \sim e^{i \phi_a (x)}$. For a closed edge with length $L$, we introduce the integer-valued total charge operator $\hat{\mathcal{N}}_a$ for type-$a$ quasiparticles, given by 
\begin{align}
    \hat{\mathcal{N}}_a \equiv \int_{0}^{L} dx K_{ab} \rho_b (x) = \frac{1}{2\pi} K_{ab}(\phi_b (L) - \phi_b (0))\,.
\end{align}
A set of the eigenvalues $n_a$ of $\hat{\mathcal{N}}_a$ forms a $d$-dimensional lattice $\Lambda = \{n_a \vec{e}_a | n_a \in \mathbb{Z} \}$~\cite{Naud_chiral_sine_2000}. An example of the lattice is shown in Fig.~\ref{Fig:lattice}  for the $\nu = 2/3$ spin-unpolarized state. The lattice fully characterizes all possible quasiparticle excitations
\begin{align}
    V_{\vec{m}} (x) \sim e^{i m_a \phi_a (x)}\,,
\end{align}
which connects two lattice points separated by vector $\vec{m}$. 
The lattice is spanned by the standard unit vectors $\hat{\vec{e}}_a = (0,...,1,...,0)$ with 1 for $a$-th component and 0 otherwise. Crucially, a set of different basis vectors $\tilde{\vec{e}}_a = W_{ab} \hat{\vec{e}}_b$ yields exactly the same lattice
provided that the transformation $W$ belongs to $\text{SL}(d, \mathbb{Z})$. Thus, the quasiparticle spectrum is invariant under this $\text{SL}(d, \mathbb{Z})$ transformation. Furthermore, invariance of the first (topological) term of the action \eqref{appeq:L0} under the $\text{SL}(d, \mathbb{Z})$ transformation implies that
the $K$-matrix in the new basis satisfies
\begin{align} \label{eq:covariantKmat}
  (\tilde{K}^{-1})_{ab} = 
  (K^{-1})_{\mu \nu} \tilde{e}_a^{\mu} \tilde{e}_b^{\nu}
  \equiv
  \tilde{\vec{e}}_a \cdot \tilde{\vec{e}}_b  \,.
\end{align}
Here, we have introduced a compact notation of a dot product, 
\begin{equation}
\vec m_1 \cdot \vec m_2 = \vec m_1^T K^{-1} \vec m_2 \,,
\label{eq:dot-product}
\end{equation} with the $K^{-1}$ matrix playing the role of a metric tensor.

Equivalently, this $\text{SL}(d, \mathbb{Z})$ basis change leads to the transformation of $K$-matrix and $\vec{t}$ vector as $K \rightarrow W^T K W$ and $\vec{t} \rightarrow W^T \vec{t}$. Thus, two phases, characterized by $(K, \vec{t})$ and $(\tilde{K}, \tilde{\vec{t}})$, respectively, are topologically equivalent if one finds a $W$ matrix, $W \in SL(d, \mathbb{Z})$, to satisfy $K = W^T \tilde{K} W$ and $\vec{t} = W^T \tilde{\vec{t}}$~\cite{Wen_Topological_1995}. Comparison with the relation \eqref{eq:covariantKmat} results in
$ \tilde{\vec{e}}_a^\mu =(W^{-1})_{a\mu} $. 

The edge may experience a localization transition by the tunneling processes between edge modes, given by 
\begin{align}\label{appeq:tunneling}
    S_\text{tun} =- \int dt dx\;  \xi(x)\exp\Big[i\sum_{a}M_a\phi_a\Big] +\text{H.c.},
\end{align}
if integers $M_a$ satisfy two conditions, (i) the charge neutrality condition $t_a (K^{-1})_{ab} M_b = 0$
and (ii) the null-vector condition $M_a (K^{-1})_{ab} M_b = 0$~\cite{Haldane_Stability_1995}.  Equivalently, these conditions can be written as (i) $\vec{t} \cdot \vec{M} = 0$ and (ii) $\vec{M} \cdot \vec{M} = 0$ by using the dot product of two vectors defined in Eq.~\eqref{eq:dot-product}.
When the tunneling term \eqref{appeq:tunneling} is relevant in the renormalization group sense, the mode $\sum_a M_a \phi_a$ is pinned to a minimum of the action \eqref{appeq:tunneling} so that it becomes massive and thus does not contribute to the low-energy transport. 

What determines the topological properties of the reduced theory after the localization transition?
The remaining modes $\sum_a v_a \phi_a$ should not be disturbed by the tunneling processes, which means that they obey the following commutation relation 
\begin{align}
    \Big [\sum_a v_a \phi_a (x), e^{i \sum_{b} M_b \phi_b (x')} \Big ] = 0\,. 
\end{align}
This implies that 
\begin{align} \label{appeq:conditionremaining}
    v_a (K^{-1})_{ab} M_b \equiv \vec{v} \cdot \vec{M} = 0\,. 
\end{align}
Note that the null-vector$\vec{M}$ also satisfies this condition, in view of $\vec{M} \cdot \vec{M} = 0$. 
We, therefore, should subtract the subspace spanned by $\vec{M}$ to find the space $\Lambda'$ on which the remaining propagating modes $\vec{v}$ reside. Therefore, the space $\Lambda'$ forms  a $d-2$ dimensional lattice 
\begin{equation} 
\Lambda' = \Lambda_1 / \Lambda_2 \,,
\end{equation}
where 
\begin{eqnarray}
\Lambda_1 &=& \{\vec{v} = v_a \vec{e}_a | v_a \in \mathbb{Z} \,\, \text{and} \,\, \vec{v}\cdot \vec{M} = 0 \} \,, \\
\Lambda_2 &=& \{ n \vec{M} | n \in \mathbb{Z} \} \,,
\end{eqnarray} 
cf. Ref.~\cite{Cano_Bulk_edge_2014}. 
We now find a set of $d-2$ basis vectors $\vec{e}_a^{\text{red}}$ to span the lattice $\Lambda'$, with an integer index $a$ running over $1\leq a \leq d-2$. From the obtained $\vec{e}_a^{\text{red}}$, we determine the $K$-matrix, $K_{\text{red}}$, for the reduced parts by using the covariant formula \eqref{eq:covariantKmat}:
\begin{align} \label{eq:covariantKmat2}
    \vec{e}_a^{\text{red}} \cdot \vec{e}_b^{\text{red}} = (e_a^{\text{red}})^{\mu} (K^{-1})_{\mu \nu}(e_b^{\text{red}})^{\nu} = (K_{\text{red}}^{-1})_{ab}\,,
\end{align}
with $1 \leq a, b\leq d-2$. Note that $K_{\text{red}}$ has rank $d-2$. 
Further, we determine the $\vec{t}$-vector $\vec{t}_{\text{red}}$ of the reduced theory by using the fact that the charge for excitations should be invariant under a basis change. In particular, for the elementary excitation $\vec{e}_a^{\text{red}}$, the corresponding charge $Q_a$ takes the form 
\begin{align} \label{eq:covarianttvec1}
    Q_a &= \vec{e}_a^{\text{red}} \cdot \vec{t} = \vec{e}_a^{\text{red}} K^{-1} \vec{t} = (K_{\text{red}}^{-1} \vec{t}_{\text{red}})_a\,, 
\end{align}
from which we find 
\begin{align}
    \vec{t}_{\text{red}} = K_{\text{red}} (W_{\text{red}} K^{-1} \vec{t})\,,
\end{align}
with $(d-2) \times d$ matrix $W_{\text{red}} = (\vec{e}_1^{\text{red}},...\vec{e}_{d-2}^{\text{red}})^T$,
$d$-dimensional matrix $K$, and ($d -$2)-dimensional matrix $K_{\text{red}}$. The original $\vec{t}$-vector can be decomposed into two parts, 
\begin{align}
 \vec{t} = \sum_{a=1}^{d - 2} t^{a}_{\text{red}}\vec{e}_a^{\text{red}} + t_{\text{loc}}\vec{M}\,,  
\end{align}
with the charge $t_{\text{loc}}$ of the localized modes. 

\begin{figure}
\includegraphics[width =\columnwidth]{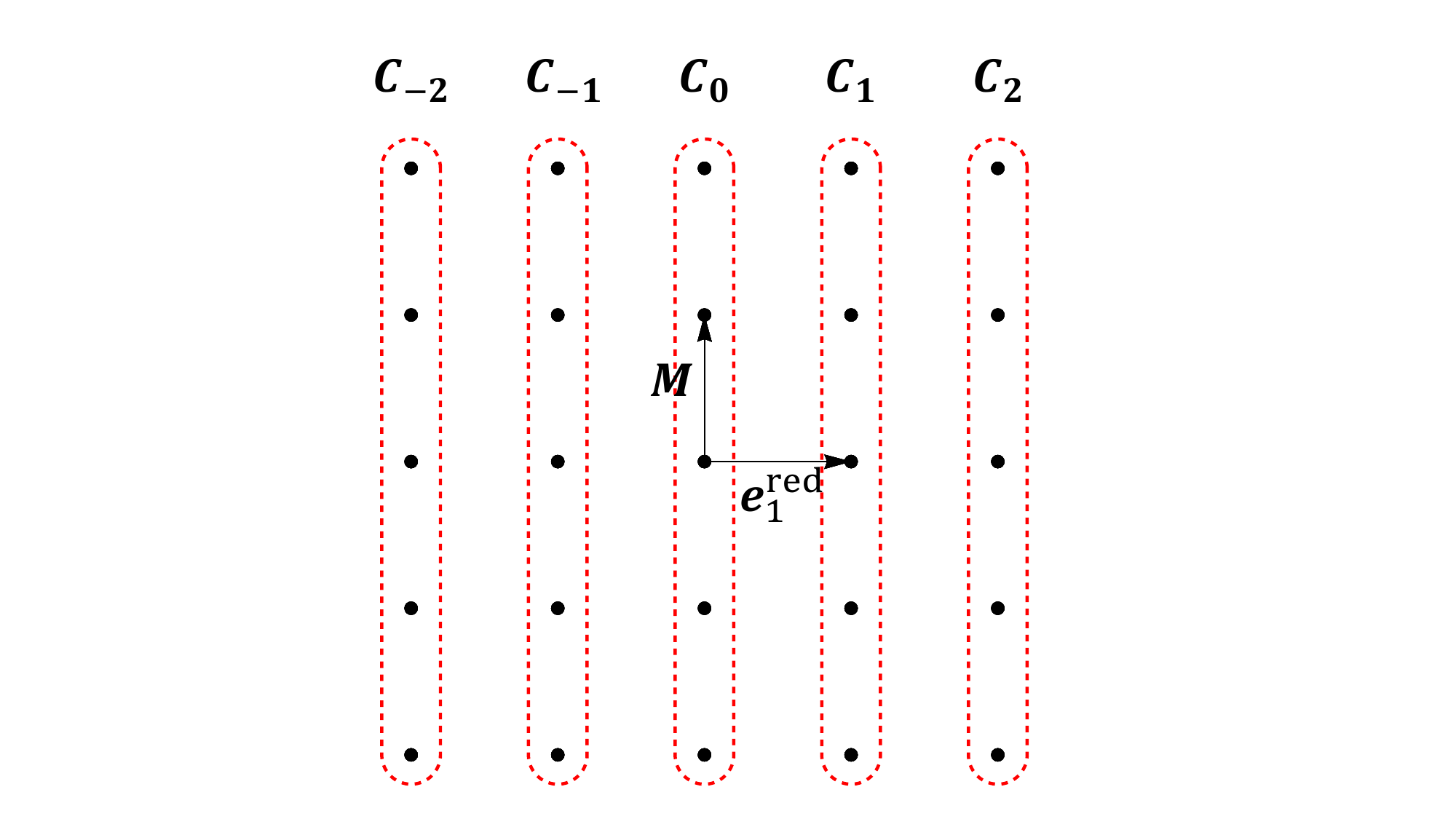}
\caption{{\bf Geographic understanding of basis vectors} after a localization transition by the null-vector$\vec{M} = (1, -2, -5)^T$ for the $\nu = 9/5$ state. Vectors $\vec{v}$ to satisfy $\vec{M} \cdot \vec{v} = 0$ form the lattice $\Lambda_1 = \mathbb{Z}_2$. Then, the cosets $\vec{v} \Lambda_1 = \{\vec{v} + n \vec{M}| n \in \mathbb{Z} \}$ depicted by the red circles form a sublattice $\subset \Lambda_1$. Basis vector
$\vec{e}_1^{\text{red}}$ of the sublattice determines the topological properties of the remaining modes after a localization. 
} \label{Fig:lattice2}
\end{figure}

To illustrate the above procedure for determining $K_{\text{red}}$ and $\vec{t}_{\text{red}}$, we consider a simple example of $\nu = 9/5$ discussed earlier in Refs.~\cite{Kao_Binding_1999, Spanslatt_binding_2023}. The $K$-matrix and $\vec{t}$-vector for this edge (before localization) are given by 
\begin{align}
   K = \begin{pmatrix}
1 & 0 & 0 \\ 0 & 1 & 0  \\ 0 & 0 & -5
\end{pmatrix}\,, \quad 
\vec{t} = \begin{pmatrix}
1 \\ 1\\ 1
\end{pmatrix}\,.
\end{align}
The null-vector is $\vec{M} = (1, -2, -5)^T$. (There is a second null-vector obtained by permuting the first two components, with identical properties.)
The original edge theory has three modes, and hence the associate lattice is $ \sim \mathbb{Z}^3$. We next find $\vec{v}$ vectors to satisfy the relation \eqref{appeq:conditionremaining}, which span a lattice  $\Lambda_1 \sim \mathbb{Z}^2$. This $\mathbb{Z}^2$ lattice is graphically shown in Fig.~\ref{Fig:lattice2}. As discussed above, this $\mathbb{Z}^2$ lattice includes the $\vec{M}$ vector and its multiples. To exclude those vectors from this lattice and thus to find the sublattice space (here, the $ \sim \mathbb{Z}$ lattice) for the remaining modes, we identify cosets $\vec{v} \Lambda_1 = \{\vec{v} + n \vec{M}| n \in \mathbb{Z} \}$ that can be graphically understood by the vertical lines of Fig.~\ref{Fig:lattice2}. Then, the remaining task is to find a set of unit vectors $\vec{e}^{\rm red}_a$ (in this example, only one vector)  that span the resulting coset space. In general, it is a non-trivial task, but we found an efficient numerical method to do this.
Specifically, we explicitly evaluate $K_{\text{red}}$ for candidate basis vectors  $\vec{e}_{a}^{\text{red}}$ by using the formula \eqref{eq:covariantKmat2}. Since $K$ can be always written in a form of Eq.~\eqref{eq:wtransformation}, 
$W^T K W = K_{\text{red}} \oplus q_1 \sigma_z \oplus ... \oplus q_m \sigma_z$, $|\text{Det} [K ]|$ should be identical to $|\text{Det} [K_\text{red}]| \prod_{1 \leq j \leq m} q_j^2 $ for odd integers $q_j$. If candidate basis vectors do not span the $\Lambda_1 / \Lambda_2$ lattice, the corresponding $K_{\text{red}}$ is not an integer-valued matrix or $|\text{Det} [K_\text{red}]|  \prod_{1 \leq j \leq m} q_j^2  \neq |\text{Det} [K ]|$. In the present example, we find $\vec{e}_1^{\text{red}}= (0, 1, 2)^T$ (up to addition of multiples of $\vec{M}$). By using the formula \eqref{eq:covariantKmat2} and \eqref{eq:covarianttvec1}, we obtain $K_{\text{red}} = 5$ and $t_{\text{red}} = 3$. The $W$-matrix can be determined in such a way that Eq.~\eqref{eq:wtransformation} is satisfied. In the example of $\nu = 9/5$, we obtain the $W$-matrix given by Eq.~\eqref{eq:winv95}.

\section{Conductance of an edge with localization} \label{app:condutanceformula}

In this Appendix, we derive the general conductance formula Eq.~\eqref{eq:twoterminalcondlocalization} of an edge where localization takes place; see Fig.~\ref{Fig:localization} for a schematic setup. We further show that the two-terminal conductance simply becomes $G = \nu$ in the case when all modes of the reduced theory (after localization) propagate in the same direction. 

We begin with Eqs.~\eqref{eq.continutity:localization2} supplemented by constraints \eqref{eq:localchemicalpotential} and \eqref{eq:localcontinuity} from the localization conditions and the continuity of the currents for non-localized modes, respectively. 
Plugging Eqs.~\eqref{eq:localchemicalpotential} and \eqref{eq:localcontinuity} into Eqs.~\eqref{eq.continutity:localization2}, we arrive at a set of $d = n_L + n_R$ independent equations
\begin{align} \label{eq:setofequations1}
    \mu_L \chi_R U K^{-1} \vec{t} &= \chi_R U K^{-1} ( \sum_{a=1}^{d-2m}\mu_{\text{red}}^{a}\vec{e}_a^{\text{red}} + \sum_{j=1}^m \mu^{L}_j\vec{M}_j )\,,  \\
    \label{eq:setofequations2}
    \mu_R \chi_L U K^{-1} \vec{t} &= \chi_L U K^{-1} ( \sum_{a=1}^{d-2m}\mu_{\text{red}}^{a}\vec{e}_a^{\text{red}} + \sum_{j=1}^m \mu^{R}_j\vec{M}_j )\,,
\end{align}
with $d$ unknown variables ($\mu^a_{\text{red}}$ for $1\leq a \leq d-2m$ and $\mu^{s}_j$ for $s = L, R$ and $1 \leq j \leq m$). 
The addition of Eqs.~\eqref{eq:setofequations1} and \eqref{eq:setofequations2} leads to 
\begin{align} \label{appeq:generalformula1}
    &\frac{1}{2}\big[(\mu_L + \mu_R) + (\mu_L - \mu_R) \Lambda \big ] U K^{-1} \vec{t}  = \sum_{a=1}^{d-2m} \mu_{\text{red}}^a  U K^{-1} \vec{e}_a^{\text{red}} 
    \nonumber \\ &+
   \frac{1}{2}\sum_{j=1}^{m} \big[ (\mu_j^L + \mu_j^R) +  (\mu_j^L - \mu_j^R) \Lambda \big] U K^{-1} \vec{M}_j \,,
\end{align}
with $\chi_{R,L} = (1 \pm \Lambda)/2$. By using the representation \eqref{eq:wtransformation} of the $\vec{t}$-vector and the form \eqref{eq:inversew} of the $W^{-1}$ matrix, we obtain
\begin{align} \label{app:tdecomp}
    \vec{t} = \sum_{a=1}^{d-2m} \vec{e}_a^{\text{red}} t^{a}_{\text{red}} +
     \sum_{j=1}^{m} t_j^{\text{loc}} \vec{M}_j\,.
\end{align}
Inserting this into Eq.~\eqref{appeq:generalformula1} yields
\begin{widetext}
\begin{align} \label{appeq:generalformula2}
   \sum_{a=1}^{d-2m} \big[t^a_{\text{red}} ((\mu_L + \mu_R) + (\mu_L - \mu_R) \Lambda) & - 2 \mu_{\text{red}}^a \big ] U K^{-1} \vec{e}_a^{\text{red}} 
 \nonumber \\  = \sum_{j=1}^{m} & \big[ (\mu_j^L + \mu_j^R) +  (\mu_j^L - \mu_j^R) \Lambda - t_{j}^{\text{loc}}((\mu_L + \mu_R) + (\mu_L - \mu_R) \Lambda) \big] U K^{-1} \vec{M}_j \,. 
\end{align}
\end{widetext}
We now multiply Eq.~\eqref{appeq:generalformula2} on the left with
$t_b^{\text{red}} (\vec{e}_b^{\text{red}})^T U^{-1}$ (no summation over $b$) with $b=1, \ldots, d-2m$, which results in the following $d - 2m $ equations:
\begin{align} \label{appeq:equations1}
    &(\mu_L - \mu_R) \sum_{a = 1}^{d-2m} t^{b}_{\text{red}} (\vec{e}_{b}^{\text{red}})^T X \vec{e}_{a}^{\text{red}} t^{a}_{\text{red}} \nonumber \\ 
     +  & \sum_{j = 1}^{m} ((\mu_L - \mu_R) t_{j}^{\text{loc}} - (\mu_j^L - \mu_j^R))
    t^{b}_{\text{red}}(\vec{e}_{b}^{\text{red}})^T X \vec{M_j} 
    \nonumber \\ 
    =  & 2 t_{\text{red}}^{b} (K^{-1}_{\text{red}} \vec{\mu_{\text{red}}})_b - t_{\text{red}}^b (K_{\text{red}}^{-1} \vec{t}_{\text{red}})_b (\mu_L + \mu_R).
\end{align}
Here, we have used the fact that basis vectors $\vec{e}_{\text{red}}^a$ of the reduced theory satisfy $(\vec{e}_{\text{red}}^a)^T K^{-1} \vec{M}_j = 0$, see Eq.~\eqref{appeq:conditionremaining}, and introduced a matrix $X \equiv U^{-1} \Lambda U K^{-1}  = (U^T U)^{-1}$. Furthermore, by acting with $\vec{M}_{j'}^T U^{-1}$ on Eq.~\eqref{appeq:generalformula2} instead, we obtain additional $m$ equations 
\begin{align} \label{appeq:equations2}
    &(\mu_L - \mu_R) \sum_{a = 1}^{d-2m} \vec{M}_{j'}^T X \vec{e}_{a}^{\text{red}} t_{a}^{\text{red}} \nonumber \\ 
    & = \sum_{j = 1}^{m} ((\mu_j^L - \mu_j^R)- (\mu_L - \mu_R) t_{j}^{\text{loc}})
    \vec{M}_{j'}^T X \vec{M}_j\,,
\end{align}
with $1 \leq j' \leq m$. We have used here the null-vector condition on $\vec{M}_j$,
Eq.~\eqref{eq:many-null-vectors-condition},
as well as \eqref{appeq:conditionremaining}. Collecting Eqs.~\eqref{appeq:equations1} with \eqref{appeq:equations2}, we find $d-m$ equations that can be presented in the form
\begin{align} \label{eq:F}
    F \vec{p} &= \vec{q}\,, \qquad \qquad F = \begin{pmatrix}
        A & B \\ B^T & C
    \end{pmatrix}\,, 
\end{align}
where $F$ is $(d-m)\times(d-m)$ block matrix consisting of  $(d-2m) \times (d-2m)$ matrix $A$, $(d-2m) \times m$ matrix $B$, and $m \times m$ matrix $C$, with matrix elements
$A_{ab} = t^{\text{red}}_a t^{\text{red}}_b (\vec{e}_a^{\text{red}})^T X \vec{e}_b^{\text{red}}$,
$B_{aj} = t^{\text{red}}_a (\vec{e}_a^{\text{red}})^T X \vec{M}_j$,
and $C_{jk} = \vec{M}_j^T X \vec{M}_k$. The $(d-m)$-component vectors $\vec{p}$ and $\vec{q}$ are decomposed into two parts, the reduced part and the localized part, as 
\begin{align}
    \vec{p} = \vec{p}_{\text{red}}  \oplus \vec{p}_{\text{loc}}\,, 
    \quad \vec{q} = \vec{q}_{\text{red}} \oplus \vec{q}_{\text{loc}}\,, 
\end{align}
with the components
\begin{align}
     p_{\text{red}}^a &=(\mu_L - \mu_R)\,, \quad 
     p_{\text{loc}}^j = 
    t^{\text{loc}}_j (\mu_L -\mu_R) - (\mu^{L}_j - \mu^R_j)
   \,,  \nonumber \\   q_{\text{red}}^a &=  2 t_{\text{red}}^a (K_{\text{red}}^{-1} \vec{\mu}_{\text{red}})_a - t_{\text{red}}^a (K_{\text{red}}^{-1} \vec{t}_{\text{red}})_a (\mu_L + \mu_R)\,, \nonumber \\ q_{\text{loc}}^j &= 0\,.
\end{align}
Applying the inverse of $F$ on  Eq.~\eqref{eq:F}, we find 
\begin{align}
    p_{\text{red}}^{a} = \sum_{b=1}^{d-2m} (F^{-1})_{ab} q_{\text{red}}^b = \sum_{b=1}^{d-2m} (A  - B C^{-1} B^T)^{-1}_{ab} q_{\text{red}}^b\,, 
\end{align}
and hence
\begin{align}
    \vec{q}_{\text{red}} = (A  - B C^{-1} B^T) \vec{p}_{\text{red}}\,,
\end{align}
from which we finally arrive at 
\begin{align} \label{eq:solution}
   &  \vec{t}_{\text{red}}^T K_{\text{red}}^{-1} \vec{\mu}_{\text{red}} = \frac{\nu}{2} (\mu_L + \mu_R) \nonumber \\ &\qquad  \quad \quad  +  \frac{\mu_L - \mu_R}{2}\sum_{a,b=1}^{d-2m} (A  - B C^{-1} B^T)_{ab}\,.
\end{align}
From Eq.~\eqref{eq:Iconleftright}, the total charge current reads
\begin{align} \label{app:totalcurrent}
    J_{\text{tot}} &= \vec{t}^T \vec{I} = \frac{1}{2\pi} \vec{t}^T K^{-1} \vec{\mu}^{L/R}_{\text{Con}} 
    \nonumber \\ 
    &= \big [ \sum_{a=1}^{d-2m} (\vec{e}_a^{\text{red}})^T t_{\text{red}}^a + \sum_{j=1}^m t_j^{\text{loc}} \vec{M}_j^T \big ] K^{-1} \nonumber 
\\ & \quad \quad \big [ \sum_{a=1}^{d-2m} \vec{e}_a^{\text{red}} \mu_{\text{red}}^a + \sum_{j=1}^m \mu_j^{L/R} \vec{M}_j] \nonumber \\ 
    &=  \frac{1}{2\pi}  \vec{t}_{\text{red}}^T K_{\text{red}}^{-1} \vec{\mu}_{\text{red}}\,. 
\end{align}
We have used the decomposition of $\vec{t}$-vector \eqref{app:tdecomp}, and the decomposition of chemical potentials \eqref{eq:localchemicalpotential} in the third equality of Eq.~\eqref{app:totalcurrent}, as well as the null-vector conditions Eqs.~\eqref{eq:many-null-vectors-condition} and \eqref{appeq:conditionremaining} in the fourth equality. By inserting Eq.~\eqref{eq:solution} and using the definition of matrices $A, B, C$, we arrive at Eqs.~\eqref{eq:totalcurrentloc} and \eqref{eq:twoterminalcondlocalization}.

We now solve Eq.~\eqref{eq:setofequations1} in a special case, i.e., the maximally localized case with $m = \text{min} (n_R, n_L)$ where the modes, which remain conducting after localization, propagate in the same direction. In this case, $K_{\text{red}}$ becomes positive definite. For definiteness, we assume that $n_R > n_L$  and thus the remaining modes flow left to right, which also means that $\nu > 0$. 
Applying $U$ to the first line of Eq.~\eqref{eq.continutity:localization}, we obtain 
\begin{align} \label{appeq:leftmovers}
    \vec{V}_{L} \equiv \chi_L \vec{C}_L =  U \left[ \frac{K^{-1}}{2\pi} ( \vec{\mu}_{\text{Con}}^{L} - \mu_L \vec{t} ) \right] \equiv U \vec{V}_R \,. 
\end{align}
We use now the same ``relativistic'' idea as in Appendix \ref{app.relat}. Crucially, $\vec{V}_L$ is a left-moving (time-like) vector with 
\begin{equation}
\vec{V}_L^T \Lambda \vec{V}_L \leq 0 \,.
\label{eq:VL-time-like}
\end{equation}
On the other hand, in view of $\vec{V}_{L} = U\vec{V}_R \equiv O_{\Upsilon} M_K \vec{V}_R$, we find  
\begin{align}\label{appeq:rightmovers}
    \vec{V}_{L}^T \Lambda \vec{V}_L &= \vec{V}_{R}^T M_K^T O_{\Upsilon}^T \Lambda  O_{\Upsilon} M_K \vec{V}_R = 
    \vec{V}_{R}^T K \vec{V}_{R} \nonumber \\ 
   &=  \frac{1}{(2\pi)^2}( \vec{\mu}_{\text{Con}}^{L} - \mu_L \vec{t})^T K^{-1} ( \vec{\mu}_{\text{Con}}^{L} - \mu_L \vec{t}) \nonumber \\ 
   & = \frac{1}{(2\pi)^2} (\mu_L \vec{t}_{\text{red}} - \vec{\mu}_{\text{red}})^T K^{-1}_{\text{red}}( \mu_L \vec{t}_{\text{red}} - \vec{\mu}_{\text{red}}) \nonumber \\ 
   &\geq 0 \,,
\end{align}
with a  positive definite matrix $K_{\text{red}}$. 
We have used  $O_{\Upsilon} \in \text{SO}(n_L, n_R)$ and $K = M_K^T \Lambda M_K$ in the second equality of Eq.~\eqref{appeq:rightmovers}, as well as Eqs.~\eqref{eq:wtransformation} and \eqref{eq:localchemicalpotential} in the fourth equality. The inequalities \eqref{eq:VL-time-like} and \eqref{appeq:rightmovers} imply that $\vec{V}_L=0$, and Eq.~\eqref{appeq:leftmovers} gives then $\vec{V}_R =0$. This  leads to
\begin{align}
    \mu_{\text{red}}^a = \mu_L t^{a}_{\text{red}}\,, \quad \mu^{s}_j = \mu_L t^{\text{loc}}_j\,.
\end{align}
Therefore, the total charge current $J_{\text{tot}} = \vec{t}^T \vec{I}$ reads 
\begin{align}
    J_{\text{tot}} = \frac{\mu_L}{2\pi} \vec{t}_{\text{red}}^T K_{\text{red}}^{-1} \vec{t}_{\text{red}} = \frac{\nu \mu_L}{2\pi} \,,
\end{align}
resulting in the two-terminal conductance $G = \nu$. This shows that $G$ becomes universal (in particular, does not depend on details of coupling to contacts) in this class of theories with localization.  
This universality of $G$ was illustrated on examples of the $\nu = 9/5$ and $\nu = 8/3$ edges in Sec.~\ref{sec:localizationexample-95} and \ref{sec:localizationexample-83}, respectively. 

\section{Topological instability with Majorana modes} \label{appen:Majorana}
In this appendix, we outline how the presence of Majorana modes can make an FQH edge topologically unstable and thus drive the localization transition. 

In many non-Abelian cases, in addition to the boson modes, the edge contains an odd number of chiral Majorana modes. They can be described by the action 
\begin{align}
    S_M = i\sum_{k=1}^{|\ell|} \int dt dx\; \gamma_k(\partial_t-\text{sgn}(\ell)v_k\partial_x)\gamma_k,
\end{align}
where depending on the sign of $\ell$, Majorana modes move to the right or left. The particularly well-known non-Abelian state with Majorana modes on edges is the $\nu = 5/2$ state; see the candidate topological orders of the $\nu =5/2$~\cite{Moore_nonabelions_1991, Levin_particle_hole_2007, Lee_particle_hole_2007, Son_is_2015} and also Refs.~\cite{Simon_equilibration_2018,Feldman_comment_2018, Feldman_equilibration_2019,Asasi_equilibration_2020,Park_noise_2020,Yutushui_Identifying_2023,Manna_Experimentally_2023,Manna_Full_2024,Hein_Thermal_2023} for discussion on the thermal conductance and other transport signatures of those candidate edge states. Aside from the $\nu=5/2$, other non-Abelian states involving Majorana edge modes have been proposed, such as, e.g., Bonderson-Slingerland state in Ref.~\cite{Bonderson_hierarchy_2008}.

The Majorana modes do not carry charge. In the absence of localization or equilibration, each Majorana mode gives only a contribution 1/2 to the central charge (and thus to the dimensionless thermal conductance), without affecting the electric conductance. 
Nevertheless, Majorana modes can be responsible for topological instability, leading to the localization of counter-propagating modes. Remarkably, in certain cases, the Majorana-assisted localization can change the charge conductance.
In particular, that was leveraged in a proposal of Ref.~\cite{MY_Identifying_2022} for an approach of experimentally determining the topological order of the $\nu=\frac{5}{2}$.  

We consider only the cases of $\ell=\pm1$, since two co-propagating Majoranas $\gamma_{1,2}$ can always be written as a single complex fermions $\psi=\gamma_1+i\gamma_2$ and further bosonized $\psi\sim e^{i\phi}$. Thus, an edge with $\left[K_0,\vec{t}_0\right]$ and $\ell=\pm(2m+1)$ Majoranas can be equivalently represented as an edge with $\left[K=\text{diag}(K_0,\pm\mathbb{1}_m),\vec{t}=(\vec{t}_0,0\ldots,0)\right]$ and $\ell=\pm1$ Majorana. 

The property of bosonizing two Majoranas can be exploited to assess the topological stability analogously to the Abelian case. Consider an edge described by $(K_0,\vec{t})$ and $\ell=\pm1$ Majorana mode $\gamma$. We assume that $\left[K_0,\vec{t}_0\right]$ is topologically stable without Majoranas. Then, we add a pair of two counter-propagating Majorana modes $\eta_{R}$ and $\eta_{L}$ and bosonize $\gamma$ with one of them. Thus, the edge is now given by $\left[K=\text{diag}(K_0,\pm1),\vec{t}=(\vec{t}_0,0)\right]$ and $\ell=\mp 1$ remaining $\eta$ Majorana mode. 

If the resulting $[K,t]$ theory is unstable according to the criteria of stability of an Abelian theory (see Sec.~\ref{sec:topo-stability}), the localization will become operative, reducing the number of modes by two: $\text{dim}(K_\text{red}) =\text{dim}(K)-2$. Then, the full reduced theory (including the remaining Majorana mode) is given by $[K_\text{red},\vec{t}_\text{red}]$ and $\ell=\mp 1$ Majorana mode. Notice that in this case, the right- and left-moving central charges (i.e., the number 
of left- and right-moving modes, with Majorana counted as one-half) are reduced by 1/2, which is precisely what is expected after the localization of a pair of Majorana modes.

\bibliography{bibliography}

\begin{thebibliography}{69}%
\makeatletter
\providecommand \@ifxundefined [1]{%
 \@ifx{#1\undefined}
}%
\providecommand \@ifnum [1]{%
 \ifnum #1\expandafter \@firstoftwo
 \else \expandafter \@secondoftwo
 \fi
}%
\providecommand \@ifx [1]{%
 \ifx #1\expandafter \@firstoftwo
 \else \expandafter \@secondoftwo
 \fi
}%
\providecommand \natexlab [1]{#1}%
\providecommand \enquote  [1]{``#1''}%
\providecommand \bibnamefont  [1]{#1}%
\providecommand \bibfnamefont [1]{#1}%
\providecommand \citenamefont [1]{#1}%
\providecommand \href@noop [0]{\@secondoftwo}%
\providecommand \href [0]{\begingroup \@sanitize@url \@href}%
\providecommand \@href[1]{\@@startlink{#1}\@@href}%
\providecommand \@@href[1]{\endgroup#1\@@endlink}%
\providecommand \@sanitize@url [0]{\catcode `\\12\catcode `\$12\catcode `\&12\catcode `\#12\catcode `\^12\catcode `\_12\catcode `\%12\relax}%
\providecommand \@@startlink[1]{}%
\providecommand \@@endlink[0]{}%
\providecommand \url  [0]{\begingroup\@sanitize@url \@url }%
\providecommand \@url [1]{\endgroup\@href {#1}{\urlprefix }}%
\providecommand \urlprefix  [0]{URL }%
\providecommand \Eprint [0]{\href }%
\providecommand \doibase [0]{https://doi.org/}%
\providecommand \selectlanguage [0]{\@gobble}%
\providecommand \bibinfo  [0]{\@secondoftwo}%
\providecommand \bibfield  [0]{\@secondoftwo}%
\providecommand \translation [1]{[#1]}%
\providecommand \BibitemOpen [0]{}%
\providecommand \bibitemStop [0]{}%
\providecommand \bibitemNoStop [0]{.\EOS\space}%
\providecommand \EOS [0]{\spacefactor3000\relax}%
\providecommand \BibitemShut  [1]{\csname bibitem#1\endcsname}%
\let\auto@bib@innerbib\@empty
\bibitem [{\citenamefont {Tsui}\ \emph {et~al.}(1982)\citenamefont {Tsui}, \citenamefont {Stormer},\ and\ \citenamefont {Gossard}}]{Tsui_fqh_1982}%
  \BibitemOpen
  \bibfield  {author} {\bibinfo {author} {\bibfnamefont {D.~C.}\ \bibnamefont {Tsui}}, \bibinfo {author} {\bibfnamefont {H.~L.}\ \bibnamefont {Stormer}},\ and\ \bibinfo {author} {\bibfnamefont {A.~C.}\ \bibnamefont {Gossard}},\ }\bibfield  {title} {\bibinfo {title} {Two-dimensional magnetotransport in the extreme quantum limit},\ }\href {https://doi.org/10.1103/PhysRevLett.48.1559} {\bibfield  {journal} {\bibinfo  {journal} {Phys. Rev. Lett.}\ }\textbf {\bibinfo {volume} {48}},\ \bibinfo {pages} {1559} (\bibinfo {year} {1982})}\BibitemShut {NoStop}%
\bibitem [{\citenamefont {Laughlin}(1983)}]{Laughlin_fqh_1983}%
  \BibitemOpen
  \bibfield  {author} {\bibinfo {author} {\bibfnamefont {R.~B.}\ \bibnamefont {Laughlin}},\ }\bibfield  {title} {\bibinfo {title} {Anomalous quantum {Hall} effect: An incompressible quantum fluid with fractionally charged excitations},\ }\href {https://doi.org/10.1103/PhysRevLett.50.1395} {\bibfield  {journal} {\bibinfo  {journal} {Phys. Rev. Lett.}\ }\textbf {\bibinfo {volume} {50}},\ \bibinfo {pages} {1395} (\bibinfo {year} {1983})}\BibitemShut {NoStop}%
\bibitem [{\citenamefont {Haldane}(1983)}]{Haldane_fqh_1983}%
  \BibitemOpen
  \bibfield  {author} {\bibinfo {author} {\bibfnamefont {F.~D.~M.}\ \bibnamefont {Haldane}},\ }\bibfield  {title} {\bibinfo {title} {Fractional quantization of the {Hall} effect: {A} hierarchy of incompressible quantum fluid states},\ }\href {https://journals.aps.org/prl/abstract/10.1103/PhysRevLett.51.605} {\bibfield  {journal} {\bibinfo  {journal} {Phys. Rev. Lett.}\ }\textbf {\bibinfo {volume} {51}},\ \bibinfo {pages} {605} (\bibinfo {year} {1983})}\BibitemShut {NoStop}%
\bibitem [{\citenamefont {Halperin}(1984)}]{Halperin1984statistics}%
  \BibitemOpen
  \bibfield  {author} {\bibinfo {author} {\bibfnamefont {B.~I.}\ \bibnamefont {Halperin}},\ }\bibfield  {title} {\bibinfo {title} {Statistics of quasiparticles and the hierarchy of fractional quantized hall states},\ }\href {https://doi.org/10.1103/PhysRevLett.52.1583} {\bibfield  {journal} {\bibinfo  {journal} {Phys. Rev. Lett.}\ }\textbf {\bibinfo {volume} {52}},\ \bibinfo {pages} {1583} (\bibinfo {year} {1984})}\BibitemShut {NoStop}%
\bibitem [{\citenamefont {Halperin}\ and\ \citenamefont {Jain}(2020)}]{Halperin_FQHE_2020}%
  \BibitemOpen
  \bibfield  {author} {\bibinfo {author} {\bibfnamefont {B.~I.}\ \bibnamefont {Halperin}}\ and\ \bibinfo {author} {\bibfnamefont {J.~K.}\ \bibnamefont {Jain}},\ }\href {https://doi.org/10.1142/11751} {\emph {\bibinfo {title} {Fractional Quantum {Hall} Effects}}}\ (\bibinfo  {publisher} {World Scientific},\ \bibinfo {year} {2020})\BibitemShut {NoStop}%
\bibitem [{\citenamefont {de~Picciotto}\ \emph {et~al.}(1997)\citenamefont {de~Picciotto}, \citenamefont {Reznikov}, \citenamefont {Heiblum}, \citenamefont {Umansky}, \citenamefont {Bunin},\ and\ \citenamefont {Mahalu}}]{Picciotto_fractional_charge_1998}%
  \BibitemOpen
  \bibfield  {author} {\bibinfo {author} {\bibfnamefont {R.}~\bibnamefont {de~Picciotto}}, \bibinfo {author} {\bibfnamefont {M.}~\bibnamefont {Reznikov}}, \bibinfo {author} {\bibfnamefont {M.}~\bibnamefont {Heiblum}}, \bibinfo {author} {\bibfnamefont {V.}~\bibnamefont {Umansky}}, \bibinfo {author} {\bibfnamefont {G.}~\bibnamefont {Bunin}},\ and\ \bibinfo {author} {\bibfnamefont {D.}~\bibnamefont {Mahalu}},\ }\bibfield  {title} {\bibinfo {title} {Direct observation of a fractional charge},\ }\href {https://doi.org/https://doi.org/10.1038/38241} {\bibfield  {journal} {\bibinfo  {journal} {Nature}\ }\textbf {\bibinfo {volume} {389}},\ \bibinfo {pages} {162} (\bibinfo {year} {1997})}\BibitemShut {NoStop}%
\bibitem [{\citenamefont {Saminadayar}\ \emph {et~al.}(1997)\citenamefont {Saminadayar}, \citenamefont {Glattli}, \citenamefont {Jin},\ and\ \citenamefont {Etienne}}]{Saminadayar_fractional_charge_1997}%
  \BibitemOpen
  \bibfield  {author} {\bibinfo {author} {\bibfnamefont {L.}~\bibnamefont {Saminadayar}}, \bibinfo {author} {\bibfnamefont {D.~C.}\ \bibnamefont {Glattli}}, \bibinfo {author} {\bibfnamefont {Y.}~\bibnamefont {Jin}},\ and\ \bibinfo {author} {\bibfnamefont {B.}~\bibnamefont {Etienne}},\ }\bibfield  {title} {\bibinfo {title} {Observation of the $\mathit{e}\mathit{/}3$ fractionally charged laughlin quasiparticle},\ }\href {https://doi.org/10.1103/PhysRevLett.79.2526} {\bibfield  {journal} {\bibinfo  {journal} {Phys. Rev. Lett.}\ }\textbf {\bibinfo {volume} {79}},\ \bibinfo {pages} {2526} (\bibinfo {year} {1997})}\BibitemShut {NoStop}%
\bibitem [{\citenamefont {{Nakamura J.}}\ \emph {et~al.}(2020)\citenamefont {{Nakamura J.}}, \citenamefont {{Liang S.}}, \citenamefont {{Gardner G. C.}},\ and\ \citenamefont {{Manfra M. J.}}}]{Nakamura2020}%
  \BibitemOpen
  \bibfield  {author} {\bibinfo {author} {\bibnamefont {{Nakamura J.}}}, \bibinfo {author} {\bibnamefont {{Liang S.}}}, \bibinfo {author} {\bibnamefont {{Gardner G. C.}}},\ and\ \bibinfo {author} {\bibnamefont {{Manfra M. J.}}},\ }\bibfield  {title} {\bibinfo {title} {{Direct observation of anyonic braiding statistics}},\ }\href {https://doi.org/https://doi.org/10.1038/s41567-020-1019-1} {\bibfield  {journal} {\bibinfo  {journal} {Nature Physics}\ }\textbf {\bibinfo {volume} {16}},\ \bibinfo {pages} {931} (\bibinfo {year} {2020})}\BibitemShut {NoStop}%
\bibitem [{\citenamefont {{Bartolomei H.}}\ \emph {et~al.}(2020)\citenamefont {{Bartolomei H.}}, \citenamefont {{Kumar M.}}, \citenamefont {{Bisognin R.}}, \citenamefont {{Marguerite A.}}, \citenamefont {{Berroir J.-M.}}, \citenamefont {{Bocquillon E.}}, \citenamefont {{Pla\c{c}ais B.}}, \citenamefont {{Cavanna A.}}, \citenamefont {{Dong Q.}}, \citenamefont {{Gennser U.}}, \citenamefont {{Jin Y.}},\ and\ \citenamefont {{F{\`e}ve G.}}}]{Bartoleomei2020}%
  \BibitemOpen
  \bibfield  {author} {\bibinfo {author} {\bibnamefont {{Bartolomei H.}}}, \bibinfo {author} {\bibnamefont {{Kumar M.}}}, \bibinfo {author} {\bibnamefont {{Bisognin R.}}}, \bibinfo {author} {\bibnamefont {{Marguerite A.}}}, \bibinfo {author} {\bibnamefont {{Berroir J.-M.}}}, \bibinfo {author} {\bibnamefont {{Bocquillon E.}}}, \bibinfo {author} {\bibnamefont {{Pla\c{c}ais B.}}}, \bibinfo {author} {\bibnamefont {{Cavanna A.}}}, \bibinfo {author} {\bibnamefont {{Dong Q.}}}, \bibinfo {author} {\bibnamefont {{Gennser U.}}}, \bibinfo {author} {\bibnamefont {{Jin Y.}}},\ and\ \bibinfo {author} {\bibnamefont {{F{\`e}ve G.}}},\ }\bibfield  {title} {\bibinfo {title} {{Fractional statistics in anyon collisions}},\ }\href {https://doi.org/https://doi.org/10.1126/science.aaz5601} {\bibfield  {journal} {\bibinfo  {journal} {Science}\ }\textbf {\bibinfo {volume} {368}},\ \bibinfo {pages} {173} (\bibinfo {year} {2020})},\ \bibinfo {note} {doi: 10.1126/science.aaz5601}\BibitemShut {NoStop}%
\bibitem [{\citenamefont {{Lee June-Young M.}}\ \emph {et~al.}(2023)\citenamefont {{Lee June-Young M.}}, \citenamefont {{Hong Changki}}, \citenamefont {{Alkalay Tomer}}, \citenamefont {{Schiller Noam}}, \citenamefont {{Umansky Vladimir}}, \citenamefont {{Heiblum Moty}}, \citenamefont {{Oreg Yuval}},\ and\ \citenamefont {{Sim H.-S.}}}]{Lee2023}%
  \BibitemOpen
  \bibfield  {author} {\bibinfo {author} {\bibnamefont {{Lee June-Young M.}}}, \bibinfo {author} {\bibnamefont {{Hong Changki}}}, \bibinfo {author} {\bibnamefont {{Alkalay Tomer}}}, \bibinfo {author} {\bibnamefont {{Schiller Noam}}}, \bibinfo {author} {\bibnamefont {{Umansky Vladimir}}}, \bibinfo {author} {\bibnamefont {{Heiblum Moty}}}, \bibinfo {author} {\bibnamefont {{Oreg Yuval}}},\ and\ \bibinfo {author} {\bibnamefont {{Sim H.-S.}}},\ }\bibfield  {title} {\bibinfo {title} {{Partitioning of diluted anyons reveals their braiding statistics}},\ }\href {https://doi.org/https://doi.org/10.1038/s41586-023-05883-2} {\bibfield  {journal} {\bibinfo  {journal} {Nature}\ }\textbf {\bibinfo {volume} {617}},\ \bibinfo {pages} {277} (\bibinfo {year} {2023})}\BibitemShut {NoStop}%
\bibitem [{\citenamefont {Heiblum}\ and\ \citenamefont {Feldman}(2020)}]{Heiblum2020edge}%
  \BibitemOpen
  \bibfield  {author} {\bibinfo {author} {\bibfnamefont {M.}~\bibnamefont {Heiblum}}\ and\ \bibinfo {author} {\bibfnamefont {D.~E.}\ \bibnamefont {Feldman}},\ }\bibfield  {title} {\bibinfo {title} {Edge probes of topological order},\ }\href@noop {} {\bibfield  {journal} {\bibinfo  {journal} {International Journal of Modern Physics A}\ }\textbf {\bibinfo {volume} {35}},\ \bibinfo {pages} {2030009} (\bibinfo {year} {2020})}\BibitemShut {NoStop}%
\bibitem [{\citenamefont {Wen}(1995)}]{Wen_Topological_1995}%
  \BibitemOpen
  \bibfield  {author} {\bibinfo {author} {\bibfnamefont {X.~G.}\ \bibnamefont {Wen}},\ }\bibfield  {title} {\bibinfo {title} {Topological orders and edge excitations in fractional quantum {Hall} states},\ }\href {https://doi.org/10.1080/00018739500101566} {\bibfield  {journal} {\bibinfo  {journal} {Advances in Physics}\ }\textbf {\bibinfo {volume} {44}},\ \bibinfo {pages} {405} (\bibinfo {year} {1995})}\BibitemShut {NoStop}%
\bibitem [{\citenamefont {Kane}\ \emph {et~al.}(1994)\citenamefont {Kane}, \citenamefont {Fisher},\ and\ \citenamefont {Polchinski}}]{Kane_Randomness_1994}%
  \BibitemOpen
  \bibfield  {author} {\bibinfo {author} {\bibfnamefont {C.~L.}\ \bibnamefont {Kane}}, \bibinfo {author} {\bibfnamefont {M.~P.~A.}\ \bibnamefont {Fisher}},\ and\ \bibinfo {author} {\bibfnamefont {J.}~\bibnamefont {Polchinski}},\ }\bibfield  {title} {\bibinfo {title} {Randomness at the edge: Theory of quantum {Hall} transport at filling \ensuremath{\nu}=2/3},\ }\href {https://doi.org/10.1103/PhysRevLett.72.4129} {\bibfield  {journal} {\bibinfo  {journal} {Phys. Rev. Lett.}\ }\textbf {\bibinfo {volume} {72}},\ \bibinfo {pages} {4129} (\bibinfo {year} {1994})}\BibitemShut {NoStop}%
\bibitem [{\citenamefont {Kane}\ and\ \citenamefont {Fisher}(1995{\natexlab{a}})}]{Kane_Impurity_1995}%
  \BibitemOpen
  \bibfield  {author} {\bibinfo {author} {\bibfnamefont {C.~L.}\ \bibnamefont {Kane}}\ and\ \bibinfo {author} {\bibfnamefont {M.~P.~A.}\ \bibnamefont {Fisher}},\ }\bibfield  {title} {\bibinfo {title} {Impurity scattering and transport of fractional quantum {Hall} edge states},\ }\href {https://doi.org/10.1103/PhysRevB.51.13449} {\bibfield  {journal} {\bibinfo  {journal} {Phys. Rev. B}\ }\textbf {\bibinfo {volume} {51}},\ \bibinfo {pages} {13449} (\bibinfo {year} {1995}{\natexlab{a}})}\BibitemShut {NoStop}%
\bibitem [{\citenamefont {Kane}\ and\ \citenamefont {Fisher}(1997)}]{kane_quantized_1997}%
  \BibitemOpen
  \bibfield  {author} {\bibinfo {author} {\bibfnamefont {C.~L.}\ \bibnamefont {Kane}}\ and\ \bibinfo {author} {\bibfnamefont {M.~P.~A.}\ \bibnamefont {Fisher}},\ }\bibfield  {title} {\bibinfo {title} {Quantized thermal transport in the fractional quantum {Hall} effect},\ }\href {https://doi.org/10.1103/PhysRevB.55.15832} {\bibfield  {journal} {\bibinfo  {journal} {Phys. Rev. B}\ }\textbf {\bibinfo {volume} {55}},\ \bibinfo {pages} {15832} (\bibinfo {year} {1997})}\BibitemShut {NoStop}%
\bibitem [{\citenamefont {Rosenow}\ and\ \citenamefont {Halperin}(2010)}]{Rosenow_signatures_2010}%
  \BibitemOpen
  \bibfield  {author} {\bibinfo {author} {\bibfnamefont {B.}~\bibnamefont {Rosenow}}\ and\ \bibinfo {author} {\bibfnamefont {B.~I.}\ \bibnamefont {Halperin}},\ }\bibfield  {title} {\bibinfo {title} {Signatures of neutral quantum {Hall} modes in transport through low-density constrictions},\ }\href {https://doi.org/10.1103/PhysRevB.81.165313} {\bibfield  {journal} {\bibinfo  {journal} {Phys. Rev. B}\ }\textbf {\bibinfo {volume} {81}},\ \bibinfo {pages} {165313} (\bibinfo {year} {2010})}\BibitemShut {NoStop}%
\bibitem [{\citenamefont {Protopopov}\ \emph {et~al.}(2017)\citenamefont {Protopopov}, \citenamefont {Gefen},\ and\ \citenamefont {Mirlin}}]{Protopopov_transport_2_3_2017}%
  \BibitemOpen
  \bibfield  {author} {\bibinfo {author} {\bibfnamefont {I.}~\bibnamefont {Protopopov}}, \bibinfo {author} {\bibfnamefont {Y.}~\bibnamefont {Gefen}},\ and\ \bibinfo {author} {\bibfnamefont {A.}~\bibnamefont {Mirlin}},\ }\bibfield  {title} {\bibinfo {title} {Transport in a disordered $\nu=2/3$ fractional quantum {Hall} junction},\ }\href {https://doi.org/https://doi.org/10.1016/j.aop.2017.07.015} {\bibfield  {journal} {\bibinfo  {journal} {Annals of Physics}\ }\textbf {\bibinfo {volume} {385}},\ \bibinfo {pages} {287 } (\bibinfo {year} {2017})}\BibitemShut {NoStop}%
\bibitem [{\citenamefont {Nosiglia}\ \emph {et~al.}(2018)\citenamefont {Nosiglia}, \citenamefont {Park}, \citenamefont {Rosenow},\ and\ \citenamefont {Gefen}}]{Nosiglia2018}%
  \BibitemOpen
  \bibfield  {author} {\bibinfo {author} {\bibfnamefont {C.}~\bibnamefont {Nosiglia}}, \bibinfo {author} {\bibfnamefont {J.}~\bibnamefont {Park}}, \bibinfo {author} {\bibfnamefont {B.}~\bibnamefont {Rosenow}},\ and\ \bibinfo {author} {\bibfnamefont {Y.}~\bibnamefont {Gefen}},\ }\bibfield  {title} {\bibinfo {title} {Incoherent transport on the $\ensuremath{\nu}=2/3$ quantum hall edge},\ }\href {https://doi.org/10.1103/PhysRevB.98.115408} {\bibfield  {journal} {\bibinfo  {journal} {Phys. Rev. B}\ }\textbf {\bibinfo {volume} {98}},\ \bibinfo {pages} {115408} (\bibinfo {year} {2018})}\BibitemShut {NoStop}%
\bibitem [{\citenamefont {Sp\aa{}nsl\"att}\ \emph {et~al.}(2021)\citenamefont {Sp\aa{}nsl\"att}, \citenamefont {Gefen}, \citenamefont {Gornyi},\ and\ \citenamefont {Polyakov}}]{Spanslatt2021}%
  \BibitemOpen
  \bibfield  {author} {\bibinfo {author} {\bibfnamefont {C.}~\bibnamefont {Sp\aa{}nsl\"att}}, \bibinfo {author} {\bibfnamefont {Y.}~\bibnamefont {Gefen}}, \bibinfo {author} {\bibfnamefont {I.~V.}\ \bibnamefont {Gornyi}},\ and\ \bibinfo {author} {\bibfnamefont {D.~G.}\ \bibnamefont {Polyakov}},\ }\bibfield  {title} {\bibinfo {title} {Contacts, equilibration, and interactions in fractional quantum hall edge transport},\ }\href {https://doi.org/10.1103/PhysRevB.104.115416} {\bibfield  {journal} {\bibinfo  {journal} {Phys. Rev. B}\ }\textbf {\bibinfo {volume} {104}},\ \bibinfo {pages} {115416} (\bibinfo {year} {2021})}\BibitemShut {NoStop}%
\bibitem [{\citenamefont {Banerjee}\ \emph {et~al.}(2017)\citenamefont {Banerjee}, \citenamefont {Heiblum}, \citenamefont {Rosenblatt}, \citenamefont {Oreg}, \citenamefont {Feldman}, \citenamefont {Stern},\ and\ \citenamefont {Umansky}}]{Banerjee_Observed_2017}%
  \BibitemOpen
  \bibfield  {author} {\bibinfo {author} {\bibfnamefont {M.}~\bibnamefont {Banerjee}}, \bibinfo {author} {\bibfnamefont {M.}~\bibnamefont {Heiblum}}, \bibinfo {author} {\bibfnamefont {A.}~\bibnamefont {Rosenblatt}}, \bibinfo {author} {\bibfnamefont {Y.}~\bibnamefont {Oreg}}, \bibinfo {author} {\bibfnamefont {D.~E.}\ \bibnamefont {Feldman}}, \bibinfo {author} {\bibfnamefont {A.}~\bibnamefont {Stern}},\ and\ \bibinfo {author} {\bibfnamefont {V.}~\bibnamefont {Umansky}},\ }\bibfield  {title} {\bibinfo {title} {Observed quantization of anyonic heat flow},\ }\href {https://doi.org/10.1038/nature22052} {\bibfield  {journal} {\bibinfo  {journal} {Nature}\ }\textbf {\bibinfo {volume} {545}},\ \bibinfo {pages} {75} (\bibinfo {year} {2017})}\BibitemShut {NoStop}%
\bibitem [{\citenamefont {Banerjee}\ \emph {et~al.}(2018)\citenamefont {Banerjee}, \citenamefont {Heiblum}, \citenamefont {Umansky}, \citenamefont {Feldman}, \citenamefont {Oreg},\ and\ \citenamefont {Stern}}]{Banerjee_observation_2018}%
  \BibitemOpen
  \bibfield  {author} {\bibinfo {author} {\bibfnamefont {M.}~\bibnamefont {Banerjee}}, \bibinfo {author} {\bibfnamefont {M.}~\bibnamefont {Heiblum}}, \bibinfo {author} {\bibfnamefont {V.}~\bibnamefont {Umansky}}, \bibinfo {author} {\bibfnamefont {D.~E.}\ \bibnamefont {Feldman}}, \bibinfo {author} {\bibfnamefont {Y.}~\bibnamefont {Oreg}},\ and\ \bibinfo {author} {\bibfnamefont {A.}~\bibnamefont {Stern}},\ }\bibfield  {title} {\bibinfo {title} {Observation of half-integer thermal {Hall} conductance},\ }\href {https://doi.org/10.1038/s41586-018-0184-1} {\bibfield  {journal} {\bibinfo  {journal} {Nature}\ }\textbf {\bibinfo {volume} {559}},\ \bibinfo {pages} {205} (\bibinfo {year} {2018})}\BibitemShut {NoStop}%
\bibitem [{\citenamefont {Melcer}\ \emph {et~al.}(2022)\citenamefont {Melcer}, \citenamefont {Dutta}, \citenamefont {Sp{\aa}nsl{\ifmmode\ddot{a}\else\"{a}\fi}tt}, \citenamefont {Park}, \citenamefont {Mirlin},\ and\ \citenamefont {Umansky}}]{Melcer_Absent_2022}%
  \BibitemOpen
  \bibfield  {author} {\bibinfo {author} {\bibfnamefont {R.~A.}\ \bibnamefont {Melcer}}, \bibinfo {author} {\bibfnamefont {B.}~\bibnamefont {Dutta}}, \bibinfo {author} {\bibfnamefont {C.}~\bibnamefont {Sp{\aa}nsl{\ifmmode\ddot{a}\else\"{a}\fi}tt}}, \bibinfo {author} {\bibfnamefont {J.}~\bibnamefont {Park}}, \bibinfo {author} {\bibfnamefont {A.~D.}\ \bibnamefont {Mirlin}},\ and\ \bibinfo {author} {\bibfnamefont {V.}~\bibnamefont {Umansky}},\ }\bibfield  {title} {\bibinfo {title} {Absent thermal equilibration on fractional quantum hall edges over macroscopic scale},\ }\href {https://doi.org/10.1038/s41467-022-28009-0} {\bibfield  {journal} {\bibinfo  {journal} {Nature Communications}\ }\textbf {\bibinfo {volume} {13}},\ \bibinfo {pages} {376} (\bibinfo {year} {2022})}\BibitemShut {NoStop}%
\bibitem [{\citenamefont {Dutta}\ \emph {et~al.}(2022{\natexlab{a}})\citenamefont {Dutta}, \citenamefont {Umansky}, \citenamefont {Banerjee},\ and\ \citenamefont {Heiblum}}]{Dutta_Isolated_2022}%
  \BibitemOpen
  \bibfield  {author} {\bibinfo {author} {\bibfnamefont {B.}~\bibnamefont {Dutta}}, \bibinfo {author} {\bibfnamefont {V.}~\bibnamefont {Umansky}}, \bibinfo {author} {\bibfnamefont {M.}~\bibnamefont {Banerjee}},\ and\ \bibinfo {author} {\bibfnamefont {M.}~\bibnamefont {Heiblum}},\ }\bibfield  {title} {\bibinfo {title} {Isolated ballistic non-{Abelian} interface channel},\ }\href {https://doi.org/10.1126/science.abm6571} {\bibfield  {journal} {\bibinfo  {journal} {Science}\ }\textbf {\bibinfo {volume} {377}},\ \bibinfo {pages} {1198} (\bibinfo {year} {2022}{\natexlab{a}})}\BibitemShut {NoStop}%
\bibitem [{\citenamefont {Srivastav}\ \emph {et~al.}(2021)\citenamefont {Srivastav}, \citenamefont {Kumar}, \citenamefont {Sp{\aa}nsl{\ifmmode\ddot{a}\else\"{a}\fi}tt}, \citenamefont {Watanabe}, \citenamefont {Taniguchi}, \citenamefont {Mirlin}, \citenamefont {Gefen},\ and\ \citenamefont {Das}}]{Srivastav2021May}%
  \BibitemOpen
  \bibfield  {author} {\bibinfo {author} {\bibfnamefont {S.~K.}\ \bibnamefont {Srivastav}}, \bibinfo {author} {\bibfnamefont {R.}~\bibnamefont {Kumar}}, \bibinfo {author} {\bibfnamefont {C.}~\bibnamefont {Sp{\aa}nsl{\ifmmode\ddot{a}\else\"{a}\fi}tt}}, \bibinfo {author} {\bibfnamefont {K.}~\bibnamefont {Watanabe}}, \bibinfo {author} {\bibfnamefont {T.}~\bibnamefont {Taniguchi}}, \bibinfo {author} {\bibfnamefont {A.~D.}\ \bibnamefont {Mirlin}}, \bibinfo {author} {\bibfnamefont {Y.}~\bibnamefont {Gefen}},\ and\ \bibinfo {author} {\bibfnamefont {A.}~\bibnamefont {Das}},\ }\bibfield  {title} {\bibinfo {title} {{Vanishing Thermal Equilibration for Hole-Conjugate Fractional Quantum {H}all States in Graphene}},\ }\href {https://doi.org/10.1103/PhysRevLett.126.216803} {\bibfield  {journal} {\bibinfo  {journal} {Phys. Rev. Lett.}\ }\textbf {\bibinfo {volume} {126}},\ \bibinfo {pages} {216803} (\bibinfo {year} {2021})}\BibitemShut {NoStop}%
\bibitem [{\citenamefont {Srivastav}\ \emph {et~al.}(2022)\citenamefont {Srivastav}, \citenamefont {Kumar}, \citenamefont {Sp{\aa}nsl{\ifmmode\ddot{a}\else\"{a}\fi}tt}, \citenamefont {Watanabe}, \citenamefont {Taniguchi}, \citenamefont {Mirlin}, \citenamefont {Gefen},\ and\ \citenamefont {Das}}]{Srivastav2022Sep}%
  \BibitemOpen
  \bibfield  {author} {\bibinfo {author} {\bibfnamefont {S.~K.}\ \bibnamefont {Srivastav}}, \bibinfo {author} {\bibfnamefont {R.}~\bibnamefont {Kumar}}, \bibinfo {author} {\bibfnamefont {C.}~\bibnamefont {Sp{\aa}nsl{\ifmmode\ddot{a}\else\"{a}\fi}tt}}, \bibinfo {author} {\bibfnamefont {K.}~\bibnamefont {Watanabe}}, \bibinfo {author} {\bibfnamefont {T.}~\bibnamefont {Taniguchi}}, \bibinfo {author} {\bibfnamefont {A.~D.}\ \bibnamefont {Mirlin}}, \bibinfo {author} {\bibfnamefont {Y.}~\bibnamefont {Gefen}},\ and\ \bibinfo {author} {\bibfnamefont {A.}~\bibnamefont {Das}},\ }\bibfield  {title} {\bibinfo {title} {{Determination of topological edge quantum numbers of fractional quantum {H}all phases by thermal conductance measurements}},\ }\href {https://doi.org/10.1038/s41467-022-32956-z} {\bibfield  {journal} {\bibinfo  {journal} {Nat. Commun.}\ }\textbf {\bibinfo {volume} {13}},\ \bibinfo {pages} {1} (\bibinfo {year} {2022})}\BibitemShut {NoStop}%
\bibitem [{\citenamefont {Le~Breton}\ \emph {et~al.}(2022)\citenamefont {Le~Breton}, \citenamefont {Delagrange}, \citenamefont {Hong}, \citenamefont {Garg}, \citenamefont {Watanabe}, \citenamefont {Taniguchi}, \citenamefont {Ribeiro-Palau}, \citenamefont {Roulleau}, \citenamefont {Roche},\ and\ \citenamefont {Parmentier}}]{Breton2022}%
  \BibitemOpen
  \bibfield  {author} {\bibinfo {author} {\bibfnamefont {G.}~\bibnamefont {Le~Breton}}, \bibinfo {author} {\bibfnamefont {R.}~\bibnamefont {Delagrange}}, \bibinfo {author} {\bibfnamefont {Y.}~\bibnamefont {Hong}}, \bibinfo {author} {\bibfnamefont {M.}~\bibnamefont {Garg}}, \bibinfo {author} {\bibfnamefont {K.}~\bibnamefont {Watanabe}}, \bibinfo {author} {\bibfnamefont {T.}~\bibnamefont {Taniguchi}}, \bibinfo {author} {\bibfnamefont {R.}~\bibnamefont {Ribeiro-Palau}}, \bibinfo {author} {\bibfnamefont {P.}~\bibnamefont {Roulleau}}, \bibinfo {author} {\bibfnamefont {P.}~\bibnamefont {Roche}},\ and\ \bibinfo {author} {\bibfnamefont {F.~D.}\ \bibnamefont {Parmentier}},\ }\bibfield  {title} {\bibinfo {title} {Heat equilibration of integer and fractional quantum {H}all edge modes in graphene},\ }\href {https://doi.org/10.1103/PhysRevLett.129.116803} {\bibfield  {journal} {\bibinfo  {journal} {Phys. Rev. Lett.}\ }\textbf {\bibinfo {volume} {129}},\ \bibinfo {pages} {116803} (\bibinfo {year} {2022})}\BibitemShut
  {NoStop}%
\bibitem [{\citenamefont {Grivnin}\ \emph {et~al.}(2014)\citenamefont {Grivnin}, \citenamefont {Inoue}, \citenamefont {Ronen}, \citenamefont {Baum}, \citenamefont {Heiblum}, \citenamefont {Umansky},\ and\ \citenamefont {Mahalu}}]{Grivnin2014}%
  \BibitemOpen
  \bibfield  {author} {\bibinfo {author} {\bibfnamefont {A.}~\bibnamefont {Grivnin}}, \bibinfo {author} {\bibfnamefont {H.}~\bibnamefont {Inoue}}, \bibinfo {author} {\bibfnamefont {Y.}~\bibnamefont {Ronen}}, \bibinfo {author} {\bibfnamefont {Y.}~\bibnamefont {Baum}}, \bibinfo {author} {\bibfnamefont {M.}~\bibnamefont {Heiblum}}, \bibinfo {author} {\bibfnamefont {V.}~\bibnamefont {Umansky}},\ and\ \bibinfo {author} {\bibfnamefont {D.}~\bibnamefont {Mahalu}},\ }\bibfield  {title} {\bibinfo {title} {Nonequilibrated counterpropagating edge modes in the fractional quantum hall regime},\ }\href {https://doi.org/10.1103/PhysRevLett.113.266803} {\bibfield  {journal} {\bibinfo  {journal} {Phys. Rev. Lett.}\ }\textbf {\bibinfo {volume} {113}},\ \bibinfo {pages} {266803} (\bibinfo {year} {2014})}\BibitemShut {NoStop}%
\bibitem [{\citenamefont {{Wang Ying}}\ \emph {et~al.}(2021)\citenamefont {{Wang Ying}}, \citenamefont {{Ponomarenko Vadim}}, \citenamefont {{Wan Zhong}}, \citenamefont {{West Kenneth W.}}, \citenamefont {{Baldwin Kirk W.}}, \citenamefont {{Pfeiffer Loren N.}}, \citenamefont {{Lyanda-Geller Yuli}},\ and\ \citenamefont {{Rokhinson Leonid P.}}}]{Wang2021}%
  \BibitemOpen
  \bibfield  {author} {\bibinfo {author} {\bibnamefont {{Wang Ying}}}, \bibinfo {author} {\bibnamefont {{Ponomarenko Vadim}}}, \bibinfo {author} {\bibnamefont {{Wan Zhong}}}, \bibinfo {author} {\bibnamefont {{West Kenneth W.}}}, \bibinfo {author} {\bibnamefont {{Baldwin Kirk W.}}}, \bibinfo {author} {\bibnamefont {{Pfeiffer Loren N.}}}, \bibinfo {author} {\bibnamefont {{Lyanda-Geller Yuli}}},\ and\ \bibinfo {author} {\bibnamefont {{Rokhinson Leonid P.}}},\ }\bibfield  {title} {\bibinfo {title} {{Transport in helical Luttinger liquids in the fractional quantum Hall regime}},\ }\href {https://doi.org/https://doi.org/10.1038/s41467-021-25631-2} {\bibfield  {journal} {\bibinfo  {journal} {Nature Communications}\ }\textbf {\bibinfo {volume} {12}},\ \bibinfo {pages} {5312} (\bibinfo {year} {2021})}\BibitemShut {NoStop}%
\bibitem [{\citenamefont {Hashisaka}\ \emph {et~al.}(2021)\citenamefont {Hashisaka}, \citenamefont {Jonckheere}, \citenamefont {Akiho}, \citenamefont {Sasaki}, \citenamefont {Rech}, \citenamefont {Martin},\ and\ \citenamefont {Muraki}}]{Hashisaka2021}%
  \BibitemOpen
  \bibfield  {author} {\bibinfo {author} {\bibfnamefont {M.}~\bibnamefont {Hashisaka}}, \bibinfo {author} {\bibfnamefont {T.}~\bibnamefont {Jonckheere}}, \bibinfo {author} {\bibfnamefont {T.}~\bibnamefont {Akiho}}, \bibinfo {author} {\bibfnamefont {S.}~\bibnamefont {Sasaki}}, \bibinfo {author} {\bibfnamefont {J.}~\bibnamefont {Rech}}, \bibinfo {author} {\bibfnamefont {T.}~\bibnamefont {Martin}},\ and\ \bibinfo {author} {\bibfnamefont {K.}~\bibnamefont {Muraki}},\ }\bibfield  {title} {\bibinfo {title} {{Andreev reflection of fractional quantum Hall quasiparticles}},\ }\href {https://doi.org/10.1038/s41467-021-23160-6} {\bibfield  {journal} {\bibinfo  {journal} {Nat. Commun.}\ }\textbf {\bibinfo {volume} {12}},\ \bibinfo {pages} {1} (\bibinfo {year} {2021})}\BibitemShut {NoStop}%
\bibitem [{\citenamefont {Dutta}\ \emph {et~al.}(2022{\natexlab{b}})\citenamefont {Dutta}, \citenamefont {Yang}, \citenamefont {Melcer}, \citenamefont {Kundu}, \citenamefont {Heiblum}, \citenamefont {Umansky}, \citenamefont {Oreg}, \citenamefont {Stern},\ and\ \citenamefont {Mross}}]{Dutta_novel_2022}%
  \BibitemOpen
  \bibfield  {author} {\bibinfo {author} {\bibfnamefont {B.}~\bibnamefont {Dutta}}, \bibinfo {author} {\bibfnamefont {W.}~\bibnamefont {Yang}}, \bibinfo {author} {\bibfnamefont {R.}~\bibnamefont {Melcer}}, \bibinfo {author} {\bibfnamefont {H.~K.}\ \bibnamefont {Kundu}}, \bibinfo {author} {\bibfnamefont {M.}~\bibnamefont {Heiblum}}, \bibinfo {author} {\bibfnamefont {V.}~\bibnamefont {Umansky}}, \bibinfo {author} {\bibfnamefont {Y.}~\bibnamefont {Oreg}}, \bibinfo {author} {\bibfnamefont {A.}~\bibnamefont {Stern}},\ and\ \bibinfo {author} {\bibfnamefont {D.}~\bibnamefont {Mross}},\ }\bibfield  {title} {\bibinfo {title} {Distinguishing between non-{Abelian} topological orders in a quantum {Hall} system},\ }\href {https://doi.org/10.1126/science.abg6116} {\bibfield  {journal} {\bibinfo  {journal} {Science}\ }\textbf {\bibinfo {volume} {375}},\ \bibinfo {pages} {193} (\bibinfo {year} {2022}{\natexlab{b}})}\BibitemShut {NoStop}%
\bibitem [{\citenamefont {Hashisaka}\ \emph {et~al.}(2023)\citenamefont {Hashisaka}, \citenamefont {Ito}, \citenamefont {Akiho}, \citenamefont {Sasaki}, \citenamefont {Kumada}, \citenamefont {Shibata},\ and\ \citenamefont {Muraki}}]{Hashisaka2023}%
  \BibitemOpen
  \bibfield  {author} {\bibinfo {author} {\bibfnamefont {M.}~\bibnamefont {Hashisaka}}, \bibinfo {author} {\bibfnamefont {T.}~\bibnamefont {Ito}}, \bibinfo {author} {\bibfnamefont {T.}~\bibnamefont {Akiho}}, \bibinfo {author} {\bibfnamefont {S.}~\bibnamefont {Sasaki}}, \bibinfo {author} {\bibfnamefont {N.}~\bibnamefont {Kumada}}, \bibinfo {author} {\bibfnamefont {N.}~\bibnamefont {Shibata}},\ and\ \bibinfo {author} {\bibfnamefont {K.}~\bibnamefont {Muraki}},\ }\bibfield  {title} {\bibinfo {title} {Coherent-incoherent crossover of charge and neutral mode transport as evidence for the disorder-dominated fractional edge phase},\ }\href {https://doi.org/10.1103/PhysRevX.13.031024} {\bibfield  {journal} {\bibinfo  {journal} {Phys. Rev. X}\ }\textbf {\bibinfo {volume} {13}},\ \bibinfo {pages} {031024} (\bibinfo {year} {2023})}\BibitemShut {NoStop}%
\bibitem [{\citenamefont {{Ronen Yuval}}\ \emph {et~al.}(2018)\citenamefont {{Ronen Yuval}}, \citenamefont {{Cohen Yonatan}}, \citenamefont {{Banitt Daniel}}, \citenamefont {{Heiblum Moty}},\ and\ \citenamefont {{Umansky Vladimir}}}]{Ronen2018}%
  \BibitemOpen
  \bibfield  {author} {\bibinfo {author} {\bibnamefont {{Ronen Yuval}}}, \bibinfo {author} {\bibnamefont {{Cohen Yonatan}}}, \bibinfo {author} {\bibnamefont {{Banitt Daniel}}}, \bibinfo {author} {\bibnamefont {{Heiblum Moty}}},\ and\ \bibinfo {author} {\bibnamefont {{Umansky Vladimir}}},\ }\bibfield  {title} {\bibinfo {title} {{Robust integer and fractional helical modes in the quantum Hall effect}},\ }\href {https://doi.org/https://doi.org/10.1038/s41567-017-0035-2} {\bibfield  {journal} {\bibinfo  {journal} {Nature Physics}\ }\textbf {\bibinfo {volume} {14}},\ \bibinfo {pages} {411} (\bibinfo {year} {2018})}\BibitemShut {NoStop}%
\bibitem [{\citenamefont {{Cohen Yonatan}}\ \emph {et~al.}(2019)\citenamefont {{Cohen Yonatan}}, \citenamefont {{Ronen Yuval}}, \citenamefont {{Yang Wenmin}}, \citenamefont {{Banitt Daniel}}, \citenamefont {{Park Jinhong}}, \citenamefont {{Heiblum Moty}}, \citenamefont {{Mirlin Alexander D.}}, \citenamefont {{Gefen Yuval}},\ and\ \citenamefont {{Umansky Vladimir}}}]{Cohen2019}%
  \BibitemOpen
  \bibfield  {author} {\bibinfo {author} {\bibnamefont {{Cohen Yonatan}}}, \bibinfo {author} {\bibnamefont {{Ronen Yuval}}}, \bibinfo {author} {\bibnamefont {{Yang Wenmin}}}, \bibinfo {author} {\bibnamefont {{Banitt Daniel}}}, \bibinfo {author} {\bibnamefont {{Park Jinhong}}}, \bibinfo {author} {\bibnamefont {{Heiblum Moty}}}, \bibinfo {author} {\bibnamefont {{Mirlin Alexander D.}}}, \bibinfo {author} {\bibnamefont {{Gefen Yuval}}},\ and\ \bibinfo {author} {\bibnamefont {{Umansky Vladimir}}},\ }\bibfield  {title} {\bibinfo {title} {{Synthesizing a ν=2/3 fractional quantum Hall effect edge state from counter-propagating ν=1 and ν=1/3 states}},\ }\href {https://doi.org/https://doi.org/10.1038/s41467-019-09920-5} {\bibfield  {journal} {\bibinfo  {journal} {Nature Communications}\ }\textbf {\bibinfo {volume} {10}},\ \bibinfo {pages} {1920} (\bibinfo {year} {2019})}\BibitemShut {NoStop}%
\bibitem [{\citenamefont {Sp\aa{}nsl\"att}\ \emph {et~al.}(2020)\citenamefont {Sp\aa{}nsl\"att}, \citenamefont {Park}, \citenamefont {Gefen},\ and\ \citenamefont {Mirlin}}]{Spanslatt_condplateau}%
  \BibitemOpen
  \bibfield  {author} {\bibinfo {author} {\bibfnamefont {C.}~\bibnamefont {Sp\aa{}nsl\"att}}, \bibinfo {author} {\bibfnamefont {J.}~\bibnamefont {Park}}, \bibinfo {author} {\bibfnamefont {Y.}~\bibnamefont {Gefen}},\ and\ \bibinfo {author} {\bibfnamefont {A.~D.}\ \bibnamefont {Mirlin}},\ }\bibfield  {title} {\bibinfo {title} {Conductance plateaus and shot noise in fractional quantum hall point contacts},\ }\href {https://doi.org/10.1103/PhysRevB.101.075308} {\bibfield  {journal} {\bibinfo  {journal} {Phys. Rev. B}\ }\textbf {\bibinfo {volume} {101}},\ \bibinfo {pages} {075308} (\bibinfo {year} {2020})}\BibitemShut {NoStop}%
\bibitem [{\citenamefont {Cohen}\ \emph {et~al.}(2023)\citenamefont {Cohen}, \citenamefont {Samuelson}, \citenamefont {Wang}, \citenamefont {Taniguchi}, \citenamefont {Watanabe}, \citenamefont {Zaletel},\ and\ \citenamefont {Young}}]{Cohen2023}%
  \BibitemOpen
  \bibfield  {author} {\bibinfo {author} {\bibfnamefont {L.~A.}\ \bibnamefont {Cohen}}, \bibinfo {author} {\bibfnamefont {N.~L.}\ \bibnamefont {Samuelson}}, \bibinfo {author} {\bibfnamefont {T.}~\bibnamefont {Wang}}, \bibinfo {author} {\bibfnamefont {T.}~\bibnamefont {Taniguchi}}, \bibinfo {author} {\bibfnamefont {K.}~\bibnamefont {Watanabe}}, \bibinfo {author} {\bibfnamefont {M.~P.}\ \bibnamefont {Zaletel}},\ and\ \bibinfo {author} {\bibfnamefont {A.~F.}\ \bibnamefont {Young}},\ }\bibfield  {title} {\bibinfo {title} {Universal chiral luttinger liquid behavior in a graphene fractional quantum hall point contact},\ }\href {https://doi.org/10.1126/science.adf9728} {\bibfield  {journal} {\bibinfo  {journal} {Science}\ }\textbf {\bibinfo {volume} {382}},\ \bibinfo {pages} {542} (\bibinfo {year} {2023})},\ \Eprint {https://arxiv.org/abs/https://www.science.org/doi/pdf/10.1126/science.adf9728} {https://www.science.org/doi/pdf/10.1126/science.adf9728} \BibitemShut {NoStop}%
\bibitem [{\citenamefont {{Ronen Yuval}}\ \emph {et~al.}(2021)\citenamefont {{Ronen Yuval}}, \citenamefont {{Werkmeister Thomas}}, \citenamefont {{Haie Najafabadi Danial}}, \citenamefont {{Pierce Andrew T.}}, \citenamefont {{Anderson Laurel E.}}, \citenamefont {{Shin Young Jae}}, \citenamefont {{Lee Si Young}}, \citenamefont {{Lee Young Hee}}, \citenamefont {{Johnson Bobae}}, \citenamefont {{Watanabe Kenji}}, \citenamefont {{Taniguchi Takashi}}, \citenamefont {{Yacoby Amir}},\ and\ \citenamefont {{Kim Philip}}}]{Ronen2021}%
  \BibitemOpen
  \bibfield  {author} {\bibinfo {author} {\bibnamefont {{Ronen Yuval}}}, \bibinfo {author} {\bibnamefont {{Werkmeister Thomas}}}, \bibinfo {author} {\bibnamefont {{Haie Najafabadi Danial}}}, \bibinfo {author} {\bibnamefont {{Pierce Andrew T.}}}, \bibinfo {author} {\bibnamefont {{Anderson Laurel E.}}}, \bibinfo {author} {\bibnamefont {{Shin Young Jae}}}, \bibinfo {author} {\bibnamefont {{Lee Si Young}}}, \bibinfo {author} {\bibnamefont {{Lee Young Hee}}}, \bibinfo {author} {\bibnamefont {{Johnson Bobae}}}, \bibinfo {author} {\bibnamefont {{Watanabe Kenji}}}, \bibinfo {author} {\bibnamefont {{Taniguchi Takashi}}}, \bibinfo {author} {\bibnamefont {{Yacoby Amir}}},\ and\ \bibinfo {author} {\bibnamefont {{Kim Philip}}},\ }\bibfield  {title} {\bibinfo {title} {{Aharonov--Bohm effect in graphene-based Fabry--P{\'e}rot quantum Hall interferometers}},\ }\href {https://doi.org/https://doi.org/10.1038/s41565-021-00861-z} {\bibfield  {journal} {\bibinfo  {journal} {Nature Nanotechnology}\ }\textbf {\bibinfo {volume}
  {16}},\ \bibinfo {pages} {563} (\bibinfo {year} {2021})}\BibitemShut {NoStop}%
\bibitem [{\citenamefont {{D{\'e}prez Corentin}}\ \emph {et~al.}(2021)\citenamefont {{D{\'e}prez Corentin}}, \citenamefont {{Veyrat Louis}}, \citenamefont {{Vignaud Hadrien}}, \citenamefont {{Nayak Goutham}}, \citenamefont {{Watanabe Kenji}}, \citenamefont {{Taniguchi Takashi}}, \citenamefont {{Gay Fr{\'e}d{\'e}ric}}, \citenamefont {{Sellier Hermann}},\ and\ \citenamefont {{Sac{\'e}p{\'e} Benjamin}}}]{Deprez2021}%
  \BibitemOpen
  \bibfield  {author} {\bibinfo {author} {\bibnamefont {{D{\'e}prez Corentin}}}, \bibinfo {author} {\bibnamefont {{Veyrat Louis}}}, \bibinfo {author} {\bibnamefont {{Vignaud Hadrien}}}, \bibinfo {author} {\bibnamefont {{Nayak Goutham}}}, \bibinfo {author} {\bibnamefont {{Watanabe Kenji}}}, \bibinfo {author} {\bibnamefont {{Taniguchi Takashi}}}, \bibinfo {author} {\bibnamefont {{Gay Fr{\'e}d{\'e}ric}}}, \bibinfo {author} {\bibnamefont {{Sellier Hermann}}},\ and\ \bibinfo {author} {\bibnamefont {{Sac{\'e}p{\'e} Benjamin}}},\ }\bibfield  {title} {\bibinfo {title} {{A tunable Fabry--P{\'e}rot quantum Hall interferometer in graphene}},\ }\href {https://doi.org/https://doi.org/10.1038/s41565-021-00847-x} {\bibfield  {journal} {\bibinfo  {journal} {Nature Nanotechnology}\ }\textbf {\bibinfo {volume} {16}},\ \bibinfo {pages} {555} (\bibinfo {year} {2021})}\BibitemShut {NoStop}%
\bibitem [{\citenamefont {Haldane}(1995)}]{Haldane_Stability_1995}%
  \BibitemOpen
  \bibfield  {author} {\bibinfo {author} {\bibfnamefont {F.~D.~M.}\ \bibnamefont {Haldane}},\ }\bibfield  {title} {\bibinfo {title} {Stability of chiral {Luttinger} liquids and abelian quantum {Hall} states},\ }\href {https://doi.org/10.1103/PhysRevLett.74.2090} {\bibfield  {journal} {\bibinfo  {journal} {Phys. Rev. Lett.}\ }\textbf {\bibinfo {volume} {74}},\ \bibinfo {pages} {2090} (\bibinfo {year} {1995})}\BibitemShut {NoStop}%
\bibitem [{\citenamefont {Kao}\ \emph {et~al.}(1999)\citenamefont {Kao}, \citenamefont {Chang},\ and\ \citenamefont {Wen}}]{Kao_Binding_1999}%
  \BibitemOpen
  \bibfield  {author} {\bibinfo {author} {\bibfnamefont {H.~C.}\ \bibnamefont {Kao}}, \bibinfo {author} {\bibfnamefont {C.~H.}\ \bibnamefont {Chang}},\ and\ \bibinfo {author} {\bibfnamefont {X.~G.}\ \bibnamefont {Wen}},\ }\bibfield  {title} {\bibinfo {title} {Binding transition in quantum {Hall} edge states},\ }\href {https://doi.org/10.1103/PhysRevLett.83.5563} {\bibfield  {journal} {\bibinfo  {journal} {Phys. Rev. Lett.}\ }\textbf {\bibinfo {volume} {83}},\ \bibinfo {pages} {5563} (\bibinfo {year} {1999})}\BibitemShut {NoStop}%
\bibitem [{\citenamefont {Sp\aa{}nsl\"att}\ \emph {et~al.}(2023)\citenamefont {Sp\aa{}nsl\"att}, \citenamefont {Stern},\ and\ \citenamefont {Mirlin}}]{Spanslatt_binding_2023}%
  \BibitemOpen
  \bibfield  {author} {\bibinfo {author} {\bibfnamefont {C.}~\bibnamefont {Sp\aa{}nsl\"att}}, \bibinfo {author} {\bibfnamefont {A.}~\bibnamefont {Stern}},\ and\ \bibinfo {author} {\bibfnamefont {A.~D.}\ \bibnamefont {Mirlin}},\ }\bibfield  {title} {\bibinfo {title} {Transport signatures of fractional quantum hall binding transitions},\ }\href {https://doi.org/10.1103/PhysRevB.107.245405} {\bibfield  {journal} {\bibinfo  {journal} {Phys. Rev. B}\ }\textbf {\bibinfo {volume} {107}},\ \bibinfo {pages} {245405} (\bibinfo {year} {2023})}\BibitemShut {NoStop}%
\bibitem [{\citenamefont {Giamarchi}\ and\ \citenamefont {Schulz}(1988)}]{Giamarchi1988Anderson}%
  \BibitemOpen
  \bibfield  {author} {\bibinfo {author} {\bibfnamefont {T.}~\bibnamefont {Giamarchi}}\ and\ \bibinfo {author} {\bibfnamefont {H.~J.}\ \bibnamefont {Schulz}},\ }\bibfield  {title} {\bibinfo {title} {Anderson localization and interactions in one-dimensional metals},\ }\href {https://doi.org/10.1103/PhysRevB.37.325} {\bibfield  {journal} {\bibinfo  {journal} {Phys. Rev. B}\ }\textbf {\bibinfo {volume} {37}},\ \bibinfo {pages} {325} (\bibinfo {year} {1988})}\BibitemShut {NoStop}%
\bibitem [{\citenamefont {Wen}(1991)}]{Wen_edge_1991}%
  \BibitemOpen
  \bibfield  {author} {\bibinfo {author} {\bibfnamefont {X.~G.}\ \bibnamefont {Wen}},\ }\bibfield  {title} {\bibinfo {title} {Edge transport properties of the fractional quantum {Hall} states and weak-impurity scattering of a one-dimensional charge-density wave},\ }\href {https://doi.org/10.1103/PhysRevB.44.5708} {\bibfield  {journal} {\bibinfo  {journal} {Phys. Rev. B}\ }\textbf {\bibinfo {volume} {44}},\ \bibinfo {pages} {5708} (\bibinfo {year} {1991})}\BibitemShut {NoStop}%
\bibitem [{\citenamefont {Moore}\ and\ \citenamefont {Wen}(1998)}]{Moore_Classification_1997}%
  \BibitemOpen
  \bibfield  {author} {\bibinfo {author} {\bibfnamefont {J.~E.}\ \bibnamefont {Moore}}\ and\ \bibinfo {author} {\bibfnamefont {X.~G.}\ \bibnamefont {Wen}},\ }\bibfield  {title} {\bibinfo {title} {Classification of disordered phases of quantum {Hall} edge states},\ }\href {https://doi.org/10.1103/PhysRevB.57.10138} {\bibfield  {journal} {\bibinfo  {journal} {Phys. Rev. B}\ }\textbf {\bibinfo {volume} {57}},\ \bibinfo {pages} {10138} (\bibinfo {year} {1998})}\BibitemShut {NoStop}%
\bibitem [{\citenamefont {Kane}\ and\ \citenamefont {Fisher}(1995{\natexlab{b}})}]{Kane1995}%
  \BibitemOpen
  \bibfield  {author} {\bibinfo {author} {\bibfnamefont {C.~L.}\ \bibnamefont {Kane}}\ and\ \bibinfo {author} {\bibfnamefont {M.~P.~A.}\ \bibnamefont {Fisher}},\ }\bibfield  {title} {\bibinfo {title} {Contacts and edge-state equilibration in the fractional quantum hall effect},\ }\href {https://doi.org/10.1103/PhysRevB.52.17393} {\bibfield  {journal} {\bibinfo  {journal} {Phys. Rev. B}\ }\textbf {\bibinfo {volume} {52}},\ \bibinfo {pages} {17393} (\bibinfo {year} {1995}{\natexlab{b}})}\BibitemShut {NoStop}%
\bibitem [{\citenamefont {Johnson}\ and\ \citenamefont {MacDonald}(1991)}]{Johnson1991composite}%
  \BibitemOpen
  \bibfield  {author} {\bibinfo {author} {\bibfnamefont {M.~D.}\ \bibnamefont {Johnson}}\ and\ \bibinfo {author} {\bibfnamefont {A.~H.}\ \bibnamefont {MacDonald}},\ }\bibfield  {title} {\bibinfo {title} {Composite edges in the \ensuremath{\nu}=2/3 fractional quantum hall effect},\ }\href {https://doi.org/10.1103/PhysRevLett.67.2060} {\bibfield  {journal} {\bibinfo  {journal} {Phys. Rev. Lett.}\ }\textbf {\bibinfo {volume} {67}},\ \bibinfo {pages} {2060} (\bibinfo {year} {1991})}\BibitemShut {NoStop}%
\bibitem [{\citenamefont {B\"uttiker}(1988)}]{Buettiker1988absence}%
  \BibitemOpen
  \bibfield  {author} {\bibinfo {author} {\bibfnamefont {M.}~\bibnamefont {B\"uttiker}},\ }\bibfield  {title} {\bibinfo {title} {Absence of backscattering in the quantum hall effect in multiprobe conductors},\ }\href {https://doi.org/10.1103/PhysRevB.38.9375} {\bibfield  {journal} {\bibinfo  {journal} {Phys. Rev. B}\ }\textbf {\bibinfo {volume} {38}},\ \bibinfo {pages} {9375} (\bibinfo {year} {1988})}\BibitemShut {NoStop}%
\bibitem [{\citenamefont {Sen}\ and\ \citenamefont {Agarwal}(2008)}]{Sen2008line}%
  \BibitemOpen
  \bibfield  {author} {\bibinfo {author} {\bibfnamefont {D.}~\bibnamefont {Sen}}\ and\ \bibinfo {author} {\bibfnamefont {A.}~\bibnamefont {Agarwal}},\ }\bibfield  {title} {\bibinfo {title} {Line junction in a quantum hall system with two filling fractions},\ }\href {https://doi.org/10.1103/PhysRevB.78.085430} {\bibfield  {journal} {\bibinfo  {journal} {Phys. Rev. B}\ }\textbf {\bibinfo {volume} {78}},\ \bibinfo {pages} {085430} (\bibinfo {year} {2008})}\BibitemShut {NoStop}%
\bibitem [{\citenamefont {Gornyi}\ \emph {et~al.}(2007)\citenamefont {Gornyi}, \citenamefont {Mirlin},\ and\ \citenamefont {Polyakov}}]{Gornyi2007electron}%
  \BibitemOpen
  \bibfield  {author} {\bibinfo {author} {\bibfnamefont {I.~V.}\ \bibnamefont {Gornyi}}, \bibinfo {author} {\bibfnamefont {A.~D.}\ \bibnamefont {Mirlin}},\ and\ \bibinfo {author} {\bibfnamefont {D.~G.}\ \bibnamefont {Polyakov}},\ }\bibfield  {title} {\bibinfo {title} {Electron transport in a disordered luttinger liquid},\ }\href {https://doi.org/10.1103/PhysRevB.75.085421} {\bibfield  {journal} {\bibinfo  {journal} {Phys. Rev. B}\ }\textbf {\bibinfo {volume} {75}},\ \bibinfo {pages} {085421} (\bibinfo {year} {2007})}\BibitemShut {NoStop}%
\bibitem [{\citenamefont {Cano}\ \emph {et~al.}(2014)\citenamefont {Cano}, \citenamefont {Cheng}, \citenamefont {Mulligan}, \citenamefont {Nayak}, \citenamefont {Plamadeala},\ and\ \citenamefont {Yard}}]{Cano_Bulk_edge_2014}%
  \BibitemOpen
  \bibfield  {author} {\bibinfo {author} {\bibfnamefont {J.}~\bibnamefont {Cano}}, \bibinfo {author} {\bibfnamefont {M.}~\bibnamefont {Cheng}}, \bibinfo {author} {\bibfnamefont {M.}~\bibnamefont {Mulligan}}, \bibinfo {author} {\bibfnamefont {C.}~\bibnamefont {Nayak}}, \bibinfo {author} {\bibfnamefont {E.}~\bibnamefont {Plamadeala}},\ and\ \bibinfo {author} {\bibfnamefont {J.}~\bibnamefont {Yard}},\ }\bibfield  {title} {\bibinfo {title} {Bulk-edge correspondence in (2 + 1)-dimensional abelian topological phases},\ }\href {https://doi.org/10.1103/PhysRevB.89.115116} {\bibfield  {journal} {\bibinfo  {journal} {Phys. Rev. B}\ }\textbf {\bibinfo {volume} {89}},\ \bibinfo {pages} {115116} (\bibinfo {year} {2014})}\BibitemShut {NoStop}%
\bibitem [{\citenamefont {Wu}\ and\ \citenamefont {Jain}(2013)}]{Wu2013}%
  \BibitemOpen
  \bibfield  {author} {\bibinfo {author} {\bibfnamefont {Y.-H.}\ \bibnamefont {Wu}}\ and\ \bibinfo {author} {\bibfnamefont {J.~K.}\ \bibnamefont {Jain}},\ }\bibfield  {title} {\bibinfo {title} {Quantum hall effect of two-component bosons at fractional and integral fillings},\ }\href {https://doi.org/10.1103/PhysRevB.87.245123} {\bibfield  {journal} {\bibinfo  {journal} {Phys. Rev. B}\ }\textbf {\bibinfo {volume} {87}},\ \bibinfo {pages} {245123} (\bibinfo {year} {2013})}\BibitemShut {NoStop}%
\bibitem [{\citenamefont {Ponomarenko}\ and\ \citenamefont {Lyanda-Geller}(2023)}]{ponomarenko2023unusual}%
  \BibitemOpen
  \bibfield  {author} {\bibinfo {author} {\bibfnamefont {V.}~\bibnamefont {Ponomarenko}}\ and\ \bibinfo {author} {\bibfnamefont {Y.}~\bibnamefont {Lyanda-Geller}},\ }\href@noop {} {\bibinfo {title} {Unusual quasiparticles and tunneling conductance in quantum point contacts in $\nu=2/3$ fractional quantum hall systems}} (\bibinfo {year} {2023}),\ \Eprint {https://arxiv.org/abs/2311.05142} {arXiv:2311.05142 [cond-mat.mes-hall]} \BibitemShut {NoStop}%
\bibitem [{\citenamefont {V\"ayrynen}\ \emph {et~al.}(2022)\citenamefont {V\"ayrynen}, \citenamefont {Goldstein},\ and\ \citenamefont {Gefen}}]{Jukka2022}%
  \BibitemOpen
  \bibfield  {author} {\bibinfo {author} {\bibfnamefont {J.~I.}\ \bibnamefont {V\"ayrynen}}, \bibinfo {author} {\bibfnamefont {M.}~\bibnamefont {Goldstein}},\ and\ \bibinfo {author} {\bibfnamefont {Y.}~\bibnamefont {Gefen}},\ }\bibfield  {title} {\bibinfo {title} {Superconductivity of neutral modes in quantum hall edges},\ }\href {https://doi.org/10.1103/PhysRevB.105.L081402} {\bibfield  {journal} {\bibinfo  {journal} {Phys. Rev. B}\ }\textbf {\bibinfo {volume} {105}},\ \bibinfo {pages} {L081402} (\bibinfo {year} {2022})}\BibitemShut {NoStop}%
\bibitem [{\citenamefont {Park}\ \emph {et~al.}()\citenamefont {Park}, \citenamefont {Goldstein}, \citenamefont {Gefen}, \citenamefont {Mirlin},\ and\ \citenamefont {V\"ayrynen}}]{Tobepublished}%
  \BibitemOpen
  \bibfield  {author} {\bibinfo {author} {\bibfnamefont {J.}~\bibnamefont {Park}}, \bibinfo {author} {\bibfnamefont {M.}~\bibnamefont {Goldstein}}, \bibinfo {author} {\bibfnamefont {Y.}~\bibnamefont {Gefen}}, \bibinfo {author} {\bibfnamefont {A.~D.}\ \bibnamefont {Mirlin}},\ and\ \bibinfo {author} {\bibfnamefont {J.~I.}\ \bibnamefont {V\"ayrynen}},\ }\href@noop {} {}\bibinfo {note} {In preparation}\BibitemShut {NoStop}%
\bibitem [{\citenamefont {Naud}\ \emph {et~al.}(2000)\citenamefont {Naud}, \citenamefont {Pryadko},\ and\ \citenamefont {Sondhi}}]{Naud_chiral_sine_2000}%
  \BibitemOpen
  \bibfield  {author} {\bibinfo {author} {\bibfnamefont {J.}~\bibnamefont {Naud}}, \bibinfo {author} {\bibfnamefont {L.~P.}\ \bibnamefont {Pryadko}},\ and\ \bibinfo {author} {\bibfnamefont {S.}~\bibnamefont {Sondhi}},\ }\bibfield  {title} {\bibinfo {title} {Quantum {Hall} bilayers and the chiral sine-gordon equation},\ }\href {https://doi.org/https://doi.org/10.1016/S0550-3213(99)00658-6} {\bibfield  {journal} {\bibinfo  {journal} {Nuclear Physics B}\ }\textbf {\bibinfo {volume} {565}},\ \bibinfo {pages} {572} (\bibinfo {year} {2000})}\BibitemShut {NoStop}%
\bibitem [{\citenamefont {Moore}\ and\ \citenamefont {Read}(1991)}]{Moore_nonabelions_1991}%
  \BibitemOpen
  \bibfield  {author} {\bibinfo {author} {\bibfnamefont {G.}~\bibnamefont {Moore}}\ and\ \bibinfo {author} {\bibfnamefont {N.}~\bibnamefont {Read}},\ }\bibfield  {title} {\bibinfo {title} {Nonabelions in the fractional quantum {Hall} effect},\ }\href {http://www.sciencedirect.com/science/article/pii/055032139190407O} {\bibfield  {journal} {\bibinfo  {journal} {Nucl. Phys. B}\ }\textbf {\bibinfo {volume} {360}},\ \bibinfo {pages} {362} (\bibinfo {year} {1991})}\BibitemShut {NoStop}%
\bibitem [{\citenamefont {Levin}\ \emph {et~al.}(2007)\citenamefont {Levin}, \citenamefont {Halperin},\ and\ \citenamefont {Rosenow}}]{Levin_particle_hole_2007}%
  \BibitemOpen
  \bibfield  {author} {\bibinfo {author} {\bibfnamefont {M.}~\bibnamefont {Levin}}, \bibinfo {author} {\bibfnamefont {B.~I.}\ \bibnamefont {Halperin}},\ and\ \bibinfo {author} {\bibfnamefont {B.}~\bibnamefont {Rosenow}},\ }\bibfield  {title} {\bibinfo {title} {Particle-hole symmetry and the {Pfaffian} state},\ }\href {https://doi.org/10.1103/PhysRevLett.99.236806} {\bibfield  {journal} {\bibinfo  {journal} {Phys. Rev. Lett.}\ }\textbf {\bibinfo {volume} {99}},\ \bibinfo {pages} {236806} (\bibinfo {year} {2007})}\BibitemShut {NoStop}%
\bibitem [{\citenamefont {Lee}\ \emph {et~al.}(2007)\citenamefont {Lee}, \citenamefont {Ryu}, \citenamefont {Nayak},\ and\ \citenamefont {Fisher}}]{Lee_particle_hole_2007}%
  \BibitemOpen
  \bibfield  {author} {\bibinfo {author} {\bibfnamefont {S.~S.}\ \bibnamefont {Lee}}, \bibinfo {author} {\bibfnamefont {S.}~\bibnamefont {Ryu}}, \bibinfo {author} {\bibfnamefont {C.}~\bibnamefont {Nayak}},\ and\ \bibinfo {author} {\bibfnamefont {M.~P.~A.}\ \bibnamefont {Fisher}},\ }\bibfield  {title} {\bibinfo {title} {Particle-hole symmetry and the $\nu=$~5/2 quantum {Hall} state},\ }\href {https://doi.org/10.1103/PhysRevLett.99.236807} {\bibfield  {journal} {\bibinfo  {journal} {Phys. Rev. Lett.}\ }\textbf {\bibinfo {volume} {99}},\ \bibinfo {pages} {236807} (\bibinfo {year} {2007})}\BibitemShut {NoStop}%
\bibitem [{\citenamefont {Son}(2015)}]{Son_is_2015}%
  \BibitemOpen
  \bibfield  {author} {\bibinfo {author} {\bibfnamefont {D.~T.}\ \bibnamefont {Son}},\ }\bibfield  {title} {\bibinfo {title} {Is the composite {Fermion} a {Dirac} particle?},\ }\href {https://doi.org/10.1103/PhysRevX.5.031027} {\bibfield  {journal} {\bibinfo  {journal} {Phys. Rev. X}\ }\textbf {\bibinfo {volume} {5}},\ \bibinfo {pages} {031027} (\bibinfo {year} {2015})}\BibitemShut {NoStop}%
\bibitem [{\citenamefont {Simon}(2018)}]{Simon_equilibration_2018}%
  \BibitemOpen
  \bibfield  {author} {\bibinfo {author} {\bibfnamefont {S.~H.}\ \bibnamefont {Simon}},\ }\bibfield  {title} {\bibinfo {title} {Interpretation of thermal conductance of the $\nu=$~5/2 edge},\ }\href {https://doi.org/10.1103/PhysRevB.97.121406} {\bibfield  {journal} {\bibinfo  {journal} {Phys. Rev. B}\ }\textbf {\bibinfo {volume} {97}},\ \bibinfo {pages} {121406(R)} (\bibinfo {year} {2018})}\BibitemShut {NoStop}%
\bibitem [{\citenamefont {Feldman}(2018)}]{Feldman_comment_2018}%
  \BibitemOpen
  \bibfield  {author} {\bibinfo {author} {\bibfnamefont {D.~E.}\ \bibnamefont {Feldman}},\ }\bibfield  {title} {\bibinfo {title} {Comment on ``{Interpretation} of thermal conductance of the $\nu=$~5/2 edge''},\ }\href {https://doi.org/10.1103/PhysRevB.98.167401} {\bibfield  {journal} {\bibinfo  {journal} {Phys. Rev. B}\ }\textbf {\bibinfo {volume} {98}},\ \bibinfo {pages} {167401} (\bibinfo {year} {2018})}\BibitemShut {NoStop}%
\bibitem [{\citenamefont {Ma}\ and\ \citenamefont {Feldman}(2019)}]{Feldman_equilibration_2019}%
  \BibitemOpen
  \bibfield  {author} {\bibinfo {author} {\bibfnamefont {K.~K.~W.}\ \bibnamefont {Ma}}\ and\ \bibinfo {author} {\bibfnamefont {D.~E.}\ \bibnamefont {Feldman}},\ }\bibfield  {title} {\bibinfo {title} {Partial equilibration of integer and fractional edge channels in the thermal quantum {Hall} effect},\ }\href {https://doi.org/10.1103/PhysRevB.99.085309} {\bibfield  {journal} {\bibinfo  {journal} {Phys. Rev. B}\ }\textbf {\bibinfo {volume} {99}},\ \bibinfo {pages} {085309} (\bibinfo {year} {2019})}\BibitemShut {NoStop}%
\bibitem [{\citenamefont {Asasi}\ and\ \citenamefont {Mulligan}(2020)}]{Asasi_equilibration_2020}%
  \BibitemOpen
  \bibfield  {author} {\bibinfo {author} {\bibfnamefont {H.}~\bibnamefont {Asasi}}\ and\ \bibinfo {author} {\bibfnamefont {M.}~\bibnamefont {Mulligan}},\ }\bibfield  {title} {\bibinfo {title} {Partial equilibration of anti-{Pfaffian} edge modes at $\ensuremath{\nu}=5/2$},\ }\href {https://doi.org/10.1103/PhysRevB.102.205104} {\bibfield  {journal} {\bibinfo  {journal} {Phys. Rev. B}\ }\textbf {\bibinfo {volume} {102}},\ \bibinfo {pages} {205104} (\bibinfo {year} {2020})}\BibitemShut {NoStop}%
\bibitem [{\citenamefont {Park}\ \emph {et~al.}(2020)\citenamefont {Park}, \citenamefont {Sp\aa{}nsl\"att}, \citenamefont {Gefen},\ and\ \citenamefont {Mirlin}}]{Park_noise_2020}%
  \BibitemOpen
  \bibfield  {author} {\bibinfo {author} {\bibfnamefont {J.}~\bibnamefont {Park}}, \bibinfo {author} {\bibfnamefont {C.}~\bibnamefont {Sp\aa{}nsl\"att}}, \bibinfo {author} {\bibfnamefont {Y.}~\bibnamefont {Gefen}},\ and\ \bibinfo {author} {\bibfnamefont {A.~D.}\ \bibnamefont {Mirlin}},\ }\bibfield  {title} {\bibinfo {title} {Noise on the non-{Abelian} $\ensuremath{\nu}=5/2$ fractional quantum {Hall} edge},\ }\href {https://doi.org/10.1103/PhysRevLett.125.157702} {\bibfield  {journal} {\bibinfo  {journal} {Phys. Rev. Lett.}\ }\textbf {\bibinfo {volume} {125}},\ \bibinfo {pages} {157702} (\bibinfo {year} {2020})}\BibitemShut {NoStop}%
\bibitem [{\citenamefont {Yutushui}\ and\ \citenamefont {Mross}(2023)}]{Yutushui_Identifying_2023}%
  \BibitemOpen
  \bibfield  {author} {\bibinfo {author} {\bibfnamefont {M.}~\bibnamefont {Yutushui}}\ and\ \bibinfo {author} {\bibfnamefont {D.~F.}\ \bibnamefont {Mross}},\ }\bibfield  {title} {\bibinfo {title} {Identifying non-abelian anyons with upstream noise},\ }\href {https://doi.org/10.1103/PhysRevB.108.L241102} {\bibfield  {journal} {\bibinfo  {journal} {Phys. Rev. B}\ }\textbf {\bibinfo {volume} {108}},\ \bibinfo {pages} {L241102} (\bibinfo {year} {2023})}\BibitemShut {NoStop}%
\bibitem [{\citenamefont {Manna}\ and\ \citenamefont {Das}(2023)}]{Manna_Experimentally_2023}%
  \BibitemOpen
  \bibfield  {author} {\bibinfo {author} {\bibfnamefont {S.}~\bibnamefont {Manna}}\ and\ \bibinfo {author} {\bibfnamefont {A.}~\bibnamefont {Das}},\ }\href@noop {} {\bibinfo {title} {Experimentally motivated order of length scales affect shot noise}} (\bibinfo {year} {2023}),\ \Eprint {https://arxiv.org/abs/2307.08264} {arXiv:2307.08264 [cond-mat.mes-hall]} \BibitemShut {NoStop}%
\bibitem [{\citenamefont {Manna}\ \emph {et~al.}(2024)\citenamefont {Manna}, \citenamefont {Das}, \citenamefont {Goldstein},\ and\ \citenamefont {Gefen}}]{Manna_Full_2024}%
  \BibitemOpen
  \bibfield  {author} {\bibinfo {author} {\bibfnamefont {S.}~\bibnamefont {Manna}}, \bibinfo {author} {\bibfnamefont {A.}~\bibnamefont {Das}}, \bibinfo {author} {\bibfnamefont {M.}~\bibnamefont {Goldstein}},\ and\ \bibinfo {author} {\bibfnamefont {Y.}~\bibnamefont {Gefen}},\ }\bibfield  {title} {\bibinfo {title} {Full classification of transport on an equilibrated $5/2$ edge via shot noise},\ }\href {https://doi.org/10.1103/PhysRevLett.132.136502} {\bibfield  {journal} {\bibinfo  {journal} {Phys. Rev. Lett.}\ }\textbf {\bibinfo {volume} {132}},\ \bibinfo {pages} {136502} (\bibinfo {year} {2024})}\BibitemShut {NoStop}%
\bibitem [{\citenamefont {Hein}\ and\ \citenamefont {Sp\aa{}nsl\"att}(2023)}]{Hein_Thermal_2023}%
  \BibitemOpen
  \bibfield  {author} {\bibinfo {author} {\bibfnamefont {M.}~\bibnamefont {Hein}}\ and\ \bibinfo {author} {\bibfnamefont {C.}~\bibnamefont {Sp\aa{}nsl\"att}},\ }\bibfield  {title} {\bibinfo {title} {Thermal conductance and noise of majorana modes along interfaced $\ensuremath{\nu}=\frac{5}{2}$ fractional quantum hall states},\ }\href {https://doi.org/10.1103/PhysRevB.107.245301} {\bibfield  {journal} {\bibinfo  {journal} {Phys. Rev. B}\ }\textbf {\bibinfo {volume} {107}},\ \bibinfo {pages} {245301} (\bibinfo {year} {2023})}\BibitemShut {NoStop}%
\bibitem [{\citenamefont {Bonderson}\ and\ \citenamefont {Slingerland}(2008)}]{Bonderson_hierarchy_2008}%
  \BibitemOpen
  \bibfield  {author} {\bibinfo {author} {\bibfnamefont {P.}~\bibnamefont {Bonderson}}\ and\ \bibinfo {author} {\bibfnamefont {J.~K.}\ \bibnamefont {Slingerland}},\ }\bibfield  {title} {\bibinfo {title} {Fractional quantum {Hall} hierarchy and the second {Landau} level},\ }\href {https://doi.org/10.1103/PhysRevB.78.125323} {\bibfield  {journal} {\bibinfo  {journal} {Phys. Rev. B}\ }\textbf {\bibinfo {volume} {78}},\ \bibinfo {pages} {125323} (\bibinfo {year} {2008})}\BibitemShut {NoStop}%
\bibitem [{\citenamefont {Yutushui}\ \emph {et~al.}(2022)\citenamefont {Yutushui}, \citenamefont {Stern},\ and\ \citenamefont {Mross}}]{MY_Identifying_2022}%
  \BibitemOpen
  \bibfield  {author} {\bibinfo {author} {\bibfnamefont {M.}~\bibnamefont {Yutushui}}, \bibinfo {author} {\bibfnamefont {A.}~\bibnamefont {Stern}},\ and\ \bibinfo {author} {\bibfnamefont {D.~F.}\ \bibnamefont {Mross}},\ }\bibfield  {title} {\bibinfo {title} {Identifying the $\ensuremath{\nu}=\frac{5}{2}$ topological order through charge transport measurements},\ }\href {https://doi.org/10.1103/PhysRevLett.128.016401} {\bibfield  {journal} {\bibinfo  {journal} {Phys. Rev. Lett.}\ }\textbf {\bibinfo {volume} {128}},\ \bibinfo {pages} {016401} (\bibinfo {year} {2022})}\BibitemShut {NoStop}%
\end{thebibliography}%

\end{document}